\documentclass[12pt]{report}

\usepackage{tamuconfig}

\usepackage{times}
\usepackage[T1]{fontenc}

\usepackage{graphicx}

\DeclareGraphicsExtensions{.PNG}

\graphicspath{ {./figures/} }

\usepackage[hidelinks]{hyperref}

\usepackage{amssymb}
\usepackage{tabularx}
\usepackage{longtable}
\usepackage{subcaption}
\usepackage{booktabs}
\usepackage{algorithm}
\usepackage{algorithmic}
\usepackage[symbol]{footmisc}
\makeatletter
\renewcommand\paragraph{%
  \@startsection{paragraph}{4}{4pt}%
  {4pt}
  {0em}%
  {\normalfont\normalsize\itshape}%
}
\makeatother

\begin{document}

\renewcommand{\tamumanuscripttitle}{Deciphering Cell Systems: Machine Learning Perspectives and Approaches for the Analysis of Single-Cell Data}

\renewcommand{\tamupapertype}{Dissertation}

\renewcommand{\tamufullname}{Yongjian Yang}

\renewcommand{\tamudegree}{Doctor of Philosophy}
\renewcommand{\tamuchairone}{James Cai}

\renewcommand{\tamumemberone}{Robert Chapkin}
\newcommand{\tamumembertwo}{Xiaoning Qian}
\newcommand{\tamumemberthree}{Aniruddha Datta}
\renewcommand{\tamudepthead}{Costas Georghiades}

\renewcommand{\tamugradmonth}{May}
\renewcommand{\tamugradyear}{2024}
\renewcommand{\tamudepartment}{Electrical Engineering}


\providecommand{\tabularnewline}{\\}

\begin{titlepage}
\begin{center}
\begin{doublespace}

\MakeUppercase{  \tamumanuscripttitle}
\end{doublespace}
\vspace{4em}

A \tamupapertype

by

\MakeUppercase{\tamufullname}

\vspace{4em}

\begin{singlespace}

Submitted to the Graduate and Professional School of \\
Texas A\&M University \\

in partial fulfillment of the requirements for the degree of \\
\end{singlespace}

\MakeUppercase{\tamudegree}
\par\end{center}
\vspace{2em}
\begin{doublespace}

\end{doublespace}
\begin{singlespace}
\begin{tabular}{ll}
 & \tabularnewline
& \cr
Chair of Committee, & \tamuchairone\tabularnewline
Committee Members, & \tamumemberone\tabularnewline
 & \tamumembertwo\tabularnewline
 & \tamumemberthree\tabularnewline
Head of Department, & \tamudepthead\tabularnewline
\end{tabular}
\end{singlespace}

\vspace{3em}

\begin{center}
\tamugradmonth \hspace{2pt} \tamugradyear

\vspace{3em}

Major Subject: \tamudepartment \par
\vspace{3em}
Copyright \tamugradyear \hspace{.5em}\tamufullname 
\par\end{center}
\end{titlepage}
\pagebreak{}



\chapter*{ABSTRACT}
\addcontentsline{toc}{chapter}{ABSTRACT} 

\pagestyle{plain} 
\pagenumbering{roman} 
\setcounter{page}{2}

\indent This doctoral dissertation delves into the application of machine learning techniques in molecular biology, exploring gene expression regulation at the single-cell level and navigating the intricacies of cellular biology. The study specifically focuses on the utilization of modern neural networks to address cell-cell communications, gene function inference, and decipher protein expression. These applications aim to elucidate the complex interactions governing cellular behavior, as evidenced by the analysis of single-cell RNA sequencing (scRNA-seq) data.

In pursuit of these goals, I have developed and implemented advanced computational methodologies that combine systems biology and modern neural networks techniques. These methods are specifically crafted to manage the high-dimensionality and complexity of single-cell data, facilitating a more nuanced comprehension of genotype-phenotype relationships.

This research makes a significant contribution to the field of computational biology by proposing the use of neural networks to tackle the longstanding optimization problem in manifold learning. Furthermore, the study investigates generative models for learning gene regulatory networks and simulates gene knockout at the single-cell resolution. Lastly, the research delves into enhancing the interpretability of black box neural network models, applying them to multimodality data.

This research also contributes to the cell biology field by first providing an in-depth analysis of cell-cell interactions, highlighting how these interactions shape cellular behavior and influence disease progression. In addition, this research investigates gene function prediction, focusing on how gene knockouts can affect cellular phenotypes and their potential therapeutic implications. Lastly, this research looks into how gene expression patterns translate into protein expression and how accurately and interpretably this translation process can be predicted. This aspect of this research yields important insights into the functional implications of gene expression, which may be applied to the understanding of disease mechanisms and drug responses.

This research serves as a valuable resource because, in addition to the three introduced tools, it provides a comprehensive overview of state-of-the-art methodologies and their respective applications in the analysis of single-cell data within the recent years. 

In conclusion, this doctoral dissertation represents a significant contribution to the field of computational biology and cellular biology by providing novel methods and insights into the genotype-phenotype relationships at the single cell level. These methods and discoveries not only improve our understanding of cellular behavior, but also pave the way for the creation of novel therapeutic strategies, thereby potentially enhancing our ability to combat a wide range of diseases.

\pagebreak{}

\chapter*{DEDICATION}
\addcontentsline{toc}{chapter}{DEDICATION}  

\begin{center}
\vspace*{\fill}
To my dear family.
\vspace*{\fill}
\end{center}

\pagebreak{}

%
%
%
%
%

\chapter*{ACKNOWLEDGMENTS}
\addcontentsline{toc}{chapter}{ACKNOWLEDGMENTS}  

\indent I would like to express my deepest gratitude to Dr. Cai for his unwavering support, encouragement and trust during my time in the Cai lab. I would also like to thank my committee members, Drs. Chapkin, Qian, and Datta, for their insightful comments, which are instrumental in improving the quality of my research. Additionally, I am grateful to former and current lab members for their friendship and contributions to our enlightening interdisciplinary discussions. Lastly, I would like to convey my heartfelt appreciation to my family and friends for their unrelenting backing, which makes my decision to return to campus feasible and fruitful.

\pagebreak{}
%
%
%
%
%

\chapter*{CONTRIBUTORS AND FUNDING SOURCES}
\addcontentsline{toc}{chapter}{CONTRIBUTORS AND FUNDING SOURCES}  

\subsection*{Contributors}
This work was supported by a doctoral thesis committee consisting of Dr. James J. Cai (J.J.C.) of Department of Veterinary Integrative Biosciences, Interdisciplinary Program of Genetics, and Department of Electrical and Computer Engineering who served as the committee chair, and Dr. Robert S. Chapkin (R.S.C.) of Department of Nutrition, Dr. Xiaoning Qian and Dr. Aniruddha Datta of Department of Electrical and Computer Engineering who served as the committee members.

In Chapter \ref{chap:xct}, J.J.C. conceived the study and designed the workflow. Yongjian Yang (Y.Y.) designed the workflow, performed data analysis, and implemented the software tool. Yan Zhong (Y.Z.) and Guanxun Li (G.L.) contributed to the workflow design. Qian Xu (Q.X.) contributed to the preparation of testing data sets. Yu-te Lin (Y.L.) contributed to the software tool implementation. Y.Y., G.L., Q.X. and R.S.C. contributed to the writing of the manuscript. R.S.C. and J.J.C. supervised the study.
In Chapter \ref{chap:genki}, Y.Y. conceived the study and designed the workflow, designed the workflow, performed data analysis, and implemented the software tool. Y.Z. and G.L. contributed to the workflow design. Bo-jia Chen (B.C.) contributed to the preparation of testing data sets. Y.Y., G.L., Q.X. B.C. and R.S.C. contributed to the writing of the manuscript. J.J.C. supervised the study.
In Chapter \ref{chap:crossmodal}, Y.Y. conceived the idea, designed and implemented the study; Y.L. performed the baseline evaluation; G.L. and Y.Z. provided related methodology support; Q.X. provided related biology support; Y.Y., G.L., Y.L., Q.X. and J.J.C. wrote the manuscript.

All other work conducted for this doctoral thesis was completed by Y.Y. independently.

\subsection*{Funding Sources}
This research was funded by Texas A\&M University 2019 X-Grants and DoD grant GW200026 for J.J.C; the National Institute of Environmental Health Sciences (P30 ES029067), the National Cancer Institute (R35 CA197707, RO1 CA245514), the Allen Endowed Chair in Nutrition \& Chronic Disease Prevention, and support from the Institute for Advancing Health through Agriculture for R.S.C.

\pagebreak{}

%
%
%
%
%

\phantomsection
\addcontentsline{toc}{chapter}{TABLE OF CONTENTS}  

\begin{singlespace}
\renewcommand\contentsname{\normalfont} {\centerline{TABLE OF CONTENTS}}

\setcounter{tocdepth}{4} 

\setlength{\cftaftertoctitleskip}{1em}
\renewcommand{\cftaftertoctitle}{%
\hfill{\normalfont {Page}\par}}

\tableofcontents

\end{singlespace}

\pagebreak{}


\phantomsection
\addcontentsline{toc}{chapter}{LIST OF FIGURES}  

\renewcommand{\cftloftitlefont}{\center\normalfont\MakeUppercase}

\setlength{\cftbeforeloftitleskip}{-12pt} 
\renewcommand{\cftafterloftitleskip}{12pt}

\renewcommand{\cftafterloftitle}{%
\\[4em]\mbox{}\hspace{2pt}FIGURE\hfill{\normalfont Page}\vskip\baselineskip}

\begingroup

\begin{center}
\begin{singlespace}
\setlength{\cftbeforechapskip}{0.4cm}
\setlength{\cftbeforesecskip}{0.30cm}
\setlength{\cftbeforesubsecskip}{0.30cm}
\setlength{\cftbeforefigskip}{0.4cm}
\setlength{\cftbeforetabskip}{0.4cm}



\listoffigures

\end{singlespace}
\end{center}

\pagebreak{}

%
\phantomsection
\addcontentsline{toc}{chapter}{LIST OF TABLES}  

\renewcommand{\cftlottitlefont}{\center\normalfont\MakeUppercase}

\setlength{\cftbeforelottitleskip}{-12pt} 

\renewcommand{\cftafterlottitleskip}{1pt}

\renewcommand{\cftafterlottitle}{%
\\[4em]\mbox{}\hspace{2pt}TABLE\hfill{\normalfont Page}\vskip\baselineskip}

\begin{center}
\begin{singlespace}

\setlength{\cftbeforechapskip}{0.4cm}
\setlength{\cftbeforesecskip}{0.30cm}
\setlength{\cftbeforesubsecskip}{0.30cm}
\setlength{\cftbeforefigskip}{0.4cm}
\setlength{\cftbeforetabskip}{0.4cm}

\listoftables 

\end{singlespace}
\end{center}
\endgroup
\pagebreak{}  


\pagestyle{plain} 
\pagenumbering{arabic} 
\setcounter{page}{1}

\chapter{INTRODUCTION}\label{chap:intro}

Single-cell technology has ushered in a new era of biomedical research, providing unprecedented insight into the complex world of cellular biology. Single-cell RNA sequencing (scRNA-seq) stands out among these revolutionary technologies as a potent technique that has revolutionized our ability to decode the transcriptomic landscape of individual cells. We can now capture gene expression profiles from tens of thousands of cells with remarkable precision and reproducibility using scRNA-seq \cite{macosko2015highly, zilionis2017single, zheng2017massively}.

This newly discovered capability for high-resolution profiling has unlocked a wealth of cellular data, allowing us to uncover previously hidden layers of tissue complexity. Using scRNA-seq data, researchers have deciphered the complexities of cellular diversity, revealing the existence of a variety of cell types and cellular states within complex biological systems \cite{aldridge2020single}. This has fundamentally changed our understanding of the heterogeneity underlying the function of tissues.

The primary objective of this study is to use the abundant single-cell data resource and apply systems biology methods along with modern machine learning techniques to gain a deeper understanding of various biological processes. Specifically, the goal of our research is to identify genes that play key roles in determining phenotype outcomes and to elucidate their contributions to three important biological processes: cell-cell communication, gene knockout, and gene-protein translation. 

Cell-Cell Communication: One aspect of this study centers on how genes mediate cell-cell communication. Understanding the genes that regulate this process is crucial for comprehending how information is exchanged within and between cells. 
We suggest employing feed-forward neural networks to tackle the manifold learning optimization problem and establish a new framework for inferring differential interactions between two conditions.

Gene Knockout: Another dimension of this study explores how genes are involved in gene knockout scenarios. By pinpointing the genes responsible for these perturbations, we gain insights into how cellular functions are impacted when specific genes are rendered non-functional.
We introduce graph neural networks to obtain effective representations of genes at the gene regulatory network level and infer potential gene functions without the necessity of actual knockout experiments.

Gene-Protein Translation: In addition, this research will attempt to decipher the genetic regulation of gene-protein translation. This process is fundamental to the production of functional proteins, and this study seeks to identify the genes that govern this critical step.
We specifically introduce a generator-discriminator framework to enhance the interpretability of our model for multimodality data.

By investigating these processes at the single cell level, we hope to reveal the complex map of interactions that governs cellular behavior. Understanding how genes orchestrate cell-cell communication, regulate essential cellular functions, and influence gene-protein translation will yield essential insights into the underlying mechanisms of biological systems.

The findings derived from this research have the potential to make substantial contributions to the field of cellular biology. By illuminating the molecular drivers of these processes, we pave the way for the development of novel therapeutic strategies. Ultimately, by combining scRNA-seq and cutting-edge machine learning techniques, this research is expected to uncover novel intervention targets and enhance our capacity to combat a wide variety of diseases.

The findings from this research also hold significant potential contributions to the field of computational biology. Through the application of modern neural networks to single-cell data, we investigate an alternative approach—utilizing neural networks instead of traditional methods—to address the manifold learning optimization problem. Moreover, we leverage the capabilities of generative models to acquire effective gene representations by incorporating graph information. We employ an interpretable approach to enhance interpretability, especially in the context of multimodality data.

The subsequent chapters of this doctoral dissertation are structured around research tasks. Over the following three chapters, I will elucidate three deep learning-based computational tools designed to address these pivotal issues.
In Chapter \ref{chap:xct}, I will present a semi-supervised approach known as scTenifoldXct, which is dedicated to predicting cell-cell interactions leveraging single-cell gene regulatory network level information. This method harnesses neural networks to tackle the manifold alignment optimization problem.
Chapter \ref{chap:genki} will introduce GenKI, an novel tool for inferring gene knockouts by learning single-cell gene regulatory networks, all without the need for real knockout data.
Chapter \ref{chap:crossmodal} will delve into the exploration of relationships and interpretations of gene-protein translation, utilizing multimodal single-cell data.
The organization of these chapters aims to provide a comprehensive understanding of the research contributions made within the context of this dissertation.

\chapter[scTenifoldXct: A SEMI-SUPERVISED METHOD FOR PREDICTING CELL-CELL INTERACTIONS]{scTenifoldXct: A SEMI-SUPERVISED METHOD FOR PREDICTING CELL-CELL INTERACTIONS\footnote{Reprinted with permission from Y. Yang et al., "scTenifoldXct: A semi-supervised method for predicting cell-cell interactions and mapping cellular communication graphs," Cell Systems, vol. 14, no. 4, pp. 302-311, 2023.}}\label{chap:xct}

\section{Introduction}
With scRNA-seq data, it is possible to study the cellular communication network through the mapping of cell-specific ligand-receptor (LR) connectivity in complex tissues \cite{turei2021integrated, armingol2021deciphering}. The evolving scRNA-seq data space has sparked the development of numerous computational tools for mining cell-to-cell communication information \cite{hou2020predicting, cabello2020singlecellsignalr, wang2019italk, jin2021inference, raredon2022computation, efremova2020cellphonedb, ramilowski2015draft, noel2021dissection, wang2020talklr, browaeys2020nichenet, cheng2021inferring, hu2021cytotalk}. Nevertheless, robust statistical confidence in detecting results has proved difficult to achieve \cite{almet2021landscape, dimitrov2022comparison}.

This chapter presents scTenifoldXct—a computational tool that incorporates intra- and inter-cellular gene networks to detect cell-cell interaction using scRNA-seq data. scTenifoldXct is semi-supervised and thus can be used with or without reference LR pairs. A redesigned crosstalk scoring metric is also introduced to estimate the interaction strength of each LR pair. The crosstalk scoring metric extends the commonly used metric, i.e., the product of ligand and receptor gene expressions, which is adopted by almost all existing methods. The built-in functionality involving single-cell gene network construction allows scTenifoldXct to couple cell-cell interactions with intracellular activities. scTenifoldXct is also able to perform differential interaction analysis, in which cell-cell interaction patterns are compared between tissue samples. When performing comparative analysis, scTenifoldXct combines and analyzes data from two samples in an integrative manner rather than processing the two samples separately, making the analysis more powerful in detecting subtle differential interactions. To demonstrate these features of scTenifoldXct, we applied scTenifoldXct to scRNA-seq data sets under the single-sample and two-sample application settings.

\section{Methods}

\subsection{Data preprocessing}

The input of scTenifoldXct is a gene-by-cell count matrix with annotated cell types. In addition, sample information is required for the comparative analysis. For all scRNA-seq data sets, we performed log-normalization using the \textit{NormalizeData} function in Seurat (v4.0.2) \cite{stuart2019comprehensive}. Highly vari able genes were selected using the \textit{FindVariableFeatures} function in Seurat (selection.method $=$ ``vst"). For each gene, Seurat computed the standardized variance of its expression across cells, controlling for the mean expression \cite{stuart2019comprehensive}. For each  data set, the top 3,000 highly variable genes were included in subsequent analyses. Cell annotations from original studies were retained and used.

\subsection{Network construction}

Given cell types $A$ and $B$, we employed PC regression to construct the intracellular gene regression networks, denoting them as $\mathbf{W}_{A}$ and $\mathbf{W}_{B}$. Specifically, suppose $\mathbf{X}\in\mathbb{R}^{n\times p}$ is the count matrix with $n$ genes and $p$ cells for cell type $A$. The gene expression level of the ${i}^{th}$ gene in all $p$ cells is represented by the $\textit{i}^{th}$ row of $\mathbf{X}$, denoted by $ X_{i}\in\mathbb{R}^{p}$. Denote $\mathbf{X}_{-i}\in\mathbb{R}^{\left(n-1\right)\times p}$ as the matrix by deleting $X_{i}$ from $\mathbf{X}$. To estimate the effects of other $n-1$ genes on the $\textit{i}^{th}$ gene, we constructed the PC regression model for $X_{i}$.
First, we applied principal component analysis (PCA) to $\mathbf{X}_{-i}^{T}$ and selected the first $M$ leading PCs to construct $\mathbf{Z}^{i} =\left(Z_{1}^{i},\cdots, Z_{M}^{i}\right)\in\mathbb{R}^{p\times M}$, where $Z_{m}^{i}\in\mathbb{R}^{p}$ is the $m^{th}$ PC of $\mathbf{X}_{-i}^{T}$, $m = 1, 2, \ldots, M$ and $M\ll min(n, p)$. Because PC regression only uses $M$ PCs as covariates, it reduces the risk of over-fitting and computation time. 
Denote $\mathbf{V}^{i}\in\mathbb{R}^{\left(n-1\right)\times M}$ as the PC loading matrix for the first $M$ leading PCs, then $\mathbf{Z}^{i} =\mathbf{X}_{-i}^{T}\mathbf{V}^{i}$, where $\mathbf{V}^{i}$ satisfies $\left(\mathbf{V}^{i}\right)^{T}\mathbf{V}^{i} =\mathbf{I}_{M}$. The next step was to generate the regression coefficients by regressing $\mathbf{X}_{i}$ on $\mathbf{Z}^{i}$ and solve the following optimization problem:
\begin{equation}
\hat{\beta}^{i} = arg\min_{\beta^{i}\in R^{M}}\left\Vert\mathbf{X}_{\mathbf{i}}-\mathbf{Z}^{i}\beta^{i}\right\Vert_{2}^{2},
\end{equation}
which can be easily solved by the ordinary least square (OLS) method. Then, the effects of the other $n-1$ genes to the \textit{i}\textsuperscript{th} gene were obtained by $\hat{\alpha}^{i} =\mathbf{V}^{i}\hat{\beta}^{i}\in\mathbb{R}^{n-1}$. Repeating this process for another $n-1$ times, with one different gene as the response gene each time, we assembled $\{\hat{\alpha}^{i}\}_{i=1}^{n}$ together and constructed an $n\times n$ weighted adjacency matrix $\mathbf{W}$ of the intracellular gene regression network. The $i^{th}$ row of $\mathbf{W}$ is $\hat{\alpha}^{i}$, and all diagonal entries of $\mathbf{W}$ are all 0. Eventually, the gene regression network $\mathbf{W}_{A}$ for cell type $A$ and gene regression network $\mathbf{W}_{B}$ for cell type $B$ were obtained. $\mathbf{W}_{A}$ and $\mathbf{W}_{B}$ could be symmetrized when necessary.

\subsection{Crosstalk score}

The product of the mean expression of known ligands and receptors has been widely used in most computational approaches to gauge cellular interactions. To account for the gene variance within cell groups, we additionally incorporated gene expression variances in the definition of the crosstalk score between gene $i$ in cell $A$ and gene $j$ in cell $B$, as shown below:
\begin{equation}
score_{iA, jB} =\left[\left(1-\lambda\right)u_{iA}^{2}+\lambda \sigma_{iA}^{2}\right]\times\left[\left(1-\lambda\right)u_{jB}^{2}+\lambda \sigma_{iB}^{2}\right],
\end{equation}
where $u_{iA}, \sigma_{iA}^{2}$ and $ u_{jB}, \sigma_{iB}^{2}$ are the expression mean and variance of gene $i$ in cell $A$ and gene $j$ in cell $B$, respectively. The hyperparameter $\lambda$ scales the relative contribution of mean and variance. By default, $ \lambda = \frac{1}{2}$, and the crosstalk score could be treated as the product of half of the second moment between gene $i$ in cell $A$ and gene $j$ in cell $B$.

\subsection{Manifold alignment}

Manifold alignment is a nonlinear feature projection method by which we embed the genes of sender and receiver cells into a unified space while simultaneously minimizing the distance between corresponding genes and preserving the original structure of gene regression networks of each cell type. Manifold alignment allows the low dimensional projections of genes to be comparable and preserves the information of gene regression networks. To summarize the strength of interactions for each pair of genes across the cell types, we defined the crosstalk score matrix $\mathbf{S}$ with $S_{i,j} = \mu  score_{iA, jB}$, where $ \mu$ is a scale factor indicating how much we want to preserve the weight of intercellular correspondences relative to the edge weight of gene regression networks. Following \cite{vu2012manifold}, we set  $ \mu  = \frac{\sum_{i, j}^{}(\mathbf{W}_{A})_{ij}+\sum_{i, j}^{}(\mathbf{W}_{B})_{ij}}{2\sum_{i, j}^{}(\mathbf{S}_{score})_{ij}}$. In this way we ensured the correspondence and gene regression networks are in a comparable scale so that the manifold alignment result is not biased toward either metric. We found that the above value setting of $\mu$ generated a more robust and unbiased detecting result than other scaling settings, including $\mu = 1$ (unscaled), $0.1\mu$, and $10\mu$ (Supplementary Figure \ref{xct_s4}(B)). The joint similarity matrix is then constructed as follows:
\begin{equation}
\mathbf{ W} = \begin{bmatrix}
\mathbf{W}_{A} &\mathbf{S} \\ 
\mathbf{S}^{\mathbf{T}} &\mathbf{W}_{B} \\ 
\end{bmatrix}     
\end{equation}
Note that $\mathbf{S}$ is asymmetric, while, when $\mathbf{W}_{A}$ and $\mathbf{W}_{B}$ are symmetric, $\mathbf{W}$ is symmetric. $\mathbf{W}_{A}$ and $\mathbf{W}_{B}$ may contain negative values when gene expressions are negatively correlated, and in such cases, the properties of the corresponding Laplacian are not well understood \cite{vu2012manifold}. We resolved this problem by adding 1 to all entries of $\mathbf{W}_{A}$ and $\mathbf{W}_{B}$, transforming the range of $\mathbf{W}_{A}$ and $\mathbf{W}_{B}$ from [-1, 1] to [0, 2]. As a result, the projected features of two genes with a positive correlation would be closer than those with a negative correlation. For convenience, we still used $\mathbf{W}_{A}$ and $\mathbf{W}_{B}$ to denote the transformed gene regression networks of the two cell types. The loss function for this manifold alignment is
\begin{align}
\mathcal{l}(\mathcal{H}) &= \sum_{i, j}^{}\left\Vert\mathcal{H}_{i} -\mathcal{H}_{j}\right\Vert^{2}\mathbf{W}_{ij} \\
&= \sum_{i, j}^{}\left\Vert\mathcal{H}_{i}^{A} -\mathcal{H}_{j}^{A}\right\Vert^{2}\mathbf{W}_{ij}^{A} + \sum_{i, j}^{}\left\Vert\mathcal{H}_{i}^{B} -\mathcal{H}_{j}^{B}\right\Vert^{2}\mathbf{W}_{ij}^{B} + 2\sum_{i, j}^{}\left\Vert\mathcal{H}_{i}^{A} -\mathcal{H}_{j}^{B}\right\Vert^{2}\mathbf{S}_{ij},
\end{align}
where $\mathcal{H} = 
\begin{bmatrix}
\mathcal{H}^{A} \\ 
\mathcal{H}^{B} \\ 
\end{bmatrix}\in\mathbb{R}^{2n\times d}$ 
is the low-dimensional representation for cell type $A$ and cell type $B$, and $d (\ll n)$ is the dimension of the latent space. Moreover, we needed an additional constraint
\begin{equation}
\mathcal{H}^{T}\mathbf{D}\mathcal{H} = \mathbf{I}_{\mathbf{d}}
\end{equation}
for this loss function to work properly, i.e., to avoid the trivial solution of mapping all instances to zero, where $\mathbf{D}$ is a diagonal matrix with $D_{ii} =  \sum_{i}^{}W_{ij}$ and $\mathbf{I}_{\mathbf{d}}$ is a $ d\times d$ identity matrix. Solutions for manifold alignment traditionally rely on the eigen decomposition, which is computationally demanding. To speed up, neural networks were used to learn the unified low-dimensional latent representation $\mathcal{H}$ \cite{nguyen2022deep}. Let $\mathcal{H} = 
\begin{bmatrix}
\mathcal{H}^{A} \\ 
\mathcal{H}^{B} \\ 
\end{bmatrix} =
\begin{bmatrix}
\mathcal{H}^{A}(\cdot, \theta^{A}) \\ 
\mathcal{H}^{B}(\cdot, \theta^{B}) \\ 
\end{bmatrix}$, 
where $\theta^{A}$ and $\theta^{B}$ are parameters for two neural networks. By minimizing the loss function $\mathcal{l}\left(\mathcal{H}; \theta_{A}, \theta_{B}\right)$, we obtained parameters $\hat{\theta }_{A}$ and $\hat{\theta }_{B}$. To guarantee that our solution $\hat{\mathcal{H}} =\mathcal{H}\left(\cdot; \hat{\theta }_{A}, \hat{\theta }_{B}\right)$ satisfies the constrain $\hat{\mathcal{H}}^{T}\mathbf{D}\hat{\mathcal{H}} =\mathbf{I}_{\mathbf{d}}$, we followed the optimization method described in \cite{nguyen2022deep}, forcing the outputs of the optimization problem on the Stiefel manifolds \cite{stiefel1935richtungsfelder}. This nonlinear method yielded the low-dimensional representation $\hat{\mathcal{H}}$, which reveals the information on both intra- and inter-cellular networks.

Neural networks have an input layer consisting of nodes of samples (cells) and two hidden layers (by default 32/16 hidden units) along with sigmoid nonlinearity, followed by a linear output layer (3 units). All the layers are fully connected. Hyperparameters such as the number of iterations and learning rate were optimized to ensure the functionality and reproducibility of the networks for single- and two-sample analyses. For each data set, we performed a random hyperparameters search of 100 trials by the module Ray Tune (v1.8.0) \cite{liaw2018tune} with the hidden neurons randomly selected from 64 and 32, the embedding dimension sampled from 2 to 10, and the learning rate sampled from a uniform distribution from 10\textsuperscript{-5} to 0.1 in log space. Supplementary Figure \ref{xct_s4}(A) shows some experimental runs including the optimal output for each data set. To implement, we initialized two neural networks with uniform weights following PyTorch (v1.9.0) recommended heuristic for linear layers and trained them from scratch. For the inflammatory skin data set of single-sample analysis, we trained the model for 1,000 iterations using the Adam optimizer with a learning rate of 0.01.

\subsection{Determining the statistical significance of interactions between cell types in a single sample}

With $\hat{\mathcal{H}}_{A} = \mathcal{H}^{A}(\cdot; \hat{\theta }_{A})$ and $\hat{\mathcal{H}}_{B} = \mathcal{H}^{B}(\cdot; \hat{\theta }_{B})\in\mathbb{R}^{n\times d}$ being the representations of genes from cell types $A$ and $B$ in the low dimensional embedding, respectively, we calculated the Euclidean distance $d_{ij}$ across cell types for every pairwise combination of gene $i$ in cell $A$ and gene $j$ in cell $B$ and denoted the square difference of projected representations between gene $i$ and gene $j$ as $d_{ij}^{2} = \left\Vert\hat{\mathcal{H}}_{i}^{A} -\hat{\mathcal{H}}_{j}^{B}\right\Vert^{2}$. 
We implemented a nonparametric test to identify significant gene pairs among all combinations under the null hypothesis that there is no LR-mediated interaction between gene pairs. The null hypothesis distribution was obtained by collecting $d_{ij}^{2}$ of all gene pairs, excluding LR pairs in the OmniPath database. Compared to gene pairs that appear in the database, those gene pairs absent in the database were considered much less likely to be gene pairs that confer LR-mediated interactions. Alternatively, the null distribution could be constructed by random shuffling of the data, which would make the test less conservative. Next, we calculated the quantiles for LR pairs under the null hypothesis distribution and set them as the original \textit{p-values}. The threshold was set to 0.05 for all data sets. We excluded the Chi-square test in this case because the left tail targeting for close gene pairs with short distances is desired and is therefore incompatible with the Chi-square test.

\subsection{Comparative manifold alignment}

Comparative biology to study the function of gene sets looks for significant differential interactions between two samples, such as healthy $H$ and disease $D$. One direct method is to walk through the above pipeline for two samples separately and obtain the difference $ d_{ij}^{H}$ and $ d_{ij}^{D}$ for different samples. Given that $ d_{ij}^{H}$ and $ d_{ij}^{D}$ were not comparable since the low-dimensional representations for two samples belonged to different latent spaces, we were asked to add constraints to make them numerically comparable so that the same genes from two samples would have similar low-dimensional representations. Following the general manifold alignment framework \cite{wang2009general}, we constructed the coupled joint similarity matrix across two samples denoted by $\mathbf{V}$ as
\begin{center}
$\mathbf{V} =\begin{bmatrix}
\mathbf{W}^{H} & \beta\mathbf{I} \\ 
\beta\mathbf{I} &\mathbf{W}^{D} \\ 
\end{bmatrix}$,
\end{center}
where $\mathbf{W}^{H}$ and $\mathbf{W}^{D}$ are the joint similarity matrices constructed by the above manifold alignment method for each sample, respectively. $\mathbf{I}$ is an $2n \times 2n$ identical matrix, and $\beta$ is a tuning hyperparameter, which is by default 0.9 times the mean value of the row sums of $\mathbf{W}^{H}$ and $\mathbf{W}^{D}$. Intuitively, a smaller factor $\beta$ would not enforce correspondence between identical genes from different samples, whereas a larger one would produce close distances between identical genes without much consideration of the given gene regression networks. We showed that within an optimal range of the scale factor in $\beta$, the aligned distances remained highly correlated (Supplementary Figure \ref{xct_s4}(C)). The loss function is defined as
\begin{equation}
\mathcal{l}\left( F\right) = \sum_{i, j}^{}\left\Vert  F\left(i,\right) -  F\left(j,\right)\right\Vert^{2}\mathbf{V}\left(i, j\right).
\end{equation}

The solution is denoted as $\hat{F} = 
[\begin{matrix}
\hat{ F }^{H} \\ 
\hat{ F }^{D} \\ 
\end{matrix}]\in\mathbb{R}^{4n\times d}$ 
where $\hat{F}^{H}$ and $\hat{F}^{D}\in\mathcal{R}^{2n\times d}$ are analogy to $\hat{\mathcal{H}}$ in single-sample analysis, which are the vertically stacked representations of genes from cell types $A$ and $B$ in each sample. In the two-sample analysis setting, four neural networks are initiated for solving the optimization problem described previously, the architecture of which are identical with ones used for the single-sample analysis. We performed a random search again for the learning rate. We trained the model for up to 3,000 iterations using the Adam optimizer with the same learning rate of 0.01 as in the single-sample analysis (Supplementary Figure \ref{xct_s4}(A)).

\subsection{Determining LR pair statistical significance of differential interactions between two samples}

From $\hat{F}^{H}$ and $\hat{F}^{D}$, we calculated the Euclidean distances $ d_{ij}^{H}$ and $ d_{ij}^{D}$ for every pairwise combination of gene $i$ in cell $A$ and gene $j$ in cell $B$ in samples $H$ and $D$, respectively, similar to the single sample scenario. We then computed the squared distance difference of each gene pair across samples as $ \Delta d_{ij}^{2} =\left\Vert d_{ij}^{H}-d_{ij}^{D}\right\Vert^{2}$. Note that this test statistic would be higher for gene pairs that exhibit a significant difference between two samples. Thus, gene pairs with larger $ \Delta d_{ij}^{2}$ are considered more differentiated. To determine the significance, we implemented the Chi-square test since our test statistic was demonstrated in the sum of square form. Here we set $df = 1$ to make a conservative selection of gene pairs with high precision. By using the right tail $P(X>x)$ of the Chi-square distribution, we assigned \textit{p-value} for each gene pair. Finally, we implemented FDR to generate the adjusted \textit{p-value} and selected significant gene pairs with adjusted \textit{p-value} < 0.05.

\subsection{Validate the predicted interactions between LR pairs}

After the Chi-square test, significant LR interactions were queried against the OmniPath database \cite{turei2021integrated}. The predicted LR pairs in the database were retained for subsequent functional enrichment analyses performed using Enrichr \cite{chen2013enrichr}.

\subsection{Visualization of integrated networks}

Python package igraph (v0.9.6) was used to generate network plots. In each of the network plots, only direct connections between TFs and the enriched LR pair of interest were shown, and the LR interaction itself was highlighted in green. The edge thickness was adjusted to be proportional to the absolute value of the coefficient between gene pairs in the intracellular networks. Positive and negative coefficients were indicated in red and blue, respectively.

\subsection{Systematic comparisons between scTenifoldXct and existing tools}

We compared scTenifoldXct with NATMI \cite{hou2020predicting}, SingleCellSignalR \cite{cabello2020singlecellsignalr}, Connectome \cite{raredon2022computation}, iTALK \cite{wang2019italk}, and CellChat \cite{jin2021inference}. The test data set was the inflammatory skin scRNA-seq data. Two cell types were fibroblasts and dendritic cells. For all the methods, the reference LR database was OmniPath \cite{turei2021integrated}. The comparison analysis was facilitated by using LIANA (v0.0.5) \cite{dimitrov2022comparison}. To show the overlap between significant results, the upset plot was generated using UpSetR (v1.4.0) \cite{conway2017upsetr}. Equal numbers of pairs ranked by each method’s default scoring metric were retained. CytoTalk (v0.99.0) \cite{hu2021cytotalk} was executed independently for reporting significant signaling pathways, and LR pairs from those pathways were subsequently used for the comparison. For plotting the ROC and precision-recall curves, a total of 160 LR pairs, for which both ligand and receptor genes were expressed in the two cell types, were included in the evaluation. Among them, 40 consensus pairs detected by all five methods except scTenifoldXct were used as positive pairs. The remaining 120 were used as negative pairs. The computation cost of scTenifoldXct was estimated on the public platform Google Colaboratory \cite{carneiro2018performance}, and the results are available in Supplementary Table \ref{xct_t5s}.

\section{Results}

\subsection{The scTenifoldXct framework for single-sample application}

\begin{figure}[h]
\centering
\includegraphics[scale=.55]{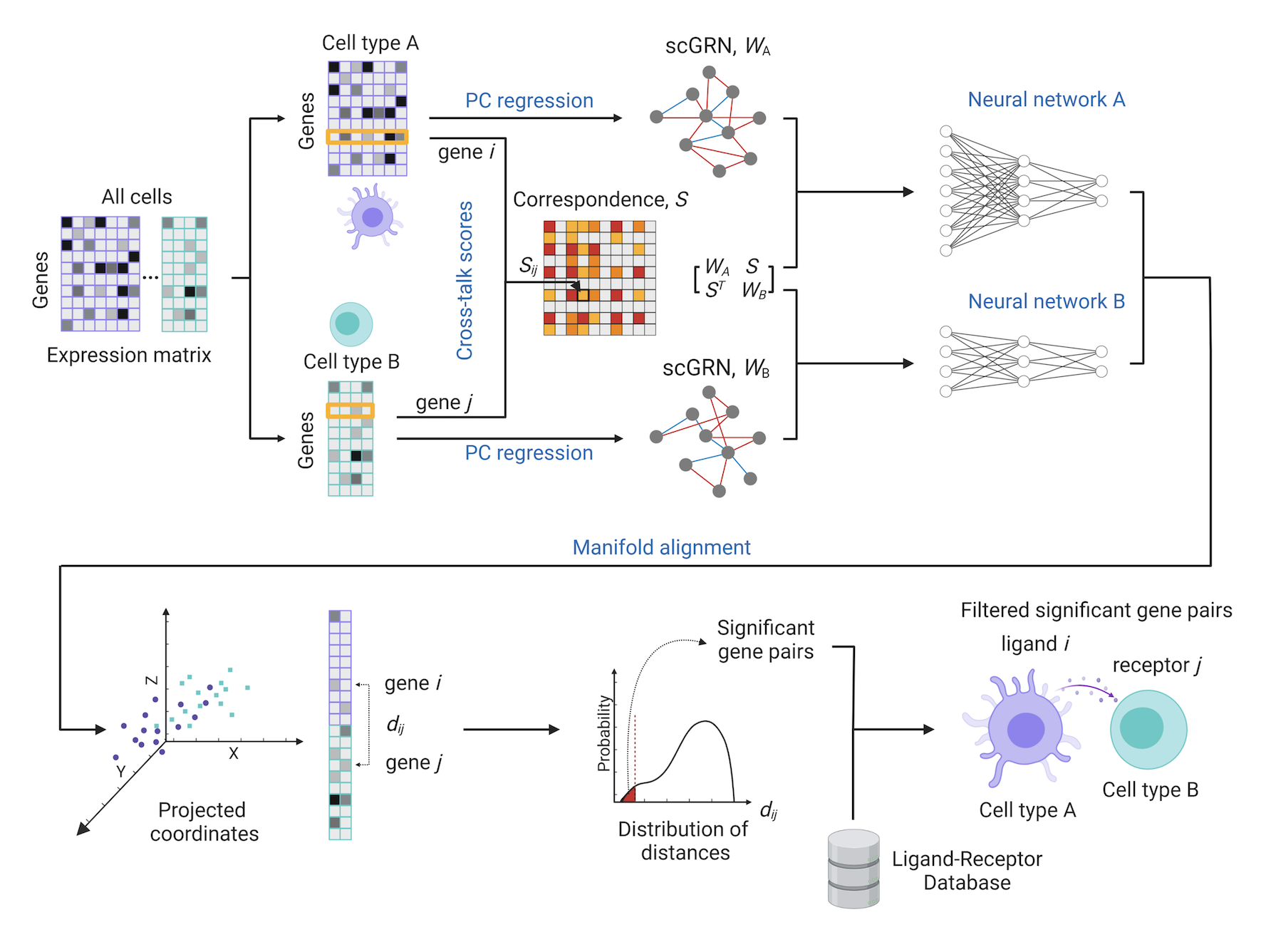}
\caption{Overview of the scTenifoldXct workflow.}
\raggedright
scTenifoldXct is designed to identify LR-mediated interactions using scRNA-seq data. The scTenifoldXct workflow involves three successive steps, namely, (i) gene similarity matrix construction, (ii) manifold alignment, and (iii) significant LR pairs detection. PC regression is used in the construction of single-cell gene regression networks (scGRNs), $\mathbf{W}_{A}$ and $\mathbf{W}_{B}$, to approximate gene regulatory networks for cell types $A$ and $B$, respectively. The joint similarity matrix is composed of $\mathbf{W}_{A}$, $\mathbf{W}_{B}$, and $\mathbf{S}$, the crosstalk score matrix. Neural networks are used to learn latent representations of each gene pair. Two genes of an LR pair are more likely to interact when their low-dimensional latent representations are more similar.
\label{xct1}
\end{figure}

The simple application for scTenifoldXct is single-sample based, i.e., an application using scRNA-seq data from one sample that contains different cell types. The goal is to detect LR pairs with gene products that interact between two given cell types. In such a single-sample application setting, scTenifoldXct requires two inputs: a gene-by-cell count matrix, $\mathbf{X}\in\mathbb{R}^{n\times m}$, and a cell type label vector, $\mathbf{c}\in\mathbb{R}^{m\times 1}$, where $n$ is the number of genes and $ m$ is the number of cells. Given any cell type $A$, let $\mathbf{X}_{A}$ be the count matrix whose cells belong to type $A$. The analytical objectives include: (1) to examine LR pairs in cell-cell interaction databases and identify those pairs that show significant activities and contribute to the LR-mediated interactions in the given sample, and (2) to discover LR pairs that are absent in database. The framework of scTenifoldXct involves three successive steps (\ref{xct1}), as briefly described below. More technical details are given in \textbf{Methods}. 

\subsubsection{Step 1 Construction of single-cell gene networks and crosstalk score matrix}

We started by creating two gene networks for the two cell types of interest, respectively (Substep 1.1). Subsequently, we computed a gene-gene crosstalk score matrix (Substep 1.2). Finally, we concatenated the two gene networks and the gene-gene crosstalk score matrix to create a joint similarity matrix (Substep 1.3), subject to manifold alignment in Step 2.

\paragraph{Substep 1.1 Constructing gene network for each of the cell types}\mbox{}\\
\indent We used principal component (PC) regression to construct single-cell gene networks as previously described \cite{osorio2020sctenifoldnet}. Briefly, for each gene, the expression of the gene was used as a response variable, and the expressions of all other genes were used as dependent variables. The constructed gene networks for cell types $A$ and $B$ were then saved as graphs with signed, weighted, and directional edge weights represented as adjacency matrices $\mathbf{W}_{A}$ and $\mathbf{W}_{B}$. Each column of an adjacency matrix stores the PC regression coefficients of a gene, indicating the regulatory relationships between this gene and all other genes. $\mathbf{W}_{A}$ and $\mathbf{W}_{B}$ were normalized separately by dividing the maximum absolute value of all entries of the matrix. Note that PC regression infers gene-gene expression relationships to approximate gene regulatory networks without requiring any information on transcription factors (TFs) and their targets or knowledge of regulatory elements such as enhancers and promoters. To avoid confusion, the networks, $\mathbf{W}_{A}$ and $\mathbf{W}_{B}$, are thereafter referred to as gene regression networks. Users can supply their own $\mathbf{W}_{A}$ and $\mathbf{W}_{B}$ at this step to replace these two gene regression networks.
\paragraph{Substep 1.2 Calculating crosstalk score between gene pairs from different cell types}\mbox{}\\
\indent We computed the crosstalk score, $score(i_{A}, j_{B})$, between gene $i$ in cell type $A$ and gene $j$ in cell type $B$. The score was designed to incorporate information from the mean and variance of expression of the two genes. The score is greater when two genes have a higher and more variable expression (see \textbf{Methods}). The crosstalk scores between pairs of all genes form the matrix $\mathbf{S}$.
\paragraph{Substep 1.3 Constructing gene similarity matrix}\mbox{}\\
\indent The joint similarity matrix $\mathbf{W}$ was composed of four blocks. The gene regression networks, $\mathbf{W}_{A}$ and $\mathbf{W}_{B}$, were set as the diagonal blocker matrices and the crosstalk score matrix $\mathbf{S}$ as the off-diagonal blocker matrix to yield the similarity matrix $\mathbf{W}$ between genes from two cell types $A$ and $B$.

\subsubsection{Step 2 Manifold alignment}

The manifold alignment method was applied to the joint similarity matrix $\mathbf{W}$ to recover latent representations of gene expression of two cell types. We used neural networks to learn the unified low-dimensional latent representation \cite{golovko2016nature}. The manifold alignment problem solved by neural networks has been shown to be computationally efficient while preserving nonlinearity properties \cite{nguyen2022deep}, and numerically  more stable than methods based on eigen decomposition \cite{vu2012manifold, wang2009general}. 

\subsubsection{Step 3 Determination of significant LR pairs}

Euclidean distance between projected genes on the aligned manifold subspace was used to determine significant LR pairs. The distances were computed between all gene pairs from cell type $A$ to $B$. Based on the distance values, gene pairs were ranked. The closer two genes are, the more likely they are to interact. To select significant LR pairs, \textit{p-value}s were calculated using a nonparametric test. A list of significant LR pairs was selected using a 5$\%$ \textit{p-value} cutoff.

\subsection{Single-sample scTenifoldXct analysis with inflammatory skin data}

We validated scTenifoldXct outcomes using published scRNA-seq data sets. The first data set was generated with lesional and non-lesional skin samples in a study of atopic dermatitis \cite{he2020single}. This inflammatory skin data set contains ten different subpopulations of cells, including three subpopulations of fibroblasts and two subpopulations of dendritic cells. Experimental evidence shows that lesional skin has enhanced CCL19-CCR7 interaction between inflammatory fibroblasts and dendritic cells. CCL19 is required for lymphocyte recirculation, homing, and migration to secondary lymphoid organs \cite{takamura2007regulatory}. Lesional skin in atopic dermatitis shows enhanced chemokine signals with higher expression of CCL19 in inflammatory fibroblasts \cite{nedoszytko2014chemokines}.

We first focused on interactions between fibroblasts and dendritic cells and identified 30 significant LR pairs (Supplementary Table \ref{xct_t1s}), including CCL19-CCR7. These significant interactions shed light on the function of inflammatory fibroblasts interacting with lymphoid cells and regulating type-2 inflammation, an inflammatory pathway involving a subpopulation of CD4+ T cells known as Th2 cells, which is consistent with the findings of the original study \cite{he2020single}. Gene ontology (GO) enrichment analysis  further suggested that significant LR genes were enriched in \textit{dendritic cell chemotaxis} and \textit{dendritic cell migration} pathways. This result is in line with another previous study \cite{saalbach2015fibroblasts}, showing that the cutaneous immune response depends on dendritic cell migration from the skin to draining lymph nodes. VCAM1-ITGB2 and CCL2-CXCR4 were identified by scTenifoldXct as the top-ranked pairs. These genes do not have a higher expression (measured by either the average expression of a single gene among cells or the product of the average expressions of two genes) than many other LR pairs. But, both VCAM1-ITGB2 and CCL2-CXCR4 are biologically relevant, i.e., VCAM1-associated type-2 inflammation causes the over-expression of cytokines (CCL2 and CCL19) in fibroblasts \cite{he2020single}, and CCL2 is known to be induced by inflammatory stimuli such as tumor necrosis factor $\alpha$ (TNF$\alpha$) \cite{gschwandtner2019more}. scTenifoldXct also identified CCL2-TNF. Again, neither of the genes was highly expressed. These examples demonstrate the capability of scTenifoldXct in detecting lowly expressed LR pairs by considering intracellular regulatory activities.

We then focused on interactions between dendritic cells and T cells and identified 33 significant LR pairs (Supplementary Table \ref{xct_t2s}). These interactions centered around two ligand genes: CCL17 and CCL22, both closely associated with the pathophysiology of atopic dermatitis \cite{mantovani2000macrophage}. Serum levels  of CCL17 and CCL22 are known to be correlated with the disease severity \cite{saeki2006thymus}. In the context of  inflamed skin, dendritic cell-derived CCL17 and CCL22 primarily attract T cells that express the cutaneous homing receptor; they bind to CCR chemokine receptors, which are preferentially expressed in T cells, resulting in inflammation \cite{hirota2011variants}. Genes of significant LR pairs also included CCR6, CXCR4, CCL17, S100A8, and S100B. The abundant expression of CCR6 in T cells in skin lesions suggests its important role in early inflammatory T cell recruitment. CXCR4 is a critical receptor involved in both homeostatic and pathological leukocyte trafficking, attracting cells to inflammatory sites and contributing to the activation of integrins required for T cell activation \cite{garcia2019role}. CCL17 binds to the receptor CCR4, which is known to be expressed on activated/memory T cells. Recent data show enhanced expression of CCL17 in the skin  lesions and serum of AD patients, leading to dendritic cell migration from the skin to the skin-draining lymph nodes \cite{stutte2010requirement}. S100 proteins  such as S100A8 and S100B are required for immunological homeostasis and inflammation and have been linked to various inflammatory skin diseases, including psoriasis and atopic dermatitis \cite{xia2018s100, broome2003s100}. S100A8 and S100B were other examples lowly expressed but were detected and ranked as the top LR pairs by scTenifoldXct.

\begin{figure}[h]
\centering
\includegraphics[scale=.40]{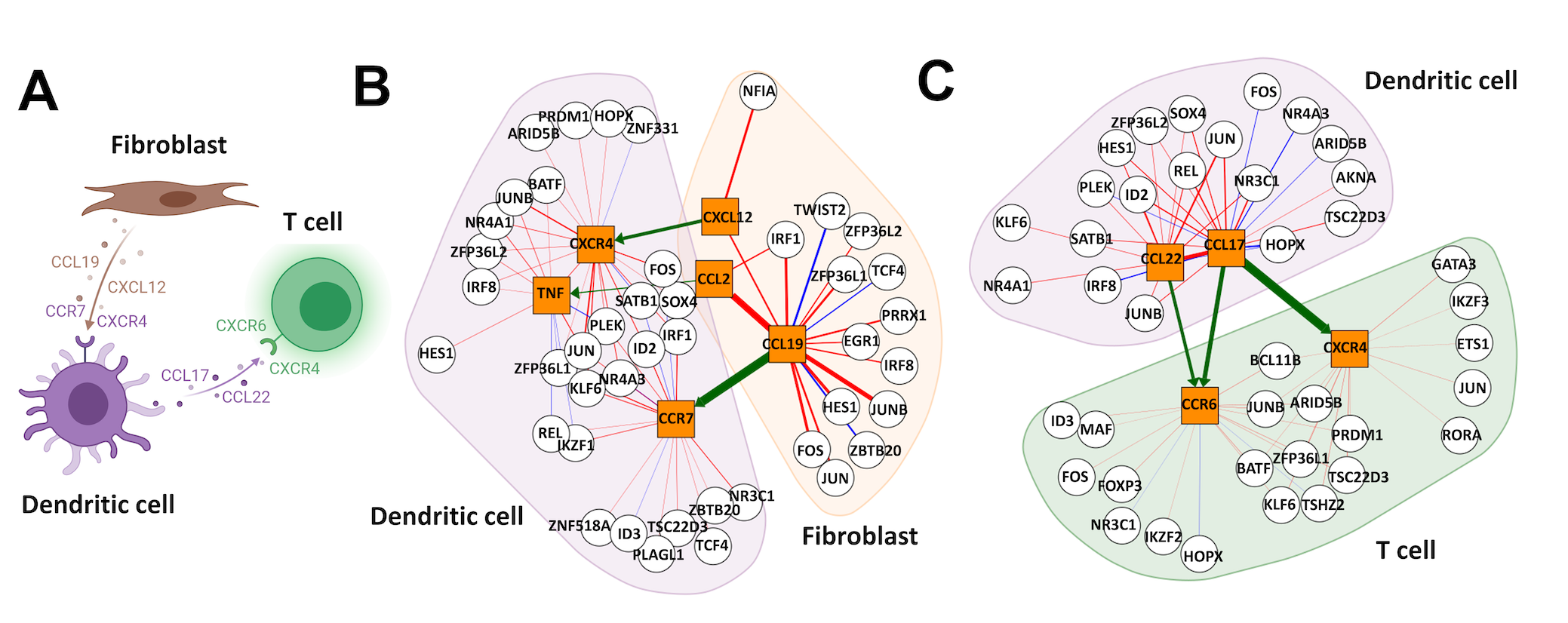}
\caption{Cell-cell interactions between fibroblasts, dendritic cells and T cells with their intracellular networks in inflamed skin.}
\raggedright
(\textbf{A}) An illustration of representative interactions between cell types in skin lesions \cite{he2020single}. (\textbf{B}) An integrated network across fibroblasts and dendritic cells with interactions: CCL19-CCR7, CXCL12-CXCR4, and CCL2-TNF (boldfaced in Supplementary Table \ref{xct_t1s}). Blue and red edges indicate negative and positive regulatory relationships between genes, respectively. Genes like HES1 and JUNB are present in both cell types and thus appear twice in the figure. (\textbf{C}) An integrated network across dendritic cells and T cells with interactions: CCL17-CCR6, CCL22-CCR6, and CCL17-CXCR4 (boldfaced in Supplementary Table \ref{xct_t2s}).
\label{xct2}
\end{figure}

Finally, we conflated the interactions across three cell types, i.e., fibroblast–dendritic cell–T cell. The combined results implied a cascade of intercellular signaling pathways where fibroblasts activate dendritic cell inflammatory responsiveness. The latter interacts with T cells to facilitate T-cell trafficking, lymphoid tissue organization, and type-2 cell recruitment (Figure \ref{xct2}(A)). Gene regression networks allowed us to approximate and examine intracellular systems, locate LR genes in the intracellular networks, and trace the upstream TFs that regulate the expression of the LR genes. Figure \ref{xct2}(B) illustrates such an integrated intra- and inter-cellular network centered around two significant interactions: CCL19-CCR7 and CXCL12-CXCR4, as predicted by scTenifoldXct. In dendritic cells, REL (Proto-oncogene c-Rel, an NF-$\kappa$B subunit) was strongly regulated by CCR7, which is consistent with the experimental results showing that CCR7 activates NF-$\kappa$B \cite{liu2011nf}. NF-$\kappa$B  is also known to be co-activated with AP1 and regulates CCR7 expression \cite{mburu2012chemokine}. JUN (the subunit of AP1) is a TF positively regulates the expression of CCR7. Figure \ref{xct2}(C) depicts another integrated network centered around ligands CCL17 and CCL22 and their receptors CCR6 and CXCR4 in dendritic and T cells. We found that AP1 subunits, JUNB and JUN, are linked with these receptors, suggesting a role of JUN/AP-1 proteins in skin inflammation \cite{schonthaler2011targeting}. In conclusion, these constructed networks were consistent with our prior knowledge about gene regulatory relationships in the studied inflammatory skin system.

\subsection{Comparison between scTenifoldXct and other existing methods}

We compared the prediction results of scTenifoldXct with those produced by five existing methods, namely CellChat \cite{jin2021inference}, Connectome \cite{raredon2022computation}, iTALK \cite{wang2019italk}, NATMI \cite{hou2020predicting}, and SingleCellSignalR \cite{cabello2020singlecellsignalr}. The prediction of each method was generated from the same input data, i.e., fibroblasts and dendritic cells in the inflammatory skin data set, and the same reference cell-cell interaction database, Omnipath \cite{turei2021integrated}, was used. For each method, we implemented them at their own optimal thresholds and retained the equal numbers of most significant LR pairs as reported (except that CellChat only produced nine significant pairs) to perform an overlap analysis using an upset plot (Supplementary Figure \ref{xct_s1}). Five interactions (CCL2-CCR7, CCL19-CCR7, CCL19-CXCR4, CXCL12-CCR7, and CXCL14-CCR7) were detected by all six methods. Predictions made by scTenifoldXct tended to overlap more with consensus inferred by other methods. For example, scTenifoldXct detected all interactions that CellChat detected. scTenifoldXct detected six interactions missed by CellChat and Connectome, but the other three methods detected them. scTenifoldXct detected three additional interactions missed by CellChat but detected by all other methods. CytoTalk \cite{hu2021cytotalk}, another method considering intracellular networks, detected four significant LR pairs (CCL19-CCR7, CCL26-CCR7, CCL2-VEGFA, and CCL26-CXCR4); all were detected by scTenifoldXct (Supplementary Table \ref{xct_t1s}).

Additionally, scTenifoldXct detected S100A8-ITGB2 and CCL26-CCR6. All other methods failed to detect these two LR pairs because this observation involved genes that were lowly expressed (Supplementary Figure \ref{xct_s2}), while the sensitivity of scTenifoldXct was augmented with the manifold alignment of intracellular networks. There is mounting evidence that shows that the two LR pairs are key factors involved in the pathogenesis of atopic dermatitis. S100A8 is an important molecule in the pathogenesis and progression of atopic dermatitis via altering cytokine and skin barrier protein expression levels \cite{kim2019effect}. S100A8 is the first S100 family member that has a potent chemokine-like activity to murine phagocytes in vitro and in vivo \cite{wang2018s100a8}. S100A8 and S100A9 induce neutrophil adhesion to fibrinogen in vitro via upregulating Mac-1 (a heterodimer of CD11b and ITGB2), indicating S100A8-ITGB2 is an important pathogenic mechanism associated with the pathogenesis and progression of atopic dermatitis. CCL26-CCR6 is another pair of molecules uniquely detected by scTenifoldXct. The lesional atopic dermatitis samples were characterized by the expansion of inflammatory dendritic cells and tissue-resident memory T cells \cite{he2020single}. CCR6, a $\beta$-chemokine receptor, mediates the migration of dendritic cells and several lymphocyte subsets to sites of epithelial inflammation \cite{phadke2007role, paradis2008essential}. It has also been reported that CCR6 is required for IL-23–induced psoriasis-like inflammation in mice \cite{hedrick2009ccr6}. CCL26 is another molecule that may serve an important role in the pathogenesis of atopic dermatitis \cite{kagami2003significant}. The expression of CCL26 is known to be increased in lesional atopic dermatitis fibroblasts \cite{he2020single}; the serum CCL26 levels in patients with atopic dermatitis tend to decrease after the treatment \cite{kagami2003significant}. These analyses demonstrate that scTenifoldXct is capable of predicting both strong and relatively weak interactions.

To further assess the performance of scTenifoldXct quantitatively, we used 40 LR pairs that were identified by all the five existing methods as the ``ground truth" and plotted the receiver operating characteristic (ROC) curve and precision-recall curve. We found that scTenifoldXct achieved the greatest area under the ROC curve (AUROC, 0.89) and the highest average precision (AP, 0.87), respectively (Supplementary Figure \ref{xct_s3}).

\begin{figure}[h]
\centering
\includegraphics[scale=.40]{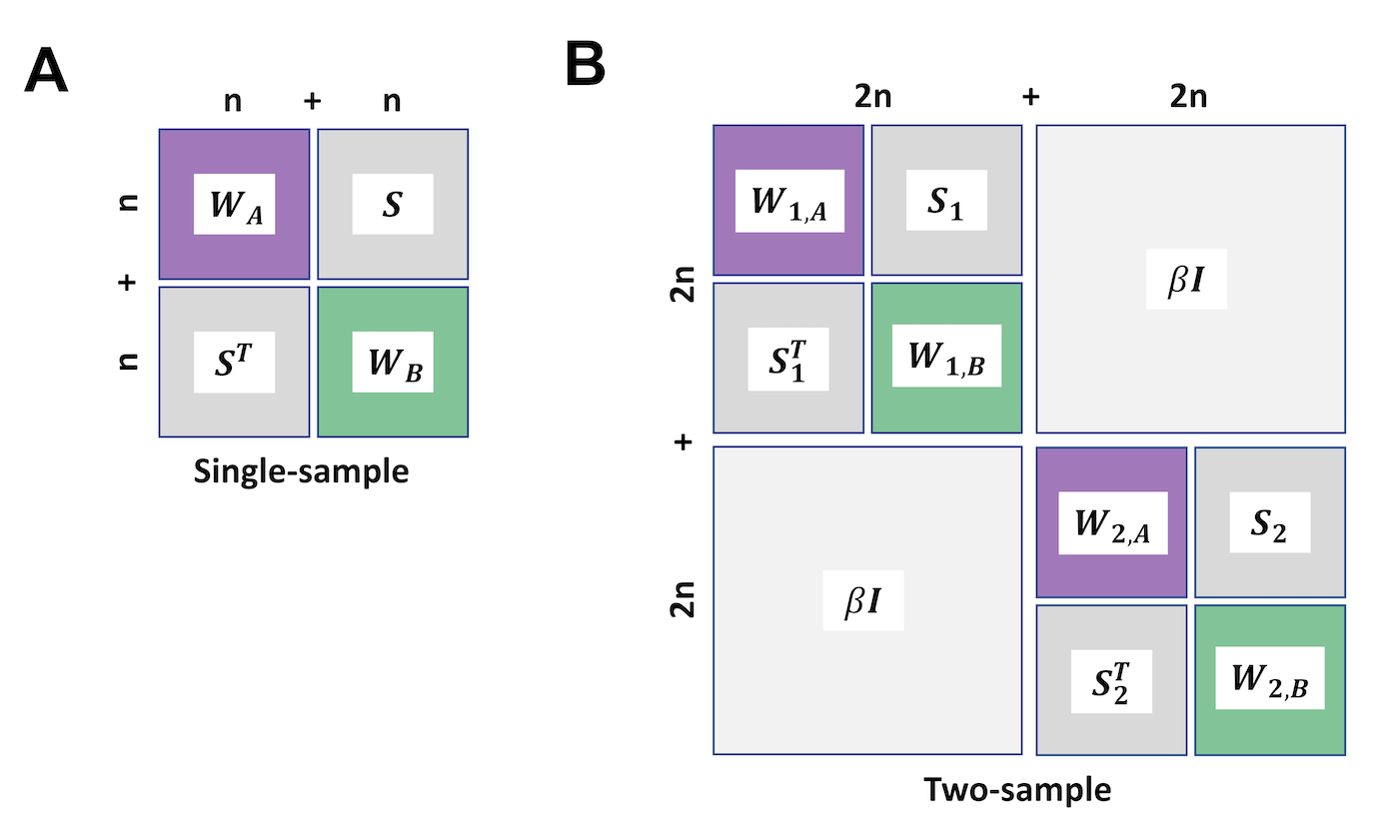}
\caption{Joint similarity matrices used in scTenifoldXct single-and two-sample analyses.}
\raggedright
(\textbf{A}) Joint similarity matrix for manifold alignment in the single-sample analysis. $\mathbf{W}_{A}$ and $\mathbf{W}_{B}$ are the gene regression networks of cell types $A$ and $B$, respectively. $\mathbf{S}$ is the crosstalk score matrix, and $n$ is the number of genes. (\textbf{B}) Coupled joint similarity matrix for manifold alignment in two-sample analysis. $\mathbf{W}_{1,A}$ and $\mathbf{W}_{1,B}$ are the gene regression networks of cell types $A$ and $B$ from sample 1, $\mathbf{W}_{2,A}$ and $\mathbf{W}_{2,B}$ from sample 2. $\mathbf{S}_1$ and $\mathbf{S}_2$ are the crosstalk matrices for samples 1 and 2, respectively. $\beta\mathbf{I}$ is an identity matrix with a tuning hyperparameter $\beta$.
\label{xct3}
\end{figure}

\subsection{The scTenifoldXct framework for two-sample application}

The basic scTenifoldXct framework for single-sample applications can be extended to the comparison between two samples, detecting LR pairs that show a significant difference in interaction strength between such as healthy and diseased samples. Similar to the basic framework, in which manifold alignment is achieved by solving a generalized eigenvalue problem with a joint matrix (Figure \ref{xct3}(A)), the extended framework applies the neural networks as in the single-sample to solve the manifold alignment problem in the two-sample. The difference is that the coupled joint matrix is formed by including four gene regression networks (from two cell types of two samples) and two crosstalk score matrices (between two cell types from each sample) (Figure \ref{xct3}(B)).

\subsubsection{Step 1 Construcing joint similarity matrices for two samples separately}

We computed a joint matrix for each sample separately using the procedure described in the single-sample application above.

\subsubsection{Step 2 Constructing a coupled joint similarity matrix}

We then placed the two joint matrices in the diagonal block of the coupled joint matrix $\mathbf{V}$ (Figure \ref{xct3}(B)). To make the low-dimensional representations of two samples numerically comparable, a constraint factor was included by setting $ \beta\mathbf{I}$ to the off-diagonal block of the joint similarity matrix, where $\mathbf{I}$ is an identity matrix that reflects the binary correspondence between genes in different samples and $\beta$ is a tuning hyperparameter.

\subsubsection{Step 3 Calculating the distance differences between two samples}

To determine significant differentially interacted LR pairs, we calculated the Euclidean distance between every pairwise combination of gene $i$ in cell type $A$ and gene $j$ in cell type $B$ for two samples, respectively. The distance difference between each gene pair was then computed across samples.

\subsubsection{Step 4 Determining significantly differential LR-mediated interactions}

We considered gene pairs with a greater distance difference between two samples to be more significantly differentiated. With this, we obtained a list of ranked gene pairs. We computed \textit{p-value}s for the gene pairs using the Chi-square test, adjusting the \textit{p-value}s with a multiple testing correction, and selected significant LR pairs using a 5$\%$ false discovery rate (FDR) cutoff against the OmniPath database \cite{turei2021integrated}.

\begin{table}[h]
\centering
\caption{Differential interactions between hepatocytes and endothelial cells.} 
\raggedright
LR pairs were ranked according to the difference in distances from the ligand gene to the receptor gene on their aligned manifold in the latent space between two cell types. The sign of difference indicates the direction of change in the interaction strength between the tumor and normal tissues.
\label{table_diff}
\begin{tabularx}{\textwidth}{|X|X|X|X|X|X|}
\hline
\textbf{LR pairs} & 
\textbf{LR distance on manifold (tumor)} & 
\textbf{LR distance on manifold (normal)} & 
\textbf{Difference in LR distances between tumor and normal} & 
\textbf{\textit{p-value}} & 
\textbf{\textit{Adj. p-value}} \\ 
\hline
\textbf{\textit{PLA2G2A-ITGA5}} & 
0.0010 & 
0.0203 & 
-0.0193 & 
3.6E-07 & 
8.0E-04 \\ 
\hline
\textit{TFF3-ACKR3} & 
0.0147 & 
0.0330 & 
-0.0183 & 
1.5E-06 & 
1.6E-03 \\ 
\hline
\textit{IGF2-PDGFRB} & 
0.0290 & 
0.0108 & 
0.0182 & 
1.8E-06 & 
1.7E-03 \\ 
\hline
\textit{IGF2-FGFR3} & 
0.0269 & 
0.0105 & 
0.0164 & 
1.6E-05 & 
6.1E-03 \\ 
\hline
\textit{MDK-PTPRB} & 
0.0041 & 
0.0193 & 
-0.0152 & 
6.6E-05 & 
1.5E-02 \\ 
\hline
\textit{CXCL3-ACKR1} & 
0.0078 & 
0.0228 & 
-0.0150 & 
8.1E-05 & 
1.8E-02 \\ 
\hline
\textbf{\textit{MDK-ITGA6}} & 
0.0059 & 
0.0206 & 
-0.0147 & 
1.1E-04 & 
2.1E-02 \\ 
\hline
\textbf{\textit{SPP1-ITGA5}} & 
0.0058 & 
0.0203 & 
-0.0145 & 
1.3E-04 & 
2.4E-02 \\ 
\hline
\textit{IGF2-ERBB2} & 
0.0275 & 
0.0135 & 
0.0140 & 
2.1E-04 & 
3.3E-02 \\ 
\hline
\textbf{\textit{SPP1-ITGB5}} & 
0.0173 & 
0.0312 & 
-0.0139 & 
2.4E-04 & 
3.6E-02 \\ 
\hline
\textit{SPP1-ITGA6} & 
0.0068 & 
0.0206 & 
-0.0138 & 
2.8E-04 & 
3.9E-02 \\ 
\hline
\end{tabularx}
\end{table}

\begin{figure}[h]
\centering
\includegraphics[scale=.40]{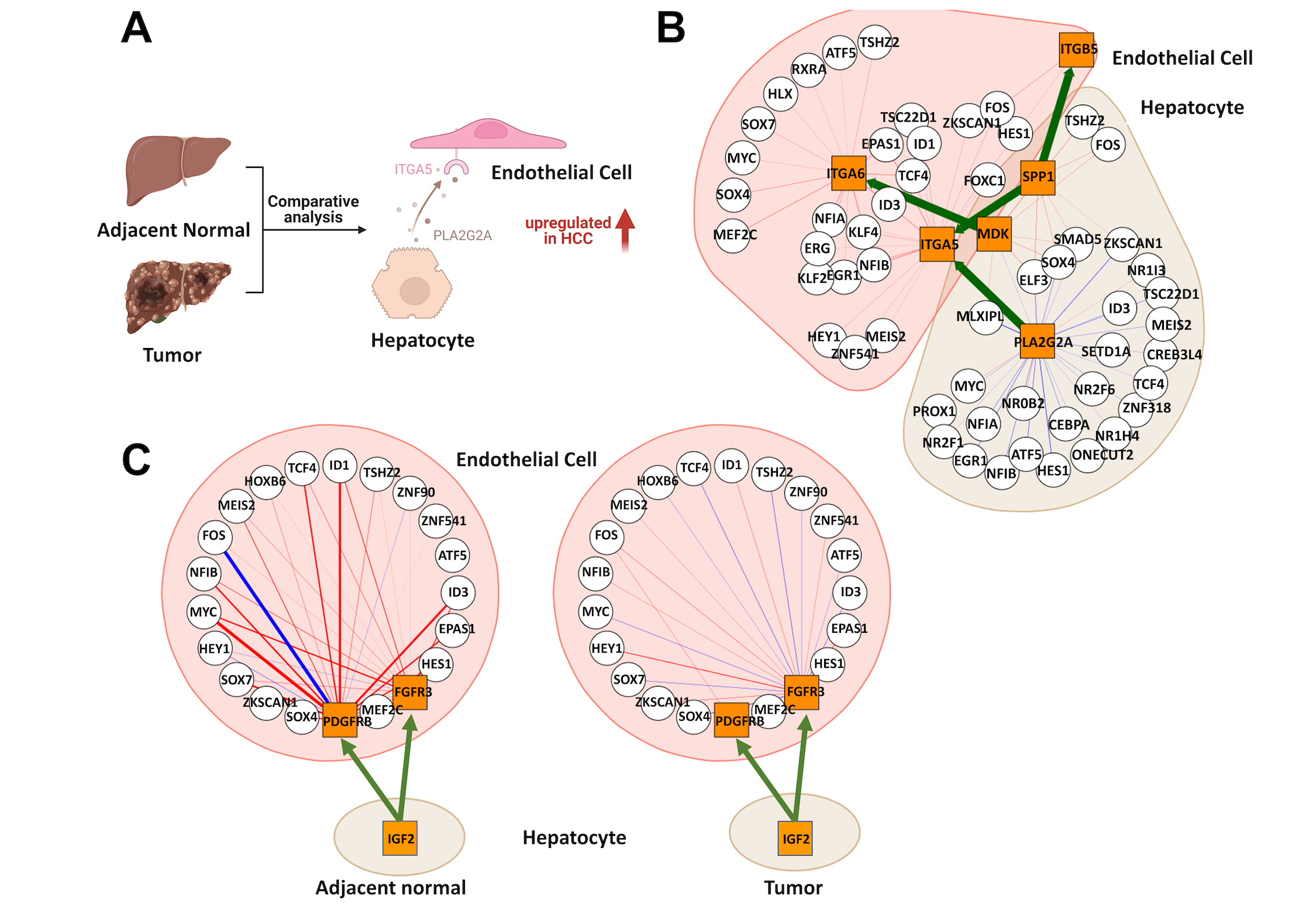}
\caption{Cell-cell interactions between hepatocytes (senders) and endothelial cells (receivers).}
\raggedright
(\textbf{A}) An illustration of representative interactions in hepatocellular carcinoma (HCC) \cite{sharma2020onco}. (B) An integrated network across hepatocytes and endothelial cells, connected by interactions PLA2G2A-ITGA5, MDK-ITGA6, SPP1-ITGA5, and SPP1-ITGB5 (boldfaced in Table \ref{table_diff}). The top 30 weighted edges in each network of the cell type are shown. (\textbf{C}) Integrated networks connected through IGF2-FGFR3 and IGF2-PDGFRB across hepatocytes and endothelial cells in the adjacent normal (left) and tumor (right) tissues.
\label{xct4}
\end{figure}

The additional applications of scTenifoldXct include multi-sample comparison. To demonstrate the performance of scTenifoldXct comparative analysis, we obtained the scRNA-seq data set from a hepatocellular carcinoma study \cite{sharma2020onco}. The data set contains cells from the tumor and adjacent normal tissues, allowing for the comparison (Figure \ref{xct4}(A)). Using scTenifoldXct to compare tumor and normal samples, we identified 11 interactions that showed significantly different strengths of hepatocyte-endothelial cell interactions (Table \ref{table_diff}). Among them, eight LR pairs interact more strongly in tumor samples. For example, PLA2G2A (secretory calcium-dependent phospholipase A2), previously implicated in host antimicrobial defense, inflammatory response, and tissue regeneration \cite{kramer1989structure, suzuki2000structures, ishizaki1999cloning}, was found to interact more strongly with ITGA5 in the tumor. GO enrichment analysis with genes in these eight LR pairs enriched \textit{beta-1 integrin cell surface interactions} (Supplementary Table \ref{xct_t3s}), highlighting the role of beta-1 integrin in the progression of hepatocellular carcinoma \cite{tian2018beta1}. Figure \ref{xct4}(B) depicts an integrated network showing four out of the eight tumor-enhanced pairs of LR-mediated interactions: PLA2G2A-ITGA5, MDK-ITGA6, SPP1-ITGA5, and SPP1-ITGB5. Corresponding intracellular networks of different cell types were included. In this integrated network, TCF4 is linked with ITGA5 and ITGA6 in the endothelial cells, indicating a functional relationship between TCF4 and the two receptor genes. TCF4 is a key component of the Wnt signaling pathway, which has been linked to the proliferation of hepatocellular carcinoma cells. It could be an effective therapeutic target for blocking the growth of hepatocellular tumors \cite{zhao2004aberrant}. Specifically, the mutation of beta-catenin (leading to its nuclear and cytoplasmic accumulation) renders it capable of bypassing APC-targeted degradation and accumulating in the nucleus to form a complex with TCF4 aberrantly activating downstream transcriptional events \cite{huang1999beta}. Similarly, ERG is also linked with ITGA5 and ITGA6 in endothelial cells. A previous study has found more ERG-positive endothelial cells in hepatocellular carcinoma tissue than in adjacent normal tissue \cite{moh2020decreased}. FOS is linked with  ITGA5 and ITGB5. FOS is an oncogene upregulated in hepatocellular carcinoma \cite{arbuthnot1991c}. Figure \ref{xct4}(C) shows two identical cellular networks of endothelial cells connected with hepatocytes via LR-mediated interactions IGF2-FGFR3 and IGF2-PDGFRB. In the tumor sample, the reduced intracellular regulation between PDGFRB and ID1 and ID3 was evident. PDGFRB has been implicated in the development and metastasis of hepatocellular carcinoma; ID1 and ID3 are associated with hepatocellular carcinoma dedifferentiation \cite{damdinsuren2005expression}.

\subsection{Differential interaction analysis between wild-type (WT) and gene knockout (KO) samples}

To test the generalizability of scTenifoldXct, especially when applying to the comparison of two samples, we performed an additional two-sample comparison analysis. We obtained scRNA-seq data derived from lung alveolar type II (AT2) cells and basophils in a study utilizing WT and IL1RL1 (IL33R) KO mice \cite{cohen2018lung}. It is known that AT2 cells produce IL33 (ligand) that binds to IL1RL1 (receptor). In IL1RL1-deficient lungs, basophils lack the expression of a large number of lung basophil-specific genes. We predicted that scTenifoldXct-based comparison analysis between the WT and KO samples should be able to identify significant differential LR pairs. Indeed, IL33-IL1RL1 ranked at the top of all LR pairs that were predicted by scTenifoldXct to be significantly differentiated (Supplementary Table \ref{xct_t4s}). In addition to IL33-IL1RL1, scTenifoldXct also identified six other significant differentially interactive LR pairs, including CSF1-TNF, CSF2-TNF, EDN1-EDNRB, and EDN3-EDNRB. CSF1 and CSF2 are CSF family cytokines that have been shown to play crucial roles in shaping the lung microenvironment \cite{shibata2001gm, guilliams2013alveolar, ginhoux2014monocytes}. The interaction strengths of CSF1-TNF and CSF2-TNF were all significantly diminished in the KO sample. This is consistent with the known functions of these genes and their interplay in lung development \cite{cohen2018lung, jones2014effect}. Similarly, the interaction strengths of EDN1-EDNRB and EDN3-EDNRB were also diminished in the KO sample, which is consistent with the observed downregulation of EDNRB in lung disease, where cells lose the ability to respond to endothelin stimulation \cite{wei2020role}.

\section{Discussion}

We have showcased the functionality and performance of scTenifoldXct. Using real-data examples, we demonstrated single-sample and two-sample applications of scTenifoldXct. A two-sample application involves comparative analysis between two samples, for which none of the existing methods adopts the integrative strategy as does scTenifoldXct. Methodologically, scTenifoldXct leverages information on the topological placements of genes in gene regression networks to explore intracellular connections of LR pairs with other genes. Since the scTenifoldXct method takes into account the networks underlying genes of LR pairs, it allows for the reconstruction of a more detailed map of biological pathways linking the two cell types of interest. Incorporating intracellular networks enables a comprehensive inference of signaling pathways across cells. To the best of our knowledge, only a few existing tools, such as NATMI and scMLnet \cite{hou2020predicting, cheng2021inferring}, incorporate intracellular network or signaling pathway information in their analyses. The difference is that scTenifoldXct requires no prior knowledge of intracellular networks, while other methods depend on signaling pathways and networks in databases such as KEGG. scTenifoldXct computes gene regression networks from the input data and operates entirely in a data-driven manner. Furthermore, in the initial stage of the analysis, scTenifoldXct assigns scores to all combinations of gene pairs without requiring prior knowledge of known LR pairs. Thus, significant LR pairs detected by scTenifoldXct are likely to be those undocumented in the databases.

scTenifoldXct adopts the redesigned crosstalk score to quantify the interaction strength between an LR pair. The product of ligand and receptor expression has been widely adopted as an indicator for this purpose, which is intuitively sound. When a ligand and its receptor interact, they are expected to be highly expressed, and the value of the product of their gene expression should be greater. The redesigned interaction score we designed for scTenifoldXct used the same principle, but we expanded its formula by incorporating gene expression variability. The latter has been shown to be equally weighted as the average gene expression in reflecting cellular functions \cite{dueck2016variation, osorio2019single}. Rather than focusing just on the expression mean, scTenifoldXct considers gene expression variability across cells and calculates the interaction score , balancing the contributions of the mean and the variability. The effectiveness of this redesigned metric has been demonstrated in our results, showing that biologically significant LR pairs with low expression can be detected.

In terms of differential interaction analysis, while many tools allow users to compare two samples \cite{hou2020predicting, cabello2020singlecellsignalr, wang2019italk, jin2021inference, raredon2022computation}, they initially infer interactions for each sample independently before comparison. The differences in obtained interaction scores are used to assess the differences between the two samples. In contrast, the scTenifoldXct comparative algorithm uses the integrated method to learn the correspondences rather than processing two samples independently. The level of differential interactions is measured by the distance differences in the joint latent space of aligned manifolds. Our manifold alignment algorithm automatically assumes that data sets share the same underlying structure across cell types. Such an assumption can be easily nullified by presenting cell type-specific cell states across heterogeneous single-cell data sets. Thus, it remains computationally challenging for manifold alignment algorithms to preserve both shared and data set-specific cellular structures across samples \cite{cao2022manifold}. Despite this concern, our common manifold-based integrative strategy has been shown to be highly effective. For example, retrospectively, we tried to apply the independent processing strategy to repeat the comparative analysis, but this strategy did not work even with scTenifoldXct—i.e., no LR pairs were detected under the independent processing.

For further development, because temporally and spatially precise cell communication is the key to cellular differentiation, we consider the next version of scTenifoldXct could be directed toward incorporating time-series and spatial transcriptomic information \cite{hu2023adept}. For time-series scRNA-seq data, existing analytical frameworks such as GraphFP and SoptSC \cite{jiang2022dynamic, wang2019cell}, have demonstrated the feasibility of simultaneous inference of cell lineages and cell-cell communications. Recent improvements in pseudotemporal ordering enable us to map the underlying regulatory networks over time \cite{qiu2017reversed, xu2022scintime}. Thus, the scTenifoldXct framework can be further updated by taking time-series or pseudotime information and integrating dynamic inference modules, enabling the assessment of changes in interaction strength throughout the processes of cellular differentiation or organ development.
\chapter[GenKI: GENE KNOCKOUT INFERENCE FROM LEARNING SINGLE-CELL GENE REGULATORY NETWORKS]{GenKI: GENE KNOCKOUT INFERENCE FROM LEARNING SINGLE-CELL GENE REGULATORY NETWORKS\footnote{Reprinted with permission from Y. Yang et al., "Gene knockout inference with variational graph autoencoder learning single-cell gene regulatory networks," Nucleic Acids Research, 2023, gkad450.}}\label{chap:genki}

\section{Introduction}
Gene perturbation experiments are a proven powerful approach to elucidate the role of a gene in a biological process. Commonly used designs include gene knockout (KO) experiments with genetically altered animals and CRISPR gene perturbations. In a KO experiment, the function of a target gene is inferred by contrasting phenotypes between KO and wild-type (WT) animals and then identifying their differences. Often, gene expression profiles serve as a quantitative phenotype at the molecular level \cite{quake2021cell}. The recent advent of single-cell RNA sequencing (scRNA-seq) \cite{hwang2018single} allows the transcriptomic information from tens of thousands of cells to be gathered in parallel, and thus it greatly improves cellular phenotyping resolution. It has become a powerful method for molecular phenotyping and comparison in KO experiments.

Conventional KO experiments, often requiring significant amounts of experimental and animal resources, are labor-intensive and time-consuming \cite{hall2009overview}. Recently developed techniques such as Perturb-seq \cite{dixit2016perturb} combine CRISPR perturbations and scRNA-seq to perform genetic screens, allowing gene function to be studied in many cells in a massively parallel manner. Nevertheless, the creation of large-scale CRISPR libraries presents a major technical challenge. For these reasons, computational tools serve as a possible alternative solution to facilitate or guide the experimental design through in-silico screening of perturbation responses. Such a computational tool would reduce the need for experimental measurements.
Indeed, several such computational tools \cite{lotfollahi2019scgen, lotfollahi2023predicting, kamimoto2023dissecting, osorio2022sctenifoldknk} have been developed (Table \ref{genki_t1}). With only one exception—scTenifoldKnk \cite{osorio2022sctenifoldknk}, all these tools require extensive input data sets including outcomes of perturbation experiments or data from other modalities. scTenifoldKnk is the only protocol that does not require such expensive input data sets. Instead, it merely requires scRNA-seq data from the WT samples as its input and considers information from the gene regulatory network (GRN). The working principle of scTenifoldKnk is to simultaneously project WT and virtual KO single-cell gene regulatory networks (scGRNs) to a joint low-dimensional space and then calculate the projection differences of genes. However, the inference of scTenifoldKnk entirely relies on the WT scGRN, which is constructed using principal component (PC) regression from the WT scRNA-seq data. It is known that constructing high-quality scGRNs is technically challenging with respect to the presence of heterogeneous sources of noise \cite{eraslan2019single}. Also, a fully connected scGRN computed by the regression-based method may not correspond to real biological processes \cite{ye2012sparse}. A method that takes full advantage of scRNA-seq expression data and tolerates “imperfect” scGRN in a robust and unbiased manner is still lacking.

\begin{table}[htbp]
\centering
\caption{Summary of existing virtual KO methods and feature comparison with GenKI.}
\label{genki_t1}
\begin{tabular}{|p{1.7cm}|p{2.4cm}|p{1.8cm}|p{2.5cm}|p{5.5cm}|p{0.8cm}|}
\hline
\textbf{Name} & \textbf{Input data required} & \textbf{Method} & \textbf{Supervised / Unsupervised} & \textbf{Description} & \textbf{Ref.} \\
\hline
scGen & scRNA-seq (WT and KO samples) & Transfer learning & Supervised & Train a variational autoencoder that learns to generalize the response of the cells in the training set of perturbations & \cite{lotfollahi2019scgen} \\
\hline
CPA & scRNA-seq (KO samples) & Generative modeling & Supervised & Train an autoencoder with adversarial that decomposes the data into a collection of embeddings associated with the cell type, perturbation, and other external covariates to study combinatorial genetic perturbation & \cite{lotfollahi2023predicting} \\
\hline
CellOracle & scRNA-seq, scATAC-seq (WT sample) & Graph-based modeling & Unsupervised & Simulate gene expressions in response to transcription factor (TF) perturbation by signal propagation through an inferred gene regulatory network & \cite{kamimoto2023dissecting} \\
\hline
scTenifold Knk & scRNA-seq (WT sample) & Manifold alignment & Unsupervised & Simultaneously project inferred WT and virtual KO gene regulatory networks to a joint low dimensional space & \cite{osorio2022sctenifoldknk} \\
\hline
GenKI & scRNA-seq (WT sample) & VGAE & Unsupervised & Train a VGAE model that learns the latent gene representations of WT sample and virtually construct a virtual KO counterpart to discern similarity & This study \\
\hline
\end{tabular}
\end{table}

Here, we present GenKI (Gene KO Inference), a virtual gene KO tool based on a variational graph autoencoder (VGAE) \cite{kipf2016variational}. GenKI simultaneously learns latent representations of scRNA-seq gene expression data of WT samples and the underlying scGRN responsible for observed phenotypes. The highly compressed representations of genes are then used for the subsequent inference. The scGRN can be constructed using the input gene expression data. GenKI propagates the transcriptomics information in the network during training and compares the WT data (including the expression data matrix and the scGRN) with its virtual KO counterpart to predict KO-responsive genes—i.e., genes functionally associated with or linked to KO gene. As a de novo inference tool, GenKI identifies KO-responsive genes without requiring prior knowledge of gene regulation or biological mechanisms.

The remainder of this paper is structured as follows: we first present an overview of the GenKI workflow and then compare its inference performance to several benchmarks using simulated data. Following these steps, we use publicly available scRNA-seq data sets (Supplementary Table \ref{genki_t1s}) to predict KO-responsive genes and compare enriched functions of them with those introduced and validated in the original studies, to highlight the performance of GenKI in real-data applications. Next, we compare GenKI to the differential expression (DE) analysis. Finally, we study the robustness and scalability of GenKI.

\section{Methods}

\subsection{Simulated data sets and evaluation}
The predefined GRNs were obtained from the GitHub repository of SERGIO \cite{dibaeinia2020sergio}. The simulated data sets contained 100, 400, and 1,200 genes (all containing 2,700 cells), respectively. Edges in the predefined GRNs were treated as the ground truth. A random classifier that ranks genes by probabilities randomly drawn from a uniform distribution between 0 and 1, a classifier that ranks genes by the Pearson correlation with the KO gene, and scTenifoldKnk, which ranks genes by FC (used for the chi-squared test), were included for benchmarking purpose. For each data set, we randomly selected a target gene with more than ten edges and virtually knocked it out using GenKI and the other three benchmarks independently. Each run outputs a gene list with scores assigned by each method. Roc\_auc\_score and average\_precision\_score function from the Python package sklearn (v.1.1.1) were used to compute the Area Under Receiver Operating Characteristic (AUROC) and the average precision (AP) at each run for each method. We repeated the procedure above ten times for each data set. The simulated BEELINE \cite{pratapa2020benchmarking} data sets were downloaded from Zenodo. GSD is the largest curated reference data set of BEELINE containing 19 genes and 2,000 cells. Its underlying GRN was used to replace the GRN construction step in this evaluation. Since the ground truth GRN was known, we divided genes into two groups based on their shortest path to the KO gene, with the close neighbors group containing all genes within the two-hop neighborhood of the KO gene and the distant neighbors group containing all other genes. To compare the inference power of GenKI and scTenifoldKnk, we virtually knocked out each gene iteratively and obtained the scores of all the genes computed by both methods. For each method, we used the Wilcoxon Rank Sum test to quantify the difference in scores between the two groups of genes. A lower \textit{p-value} indicates a larger difference, thus implying greater inference power of the method for detecting KO-responsive genes.

\subsection{Processing of real data sets}
The specifics and source of real scRNA-seq data sets used in this paper can be found in Supplementary Table \ref{genki_t1s}. We performed regular preprocessing for all scRNA-seq data sets using Seurat (v.4.0.2) package \cite{hao2021integrated}. We first performed log normalization using the NormalizeData function. Highly variable genes were selected using the FindVariableFeatures function (selection.method = “vst”) and by default, the top 3,000 highly variable genes were included in subsequent analyses. We then standardized the data by the ScaleData function, and the resulting transformed data served as the gene expression profile for the GenKI input. Cell annotations from original studies were retained and used if provided.

\subsection{Gene regulatory network construction}
We constructed scGRNs using the PC regression method, which was first proposed in scTenifoldNet \cite{osorio2020sctenifoldnet}. Let $X \in \mathbb{R}^{p \times n}$ represent the scRNA-seq gene expression matrix of the WT samples, which contained gene expression levels for $p$ genes in $n$ cells. We used the PC regression method to build the scGRN denoted with its adjacency matrix $A$. Specifically, each time one gene was selected as the response variable, while the remaining genes served as explanatory variables. Principal component analysis \cite{jolliffe2016principal} was performed on the explanatory variables, and then we regressed the response variable on the first $d \ll n$ leading principal components. Next, we transformed the obtained regression coefficients of the $d$ leading PCs into the coefficients of the original explanatory variables, which should reflect the interaction strengths between the response gene and all other genes. In the final step, we assembled the coefficients of $p$ regression models into a $p \times p$ adjacency matrix $A$, where the $(i,j)$ entry represents the regression coefficient of the $i$-th gene on the $j$-th gene. Therefore, $A$ accumulates the interaction strength between each pair of genes.

Note that the output of this PC regression method is a fully connected scGRN, in which some links between genes might not correspond to real biological interactions, as in general, there are very few connections between TFs and genes \cite{ye2012sparse}. Therefore, for such an scGRN, we assumed that the edge is activated if the absolute value of its weight is greater than a certain threshold, i.e., edges with a greater weight are more likely to be the true regulatory relationships between genes than those with a lower weight. The average absolute weight between TF-target gene pairs constructed scGRNs was indeed significantly greater than that between random gene pairs, as described in \cite{osorio2020sctenifoldnet}. Based on these findings, for a particular scGRN, we filtered edges and, by default, conservatively only kept the top 15\% of edges. A more thorough evaluation of the cutoff selection can be found in Supplementary Figure \ref{genki_s1}, which shows a heatmap of Spearman correlation coefficients between scores of Kullback–Leibler (KL) divergence given by GenKI across four different cutoffs. Within an optimal range of the cutoff, the ranking results given by GenKI were found to be highly consistent. However, we contend that extremely conservative choices of the cutoff would overlook potential links. Notably, we allow users to modify this default setting to accommodate their own biological scenarios. For example, those who believe their gene regulatory networks are scale-free are encouraged to use the poweRlaw package \cite{alstott2014powerlaw} to determine the best-fit threshold. Next, we converted the scGRN into an adjacent Boolean matrix as the input requested for the VGAE model of GenKI. As a result, although obtained without any information on TFs and their targets or knowledge of regulatory elements, these remaining edges could be deemed biologically responsive. By abuse of notations, we still denoted this new scGRN as A and we referred to it as the thresholded scGRN for later use. Although the filter step removed potential false positive edges, it inevitably introduced false negative findings, i.e., missing some truly connected edges. Therefore, we treated this thresholded scGRN as an incomplete network, and our goal was to reconstruct an scGRN from this incomplete network to learn the latent embeddings of nodes, namely, genes in our setting. This can be interpreted as a transductive link prediction task \cite{ravindra2020disease}. Alternatively, users can supply their own GRN at this step to replace the PC regression-derived network.

\subsection{Variational graph autoencoder (VGAE) model}
The VGAE model used in GenKI is similar to the framework described in \cite{kipf2016variational}. It is made up of a two-layer graph convolutional network (GCN) encoder and an inner product decoder. We utilized a two-layer GCN architecture because deeper graph convolutional networks are prone to over-smoothing \cite{yang2020revisiting}. Recall that $X$ is the gene expression matrix and $A$ is the adjacency matrix, and we denoted the normalized adjacency matrix as $\widetilde{A} = D^{-\frac{1}{2}}AD^{-\frac{1}{2}}$, where $D = \text{diag}(d_{11},d_{22},\ldots,d_{pp})$ is a diagonal matrix with entries $d_{ii} = \sum_{i=1}^{p} A_{ij}$, where $A_{ij}$ is the $(i,j)$-th entry of the matrix $A$. Then, the two-layer GCN is defined as:
\[
\text{GCN}(X,A) = \widetilde{A} \text{ReLU}(\widetilde{A}XW_{0})W_{1}
\]
where $\text{ReLU}(x) = \max(0,x)$ is the activation function introduced in the first GCN layer, and $W_{0}$ and $W_{1}$ are parameters of the neural networks. We assumed that the data were generated by certain random processes involving an unobserved latent continuous random variable $Z$. Let $p(Z)$ be the prior distribution of $Z$, for which we chose a bivariate Gaussian distribution for convenience. For the encoder part, we introduced a recognition model $q(Z|X,A) = \prod_{i=1}^{p} q(z_{i} |X,A)$, where $q(z_{i}|X,A) \sim \mathcal{N}(\mu_{i},\Sigma_{i})$, $\Sigma_{i} = \text{diag}(\sigma_{i1}^2,\sigma_{i2}^2)$ is a diagonal covariance matrix, and
\[
\mu = (\mu_{1}^{T},\ldots,\mu_{p}^{T})^{T} = \text{GCN}_{\mu}(X,A), 
\log(\Sigma) = \log((\sigma_{1}^2,\ldots,\sigma_{p}^2)) = \text{GCN}_{\sigma^2}(X,A),
\]
where $\sigma_{i}^2 = [\sigma_{i1}^2,\sigma_{i2}^2]^{T}$. For the decoder part, we used the inner product to reconstruct the scGRN $A$ by
\[
P(A_{ij} = A_{ji} = 1) = \text{sigmoid}(z_{i}^{T}z_{j}).
\]
Here, by abuse of notation, $z_{i}$ is the latent representation of the $i$-th gene.

For any two distribution functions $p$ and $q$, let $\text{KL}(p\|q) = \int p(x)\log\frac{p(x)}{q(x)}\,dx$ be the KL divergence between $p$ and $q$. The objective of the VGAE model is to maximize the evidence lower bound (ELBO):
\[
\mathbb{L} = E_{q(Z|X,A)}\log p(A|Z) - \beta\cdot\text{KL}(q(Z|X,A)\|p(Z)),
\]
where $\beta$ is an adjustable hyperparameter that balances the independent constraints and reconstruction accuracy. Notice that here we adapted the loss from beta-VAE \cite{higgins2016beta}, and $\mathbb{L}$ would represent the standard ELBO when $\beta = 1$.

\subsection{Hyperparameters, metrics and implementation}
We randomly split the edges of a Boolean scGRN into three data sets for training (75\%), validation (5\%), and testing (20\%). We labeled them as positive edges. Equal numbers of negative edges, composed of a set of “fake” edges not presented in the scGRN, were sampled for data balancing purposes. We used AUROC and AP to evaluate the model performance. We expected positive edges to have higher interaction probabilities compared to negative edges. Thus, the higher value of AP or AUROC would indicate better performance of training. To tune the hyperparameters, we performed random hyperparameters search of 100 trials by using the Tune module from the Python package Ray \cite{liaw2018tune} (v.1.13.0). Specifically, the logarithm base 10 of hyperparameter $\beta$ was sampled from a uniform distribution from \{-5, -4, …, -1\}, the learning rate was sampled from a uniform distribution from \{-4, -3, …, -1\}, and the weight decay of optimizer was sampled from a uniform distribution from \{-7, -6, …, -3\}. To make our sampled hyperparameters more accurate, we multiplied each one by a scale factor randomly selected from integers 1 to 9. For each set of hyperparameters, we evaluated the model performance on the validation set and selected the hyperparameter set with the best performance based on the metrics AUROC and AP. Based on our experimental results, we set $\beta$ of 1E-4 and weight decay of 9E-4 for all the data sets, and set learning rate of 7E-4 for the microglia, lung, intestine data set, 5E-3 for the COVID-19 data set. The maximum iteration number was set to 100, and early stopping was added when AP reached the maximum and began to decrease. The Adam optimizer \cite{kingma2014adam} was used for all the trainings, and Xavier initialization \cite{glorot2010understanding} was used to initialize all the weights.

\subsection{Determination of the rank of KO-responsive genes}
After training the VGAE model using the WT data, for each fixed gene $g$, we obtained its latent distribution $\mathcal{N}(\hat{\mu}_g, \hat{\Sigma}_g)$, where $\hat{\mu}_g$ and $\hat{\Sigma}_g$ were the latent mean and covariance fitted by the VGAE model. We next fed the trained VGAE model with the virtual KO data and obtained the latent distribution of the $g$-th gene for the KO samples. Then, we calculated the Kullback-Leibler (KL) divergence between these two normal distributions. The procedure was repeated for all genes. The top 5\% of genes ranked by the KL divergence were preserved.

Instead of using the raw ranks, we proposed a bagging-based method to improve the stability and accuracy of our inference. Specifically, each time we permuted the cell order of the WT gene expression matrix and obtained its corresponding virtual KO data. Without training a new model, we fed this pair of permuted WT and virtual KO data into our fitted VGAE model, calculated the KL divergence value for each gene, and bagged the top 5\% of genes. We repeated this procedure 1,000 times and compiled the genes that were bagged more than 95\% of the time as KO-responsive genes.

\subsection{Benchmarking GenKI’s tolerance to random noise in gene expression profiles}
To show the robustness of our method, we generated random noise in the log space, added it to gene expression profiles, and evaluated the training performance of GenKI. Specifically, for gene $i$ in cell $j$, the regenerated expression $x_{i,j}'$ was defined as:
\[
\frac{x_{i,j}'}{x_{i,j}} = 2^\gamma
\]
where $\gamma \sim \mathcal{N}(0,\sigma)$ and $x_{i,j}$ represents the original expression. The fold change $\gamma$ was used to approximate the noise level, which followed the normal distribution $\mathcal{N}(0,\sigma)$, whereas different $\sigma$ values would result in different levels of random noise. We conducted 30 independent runs with random splits of the dataset at different noise levels.

\subsection{Gene function annotation and function enrichment tests}
Enrichr \cite{kuleshov2016enrichr} with default setting was used for gene functional enrichment analyses. The protein-protein interaction enrichment tests were performed using the web tool of the STRING database \cite{von2005string}. In the STRING network plots, isolated nodes were removed, and only edges labeled with confidence greater than the medium level were retrieved and shown. Enrichment \textit{p-value}s, which indicate whether input proteins have more interactions among themselves than what would be expected for a random set of protein-coding genes of the same size and degree distribution drawn from the genome, were computed with the default setting.

\subsection{Prediction of KO gene’s expression from WT cells with linear regression}
For the microglia data set, a simple multivariate linear regression model was applied to evaluate the relationship between the KO gene Trem2 and other KO-responsive genes. Specifically, microglia cells’ Trem2 expression profile was used as the response variable and the expression profiles of other genes as explanatory variables. The adjusted $R^2$ (coefficient of determination) was used to quantify how much variance of the KO gene can be explained by the other KO-responsive genes. In comparison, an equal number of the KO-responsive genes were randomly sampled as explanatory variables, and their $R^2$ was also calculated. This evaluation was repeated 30 times with different splits of the data set and random gene selections.

\subsection{Differential gene expression analysis}
DE analysis was performed using Scanpy \cite{wolf2018scanpy}(v.1.9.1) function rank\_genes\_groups with the Wilcoxon rank-sum test. All parameters were set to default. Adjusted \textit{p-value}s were obtained after the Benjamini–Hochberg adjustment \cite{benjamini1995controlling}. DE genes were determined based on the condition of adjusted \textit{p-value} < 0.05 and absolute log2(fold change) > 0.25. DE ranks of the DE genes were determined based on their adjusted \textit{p-value}. To examine the expression level changes, for each data set, the KO-responsive genes and an equal number of randomly chosen unperturbed genes were used and their fold change (FC) of WT/KO was calculated. The absolute log2-transformed FC values of the KO-responsive genes and the unperturbed genes were used to perform the one-sided t-test.

\section{Results}

\subsection{The GenKI framework}
\begin{figure}[h]
\centering
\includegraphics[scale=.80]{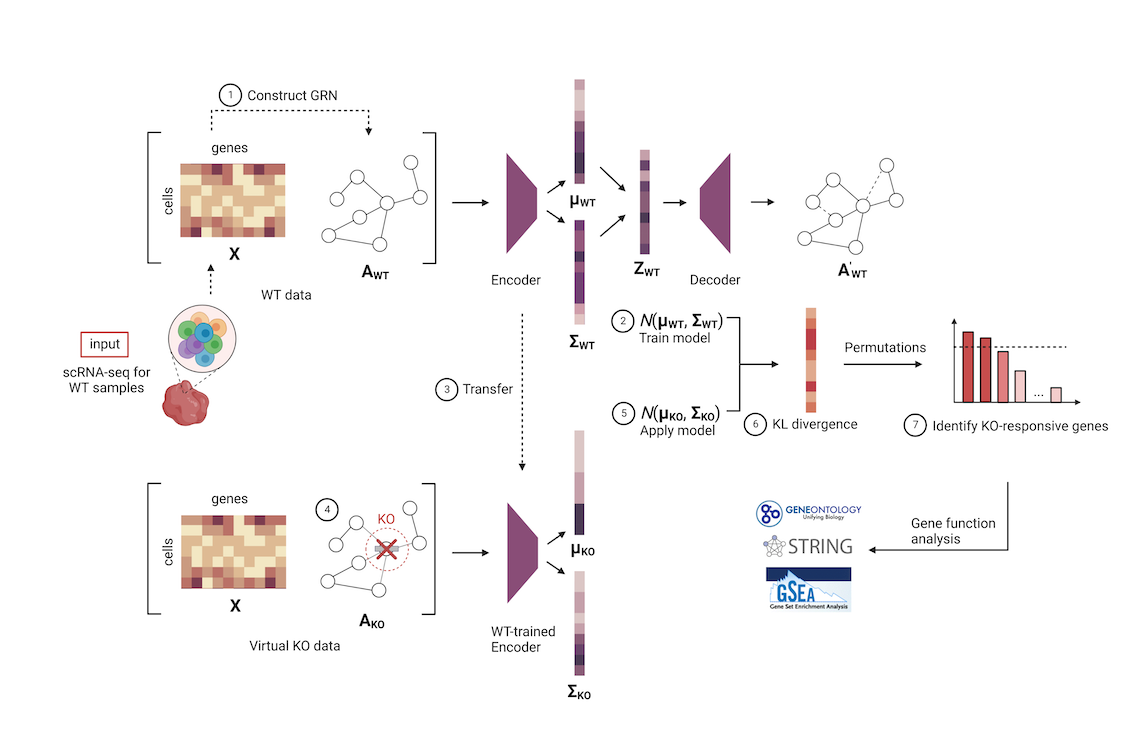}
\caption{The GenKI framework of gene KO inference for gene function prediction.}
\raggedright
The pipeline contains seven steps: (1) construction of WT scGRN, (2) training VGAE model, (3) transfer the trained VGAE model, (4) construction of virtual KO data, (5) latent embeddings of WT and virtual KO data, (6) calculation of KL divergence, and (7) identification of KO-responsive genes for function annotation and analysis.
\label{genki_1}
\end{figure}

The framework of GenKI is depicted in Figure \ref{genki_1}. The pipeline starts with a single input, that is, the scRNA-seq gene expression matrix from WT samples of interest. For each virtual KO application, GenKI first constructs an scGRN from the WT gene expression data. The WT gene expression data matrix and the constructed WT scGRN are then used as input of WT data to train a VGAE model, which is a two-layer GCN encoder with an inner product decoder. The latent embedding of each node is defined to follow a bivariate Gaussian distribution. After training, the latent representations of genes under the WT setting are collected and the model with its weights is transferred. Next, to generate virtual KO data, the WT data is “copied”. From the WT scGRN copy, the KO gene—i.e., the gene being knocked out for functional study—is virtually deleted. The deletion is achieved by setting the weight of all edges from and to the KO gene to zero. After the virtual deletion, the virtual KO data is generated, while the original WT scGRN remains untouched. The transferred model is fed with the virtual KO data to obtain the latent representations of genes under the KO setting. Two parameters, mean and covariance of each gene’s latent distribution from the WT and KO settings are then collected to calculate the Kullback–Leibler (KL) divergence between these two distributions. The higher the KL divergence value of a gene, the greater the impact of the KO on the gene. Finally, a bagging-based method is used to determine genes that tend to be significantly perturbed by the deletion of the KO gene. The enriched functions of these significantly perturbed genes (i.e., KO-responsive genes) are used to give prediction of the KO gene functions.

\subsection{Performance of GenKI with simulated data}
\begin{figure}[h]
\centering
\includegraphics[scale=.70]{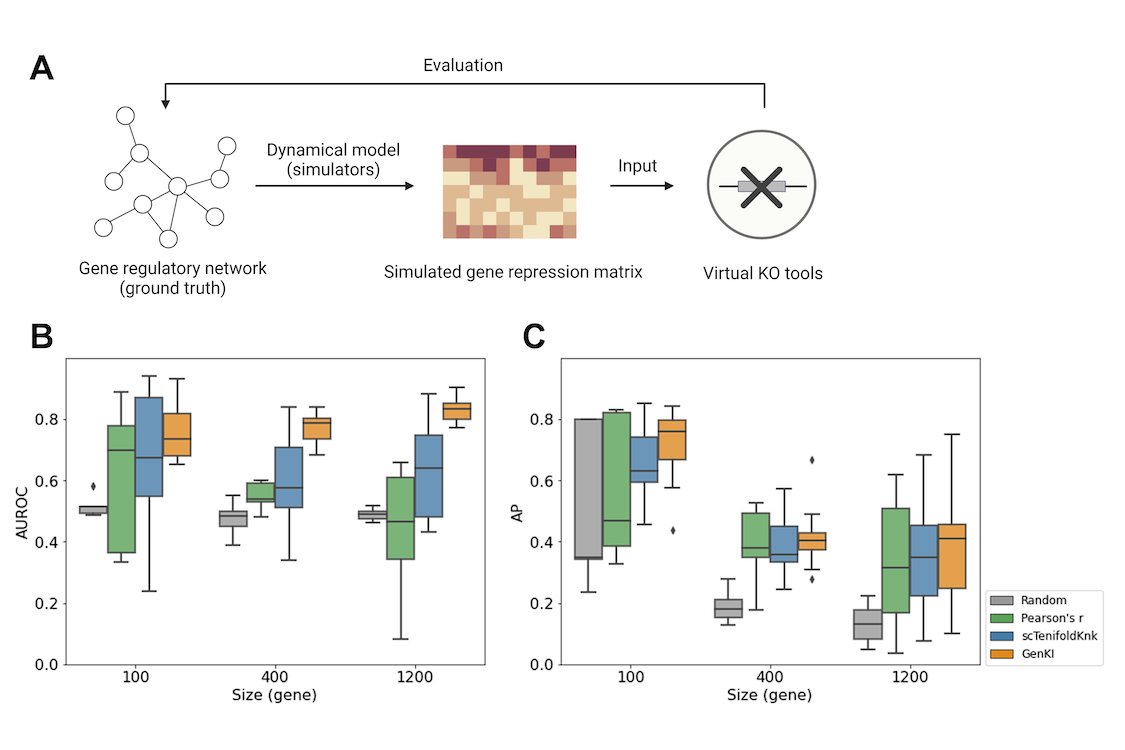}
\caption{Comparing prediction performance between GenKI and other virtual KO methods with simulated data sets.}
\raggedright
Three methods were included in the comparison including scTenifoldKnk and two baseline predictors, which are based on random rankings and Pearson’s correlation, respectively (Materials and Methods). (\textbf{A}) The procedure of assessment of virtual KO tools using simulated data sets. (\textbf{B}) The levels of AUROC of virtual KO experiments using three simulated SERGIO data sets. (\textbf{C}) The levels of AP of virtual KO experiments using three simulated SERGIO data sets. Size represents the number of genes in each data set.
\label{genki_2}
\end{figure}

We used simulated data to evaluate the performance of our method (Figure \ref{genki_2}(A)). To do so, we generated scRNA-seq data sets of different sizes (2,700 cells with 200, 400, and 1,200 genes, respectively) using single-cell expression simulator SERGIO \cite{dibaeinia2020sergio}. SERGIO’s simulations were guided by predefined GRNs; therefore, the simulated scRNA-seq data sets had their underlying GRNs. Knowing these ground truths GRNs facilitated the performance evaluation of virtual KO methods, as genes linked with the KO gene were supposed to be perturbed by the KO and more likely to be KO-responsive genes. A good virtual KO tool should preferably identify those genes linked with the KO gene in the given GRN. For each of the simulated data sets, we applied GenKI and three other benchmarking methods, including scTenifoldKnk, with the same KO genes being knocked out (Materials and Methods). All the methods produced a ranked list of KO-responsive genes. Figure \ref{genki_2}(B) shows the levels of AUROC for GenKI and other benchmarking methods. Figure \ref{genki_2}(C) shows the levels of AP resulted from the same KO genes. Three additional ROC curves as examples of virtual KO experiments performed by GenKI and scTenifoldKnk for each data set are presented in Supplementary Figure \ref{genki_s2}. We found that GenKI outperformed all the other benchmark methods, including scTenifoldKnk, across all the data sets evaluated. We believe this is because GenKI incorporates information from both the gene expression matrix and GRN.

To demonstrate that GenKI learns higher-order neighborhood information from the underlying GRN through the VGAE model, which contributes to its greater performance than scTenifoldKnk, we systematically knocked out each of the 19 genes in the GSD network of BEELINE \cite{pratapa2020benchmarking}. In each virtual KO experiment, we obtained the perturbation scores of all genes. For a given KO gene, we used the Wilcoxon rank sum test to compare the difference in perturbation scores between the KO gene’s two-hop neighbor genes and all the other distant genes (Materials and Methods). A smaller \textit{p-value} indicates a greater inference power of the method for differentiation between these two groups. It is rational to expect that close neighbor genes have high perturbation scores. Compared to scTenifoldKnk, as expected, \textit{p-value}s obtained in GenKI are significantly lower (Supplementary Figure \ref{genki_s3}, Wilcoxon Rank Sum test, \textit{p-value} < 0.05). This is attributed to manifold alignment in scTenifoldKnk only keeps track of the similarities between genes in the first-order neighborhood of GRN, while GenKI’s two-layer GCN looks at similarities between genes up to the second-order neighborhood. This simulation study using the BEELINE network data also demonstrated that GenKI can take user input GRN as an optional rather than reconstructing GRN by its own.

\subsection{Real-data GenKI analysis recapitulates findings of the Trem2 KO experiment}
\begin{figure}[h]
\centering
\includegraphics[scale=.80]{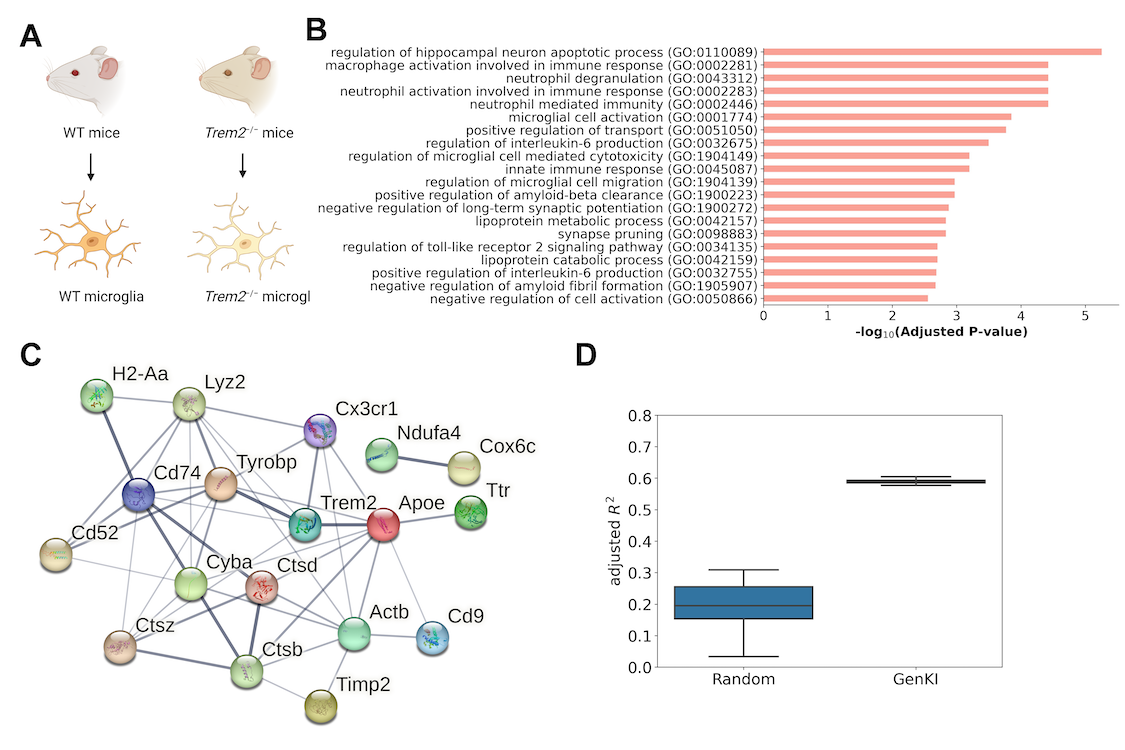}
\caption{Trem2-KO responsive genes inferred by GenKI.}
\raggedright
(\textbf{A}) Illustration of Trem2-KO experiment generating the microglia data set. (\textbf{B}) GO terms significantly enriched in functions of Trem2-KO responsive genes. The -log10-transformed adjusted \textit{p-value} indicates the strength of enrichment for each term. (\textbf{C}) STRING network of Trem2-KO responsive genes. Edge thickness indicates the strength of data support. (\textbf{D}) Adjusted R2 score of the regression of expression levels by setting the Trem2 as a response variable and other KO-responsive genes as explanatory variables, compared to that of randomly selected genes as the explanatory variables.
\label{genki_3}
\end{figure}

GenKI, as a virtual KO tool, is expected to recapitulate the overall discoveries of real KO experiments. To validate its performance, we applied GenKI to several publicly available scRNA-seq data sets. The first data set was from the KO experiment conducted by Nugent et al. \cite{nugent2020trem2}, in which scRNA-seq was performed with microglial cells isolated from Trem2$^{+/+}$ and Trem2$^{-/-}$ mice (Figure \ref{genki_3}(A)). The study reported that Trem2 upregulates apolipoprotein E (APOE) and other genes involved in cholesterol transport and metabolism, causing robust intracellular accumulation of a storage form of cholesterol upon chronic phagocytic activities \cite{nugent2020trem2}. Trem2 is also known to regulate the expression of genes associated with cell damage response, lysosome and phagosome function, Alzheimer’s disease, and oxidative phosphorylation \cite{shi2018interplay}. With this data set, we used the WT gene expression profile of 648 microglial cells as the input for GenKI and fed it along with the constructed scGRN to the VGAE model of GenKI.

We first evaluated the robustness of our model before performing prediction. The model robustness evaluation was performed to test the tolerance of the model by artificially adding different levels of random noise to the WT gene expression profile (Materials and Methods). A robust model would correctly capture the latent embeddings of genes, and thus more confidence for the inference regarding differences between WT and virtual KO samples. AUROC and AP were used to evaluate the reconstruction performance of the model. As shown in Supplementary Figure \ref{genki_s4}, our model was not compromised by high levels of noise $\sigma=1.5$, indicating the robustness of GenKI to the technical noise that naturally existed in the scRNA-seq data. We observed poorer performance under the conditions of very high levels of noise $\sigma\geq3$, which was expected as highly noisy gene expression profiles would mislead the training, and thus, the model could not be generalized to the testing data set. These results also indicated the lower bound of noiseless gene expression information needed to correctly reconstruct the scGRN and eventually infer the latent embeddings of genes.

After the model robustness evaluation, we then trained the model and performed the virtual KO experiment. Specifically, we virtually knocked out Trem2 by removing all its edges in the scGRN of microglial cells and compared profiles of genes in the latent space between WT and virtual KO samples using KL divergence (Materials and Methods). The results of the analysis showed that 20 genes, including Trem2 itself, were detected as Trem2-KO responsive genes (Supplementary Table \ref{genki_t2s}). Trem2 was ranked at the top of the KO-responsive genes, followed by CTSD, the gene associated with lysosomal dysfunction \cite{reifschneider2022loss}, and APOE, the key lipid transporter gene expressed in both the central nervous system and the periphery \cite{li2022trem2}. Pathway enrichment analysis based on Enrichr showed that Trem2-KO responsive genes were enriched with genes associated with interleukin-2 signaling pathway, lysosome, and Alzheimer’s disease (Supplementary Table \ref{genki_t3s}). Gene ontology (GO) enrichment analysis further ranked several enriched terms, including macrophage activation involved in immune response and lipoprotein metabolic process, on the top (Figure \ref{genki_3}(B) and Supplementary Table \ref{genki_t4s}). By modulating the macrophage transcriptome in adipose tissue, Trem2 was found to regulate blood cholesterol metabolism in obese mice, thereby indicating a connection between Trem2 and lipid metabolism \cite{jaitin2019lipid}. The overall results of our enrichment analyses revealed these functions of Trem2 with consistency. In addition, the Trem2-KO responsive genes were found to be biologically connected, as shown by the STRING interaction network (Figure \ref{genki_3}(C), \textit{p-value} < 0.01, STRING interaction enrichment test). Note that links in STRING interaction networks represent functional associations between genes. These associations include direct regulations as well as indirect interactions between genes or their products. Thus, our results suggest abundant functional connectivity between KO-responsive genes.

Next, we investigated whether Trem2's measurable gene expression was intrinsically interpreted by other KO-responsive genes. Indeed, the variance of Trem2 expression across cells could be substantially explained by the remainder of the KO-responsive genes (Figure \ref{genki_3}(D)). We fitted a multivariable linear regression model by setting Trem2 as the response variable (Materials and Methods) and found that when using KO-responsive genes as explanatory variables, the adjusted $R^2$ of the model was significantly higher than when using an equal number of randomly selected genes as explanatory variables (\textit{p-value} < 0.01, one-sided t-test). This finding suggests the KO gene and its KO-responsive genes predicated by GenKI tend to be transcriptionally associated.
Finally, we showed that one could not simply obtain the ranked gene list inferred by GenKI to identify KO-responsive genes using naïve network analysis metrics. We presented that, as an example, the KO-responsive genes could not be simply inferred either from ranking their gene expression or edge weight associated with the KO gene Trem2 in the inferred scGRN (Supplementary Figure \ref{genki_s5}). The GenKI model nonlinearly learns both gene expression and edge weight information and infers from compressed embeddings of genes that it has learned. Thus, it ranks and infers the perturbed genes in a more comprehensive way than ranking methods based on any single observable property.

Collectively, our results shed light on Trem2-related functions by annotating the perturbed genes following its deletion. We showed that the inferred genes were functionally connected and, more importantly, predicted functions were consistent with those reported in the Trem2 studies.

\subsection{Real-data GenKI analysis recapitulates findings of the Nkx2-1 KO experiment}
NK homeobox 2-1 (Nkx2-1) is highly expressed in lung epithelial cells and plays a crucial role in alveolar type 1 (AT1) cell development and maintenance \cite{liebler2016combinations}. We collected the second scRNA-seq data from an in vivo KO experiment performed with lung epithelial cells of AT1 isolated from WT and Nkx2-1$^{-/-}$ mice. The study reported that the Nkx2-1 knocked-out AT1 cells lost their characteristics and abnormally turned into gastrointestinal fate \cite{little2019transcriptional}. The study concluded that without Nkx2-1, developing AT1 cells lose three defining features—molecular markers, expansive morphology, and cellular quiescence—leading to alveolar simplification and lethality.

With this data set, we used the WT gene expression profile of 624 AT1 cells as the input for GenKI and virtually knocked out Nkx2-1 following the methods described above. The GenKI analysis discovered 82 KO-responsive genes (Supplementary Table \ref{genki_t5s}). The KO gene, Nkx2-1, topped the gene list, followed by 13 marker genes of AT1 and AT2 cells offered by PanglaoDB \cite{franzen2019panglaodb}, consistent with their downregulation in the Nkx2-1 mutant cells from the bulk RNA-seq experiment introduced in the original study. Previous research \cite{lee2016epithelial, bruggeman2007expression, lopez2016tead1, weisenhaus2010mutations} discovered that Nkx2-1 binds to a group of AT1 cell-specific genes that regulate the cytoskeleton, membrane composition, and extracellular matrix. We found that Pdlim1, Clic5, Tuba1a, Krt8, Actn4, and Clu, which encode cytoplasmic proteins associated with the cytoskeleton, were highly ranked in our list. Ctsh, a gene involved in epithelial tube branching and lung morphogenesis \cite{chang2013lung}, and a great number of genes related to membrane composition, such as Anxa1, were also observed among the KO-responsive genes. Two other significant genes, Napsa and Sftpc, collaborate with Ctsh to perform functions related to the collagen-containing extracellular matrix and alveolar lamellar body. Cldn33, Cldn7, and Epcam, which were shown to be involved in the apical junction complex \cite{wu2013epithelial}, are in agreement with the observation that mutant AT1 cells form dense microvilli-like structures apically concluded in the original study.

\begin{figure}[h]
\centering
\includegraphics[scale=.80]{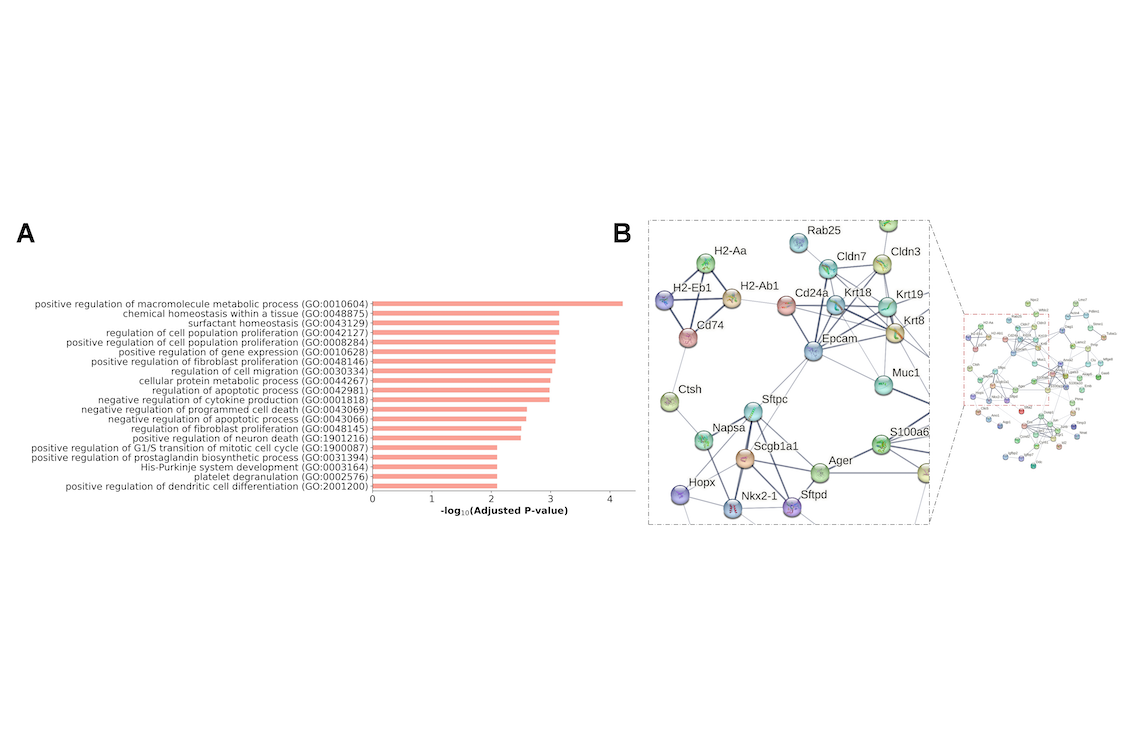}
\caption{Nkx2-1-KO responsive genes inferred by GenKI.}
\raggedright
(\textbf{A}) GO terms significantly enriched in functions of Nkx2-1-KO responsive genes. The -log10-transformed adjusted \textit{p-value} indicates the strength of enrichment for each term. (\textbf{B}) STRING network consists of Nkx2-1-KO responsive genes. The zoomed inset demonstrates a subnetwork module containing the KO gene.
\label{genki_4}
\end{figure}

GO enrichment analysis indicates these genes were enriched for functional categories led by surfactant homeostasis and positive regulation of cell population proliferation (Figure \ref{genki_4}(A), Supplementary Table \ref{genki_t6s}), suggesting the role of Nkx2-1 in regulating surfactant production and suppressing AT1 cell proliferation validated in the study. HDAC3-dependent TGF-beta signaling is required for proper epithelium expansion and AT1 cell spacing \cite{wang2016hdac3, wang2016expression}, disruption of which significantly perturbed 13 genes from the list related to TGF-beta regulation of extracellular matrix. Additionally, due to mutant cells undergoing apoptosis, which was validated by staining in the original study, a few terms indicating the apoptotic process were observed. Many other GO terms, which are significant but not shown in Figure \ref{genki_4}(A), such as epithelial tube branching involved in lung morphogenesis and epithelial cell morphogenesis demonstrate the conclusion that NKX2-1 defines the cell morphology of developing AT1 cells. The STRING interaction network of these 82 KO-responsive genes is shown in Figure \ref{genki_4}(B), suggesting that they tend to be biologically connected with a closely related functional relationship (\textit{p-value} < 0.01, STRING interaction enrichment test).

\subsection{Real-data GenKI analysis recapitulates findings of the Hnf4a-Smad4 double KO experiment}
Using two real scRNA-seq datasets in which a single KO gene was knocked out, we have demonstrated the general performance of GenKI. Next, we investigated whether GenKI is able to virtually predict the effects of double KO (DKO). To accomplish this, we obtained a scRNA-seq data set performed with enterocytes isolated from WT and Hnf4aKO-Smad4KO mice \cite{chen2019reinforcing}. The study reported that Smad4 and Hnf4 work together in a feed-forward loop to activate one another's expression and co-bind to differentiation gene regulatory regions. This feed-forward regulatory module supports and maintains enterocyte cell identity. Loss of this regulatory loop could impair enterocyte differentiation and destabilize enterocyte identity. This intersection of signaling and transcriptional regulation provides a framework for understanding the cellular plasticity of the regeneratable tissue \cite{kim2014broadly}.
In this experiment, we used the WT gene expression profile of 502 enterocytes as the input for GenKI and virtually knocked out Hnf4a and Smad4 simultaneously. 14 KO-responsive genes were reported by GenKI (Supplementary Table \ref{genki_t7s}). The two KO genes, Hnf4a and Smad4, topped the gene list, followed by regenerating islet-derived 1 (Reg1), a regulator of cell growth that is required to generate and maintain the villous structure of the small intestine \cite{ose2007reg}. Hnf4a regulates intestinal epithelium homeostasis and intestinal absorption of dietary lipids \cite{baraille2015glucose}. Loss of this gene is likely to disrupt glucose metabolism, which is regulated by intestinal Reg3b \cite{bluemel2018role}, another significant gene. Also included was Gcg, a gene that may modulate gastric acid secretion and gastro-pyloro-duodenal activity \cite{uniprot2023uniprot}.

\begin{figure}[h]
\centering
\includegraphics[scale=.80]{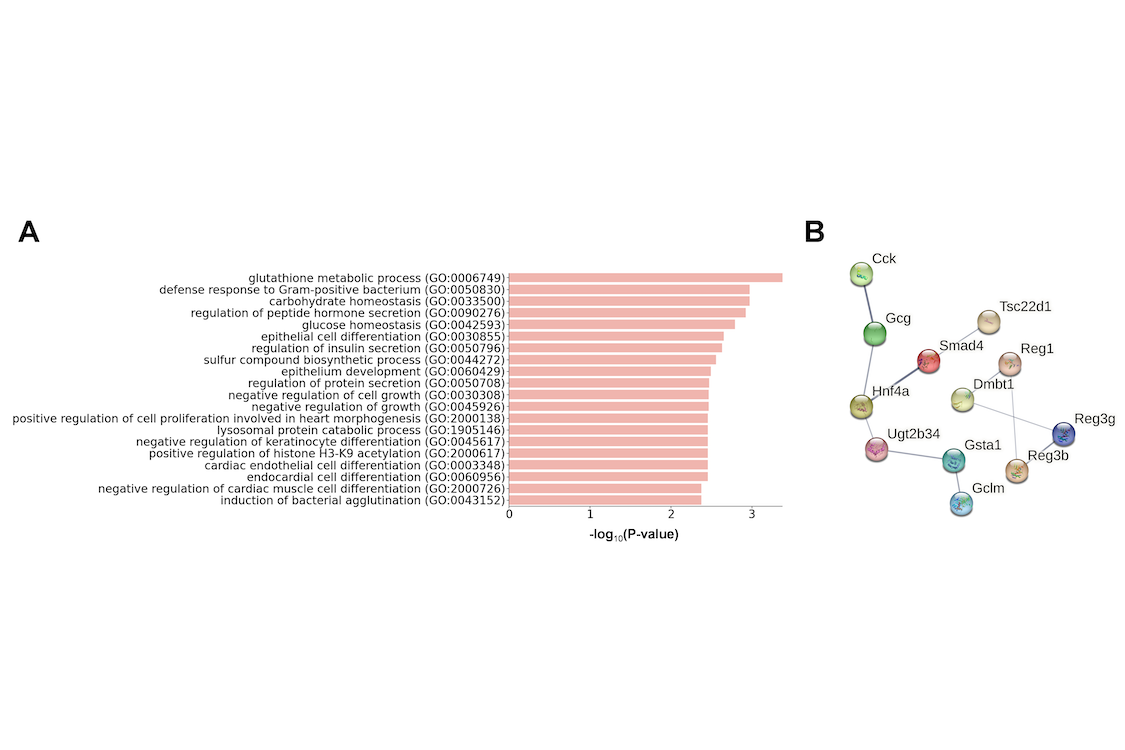}
\caption{Hnf4a \& Smad4-KO responsive genes inferred by GenKI.}
\raggedright
(\textbf{A}) GO terms significantly enriched in functions of inferred Hnf4a-Smad4-KO responsive genes. The -log10-transformed \textit{p-value} indicates the strength of enrichment for each term. (\textbf{B}) STRING subnetwork consists of Hnf4a-Smad4-KO responsive genes.
\label{genki_5}
\end{figure}

Figure \ref{genki_5}(B) depicts the STRING interaction network of these KO-responsive genes. Despite the network being split into two parts under the default setting, we found two disconnected genes, Dmbt1 and Gsta1, were indeed functionally connected—GO enrichment analysis indicates that these two genes were enriched for epithelium cell differentiation (Figure \ref{genki_5}(A), Supplementary Table \ref{genki_t8s}), indicating the loss of enterocytes differentiation after the DKO measure discovered in the original study. Thus, these genes are statistically (\textit{p-value} < 0.01, STRING interaction enrichment test) and biologically linked. Other significant GO terms, such as negative regulation of cell growth and carbohydrate homeostasis correlated with results of the enterocytes study, have also been illustrated in our analysis. Together, this virtual DKO experiment demonstrates that perturbation effects from multiple KO genes are nonlinearly accumulable and can be recapitulated by GenKI.

\subsection{KO-responsive genes are differentially expressed}
We next set out to answer the following question: do KO-responsive genes exhibit differential expression? We first analyzed the expression level changes of predicted KO-responsive genes by comparing them to unperturbed genes across data sets (Materials and Methods). We discovered that the KO-responsive genes predicted by GenKI tend to have greater absolute FC values than unperturbed genes (Supplementary Figure \ref{genki_s6}, \textit{p-value} < 0.05, one-sided t-test). Thus, we came to the conclusion that KO-responsive genes predicted by GenKI are more likely to be differentially expressed.

\begin{figure}[h]
\centering
\includegraphics[scale=.80]{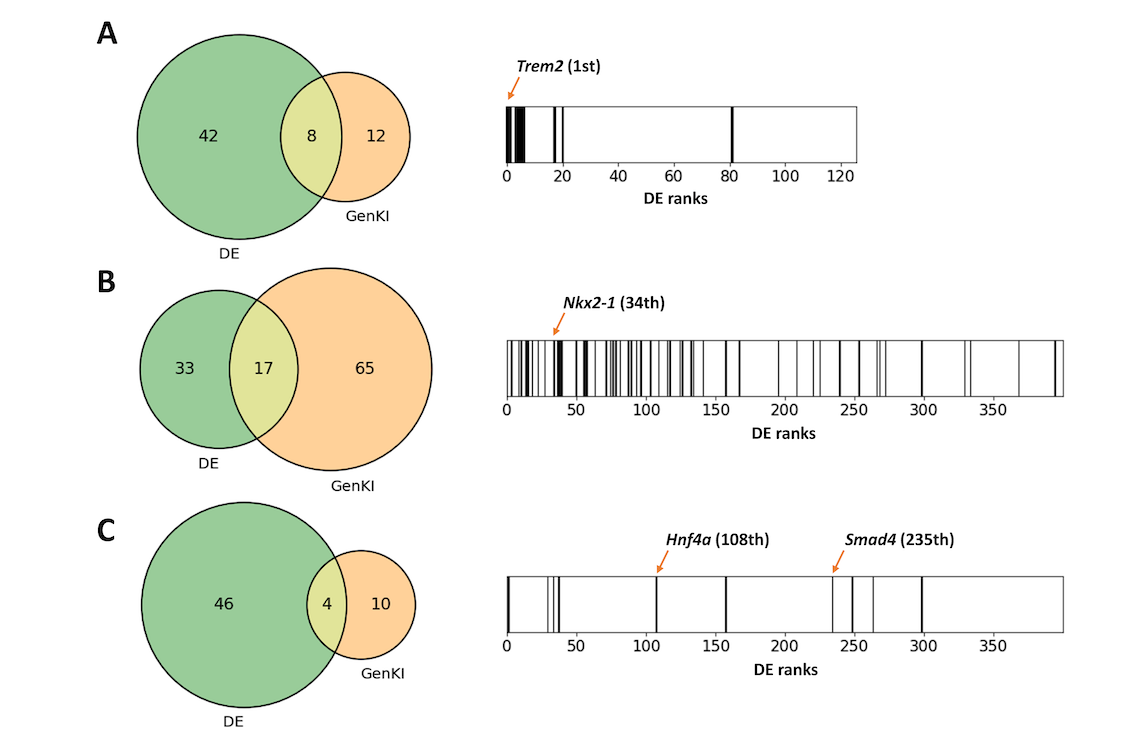}
\caption{Venn diagrams and barcode enrichment plots showing the intersection and differences between the KO responsive genes given by GenKI and DE genes.}
\raggedright
Venn diagram and barcode enrichment plot of (\textbf{A}) microglia data set, (\textbf{B}) lung data set, and (\textbf{C}) intestine data set. All the numbers of overlapped genes were significantly greater than random expectations (\textit{p-value} < 10E-05, hypergeometric test).
\label{genki_6}
\end{figure}

Next we showed that GenKI analysis is different from the DE analysis: KO-responsive genes are not necessarily DE genes. We examined this by comparing the real KO data of each data set to their WT, where 126, 1129 and 1215 DE genes were identified, respectively (Materials and Methods). The overlap between the predicted KO-responsive genes and the top-ranked 50 DE genes in each data set is shown with a Venn diagram in Figure \ref{genki_6} left panel.

The eight overlapping genes of the microglia data set includes Trem2 and other lipoproteins-forming genes like Apoe (Figure \ref{genki_6}(A) left). The 17 intersection genes of the lung data set contain Nkx2-1, the pulmonary surfactant Sttpc and several AT1 and AT2 cell markers (Figure \ref{genki_6}(B) left). Thus, GenKI could be used to predict some of the DE genes. In addition, GenKI identified KO-responsive genes that are not ranked highly by the DE method. By using a barcode enrichment plot (Figure \ref{genki_6} right panel), we were able to visualize the exact locations of the KO-responsive genes across the DE ranks, with each black stick denoting a "hit" of the KO-responsive genes. H2-Aa, a recognized DE gene but not ranked highly (82nd shown in Figure \ref{genki_6}(A) right), is known to function with other genes such as Cd74, Ctsb, and Ctsd in histocompatibility complex (MHC) class $I\!I$ presentation \cite{frigerio2019major}. Napsa, which functions together with Nkx2-1 and Ctsh in the processing of pneumocyte surfactant precursors, was likely to be underestimated (763rd, out of scope in Figure \ref{genki_6}(B) right). The double KO genes Hnf4a and Smad4, which were not included in the intersection of Figure \ref{genki_6}(C) left, weakly ranked 108th and 235th, respectively (Figure \ref{genki_6}(C) right). These perturbed genes were prioritized by GenKI, whereas the DE analysis did not. GenKI further identified KO-responsive genes that are not DE genes. These genes are likely to be at least as important as the DE genes, if not more. For example, concerning the microglia data set, Ctsd is one leading gene involved in cholesterol metabolism \cite{deczkowska2018disease}, and Cx3cr1 and Tyrobp play an important role in macrophage activation \cite{burgess2019cx3cr1, liang2021tyrobp, dang2020computational}. All of them were not the DE genes.

Do DE genes appear more adjacent to KO-responsive genes in a scGRN? To answer this question, we performed the STRING network analysis by combining the top-ranked DE genes with the KO-responsive genes using the microglia data set as an example. The outcome is depicted in Supplementary Figure \ref{genki_s7}, showing that 23 out of 42 DE genes are directly or indirectly linked to the KO genes. That is to say, in this given case, more than half of DE genes might be functionally involved in the perturbed KO gene network identified by GenKI.

Utilizing DE and GenKI analyses in a complementary manner might be a good idea. To illustrate our point, we applied seven different DE analysis methods and settings to the lung data and summarized the number of DE genes detected and their intersection with GenKI-identified genes (Supplementary Table \ref{genki_t9s}). We found that the result of a DE analysis were largely depend on what method was selected to use and what fold-change and \textit{p-value} cutoffs were set, and the functional interpretation of the DE analysis results was also depended whether up- and down-regulated genes are pooled together. In general, we found different DE methods with varying model assumptions and thresholds could not converge to a consensus set of DE genes. The number of DE genes and their rankings changed greatly depending on many technical factors as mentioned. Furthermore, most DE methods with default settings produce excessive numbers of DE genes, making downstream functional enrichment analysis difficult and obscuring true signals caused by the perturbation itself to be detected. GenKI, on the other hand, as a method independent of DE methods, provides additional evidence for gene functions. Most of GenKI's KO-responsive genes overlapped with DE genes regardless of the DE method. With the default setting, GenKI produced fewer significant genes than DE methods, which may improve the interpretability of gene function. In this sense, we are not developing an alternative to DE, but rather a complementary technique that produces more targeted results.

\subsection{Real-data GenKI analysis predicts function of key transcriptional factor STAT1}
Above we have validated GenKI performance by comparing the inference results to DE genes using three scRNA-seq data sets that all included WT and KO groups. We questioned whether GenKI is able to reveal gene functions of any target gene from a standalone WT scRNA-seq data set without pairing it with a KO counterpart, which should be a more common occurrence when using virtual KO tools. We obtained a data set from a study of 19 patients with severe coronavirus disease 2019 (COVID-19) \cite{chua2020covid}. It contains 8,920 cells collected from nasopharyngeal and bronchial samples. The study found that epithelial cells of COVID-19 patients showed an average three-fold increase in expression of the SARS-CoV-2 entry receptor ACE2, and signal transducer and activator of transcription 1 (STAT1), a central transcription factor of the interferon response, was among the top predictors for ACE2 expression. Previous research also shows that STAT1 is critical for virus clearance and disease resolution, and STAT1 KO mice have impaired interferon gamma (IFNG) signaling \cite{sun2020generation}. In this virtual KO experiment, we focused on a subpopulation of pulmonary epithelial cells differentiating from immature secretory cells to ciliated cells. The original study demonstrated an alternative differentiation pathway leading from immature secretory cells directly into ciliated cells mediated by these IFNG-responsive epithelial cells, suggesting that this direct differentiation pathway is dependent on the interferon response \cite{chua2020covid}.

We virtually knocked out STAT1 in these epithelial cells. Firstly, we validated the robustness of our model by artificially adding different levels of random noise to the gene expression profile (Supplementary Figure \ref{genki_s8}). The GenKI analysis identified 28 STAT1-KO responsive genes (Supplementary Table \ref{genki_t10s}). STAT1 was ranked at the top, followed by three human leukocyte antigen (HLA) genes (HLA-DRA, HLA-DRB1, HLA-DPA1), which are known to encode Class $I\!I$ major histocompatibility complex (class $I\!I$ MHC). Class $I\!I$ MHC, which are reported to be highly expressed only in antigen-presenting cells (APC), is induced in other cell types as well by inflammation or IFNG \cite{muhlethaler1997expression}. Moreover, lysosomes are required for lysis of the protein into peptides for class $I\!I$ MHC presentation to the immune cells \cite{roche2015ins}. In our inferred gene list, the lysosome-related genes CTSB, CTSD, and CSTB were included, and were related to the antigen-presenting process. Previous research indicates that the nuclear factor-$\kappa$B (NF-$\kappa$B) can be activated by IFNG \cite{pfeffer2011role}. This is consistent with genes in the list believed to participate in NF-$\kappa$B-related pathways and inflammation. For example, ANXA1 is reported to have anti-inflammation activity in lung endothelial cells and is able to prevent lung fibrosis \cite{damazo2011endogenous}. GPX1 participates in the NF-$\kappa$B pathway and is crucial for respiratory virus infection \cite{seale2020role}. S100 family proteins are well-characterized for their function in inflammation and innate immunity \cite{singh2022multifunctional}. Additionally, S100 proteins are damage-associated molecular patterns (DAMP) that promote inflammation by binding to the pattern-recognition protein (PRR) \cite{yang2005cytokine}. HSPB1 and HSPB5 belonging to DAMP are also listed.

\begin{figure}[h]
\centering
\includegraphics[scale=.90]{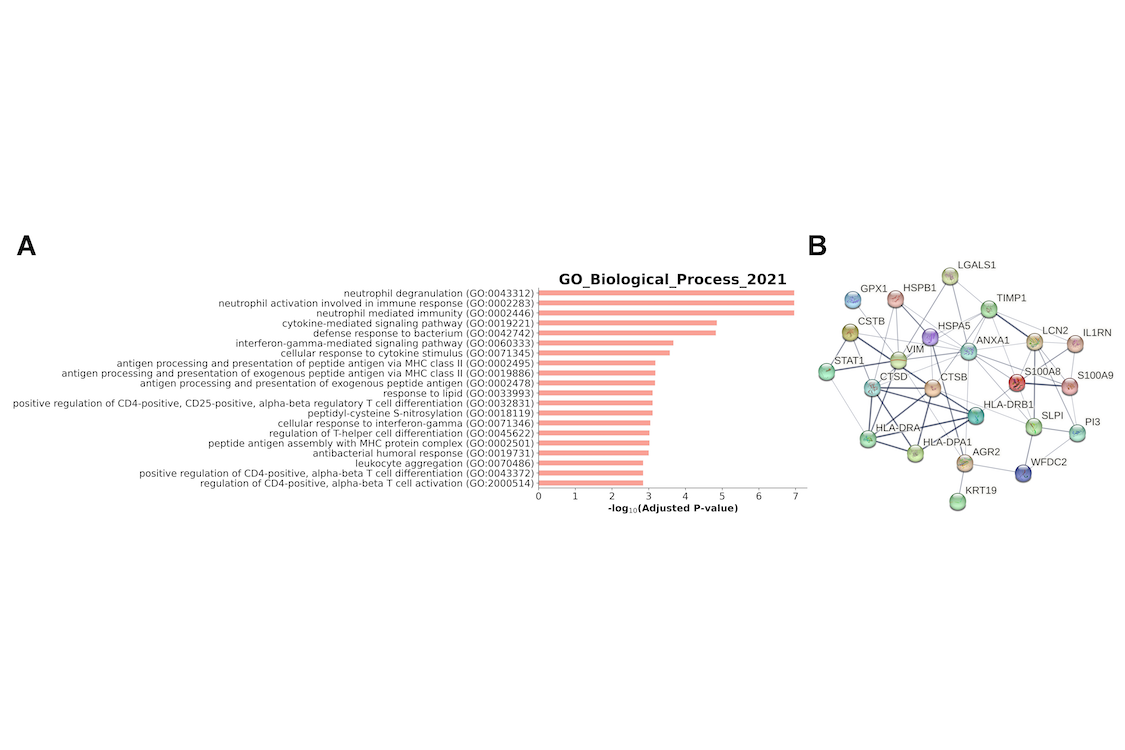}
\caption{STAT1-KO responsive genes inferred by GenKI.}
\raggedright
(\textbf{A}) GO terms significantly enriched in functions of STAT1-KO responsive genes. The -log10-transformed adjusted \textit{p-value} indicates the strength of enrichment for each term. (\textbf{B}) STRING network consists of STAT1-KO responsive genes.
\label{genki_7}
\end{figure}

The result of GO enrichment analysis is presented in Figure \ref{genki_7}(A) and Supplementary Table \ref{genki_t11s}. The neutrophil-related pathways were ranked at the very top in the enrichment analysis, suggesting the communication between IRC and neutrophils, which is in agreement with the finding of the original study \cite{chua2020covid}. The interferon-gamma, class $I\!I$ MHC antigen-presenting, NF-$\kappa$B, and innate immunity-related pathways were also detected by the enrichment analysis. Thus, these results strongly suggest that the GenKI is able to accurately predict the potential perturbed genes and their shared functions. We further analyzed 28 genes using STRING to understand their interaction (Figure \ref{genki_7}(B)). The resulting subnetwork, which contains significantly more interactions than expected (\textit{p-value} < 0.01, STRING interaction enrichment test), again suggests that these genes are closely connected due to their shared biological functions. This virtual KO study demonstrates that GenKI can reliably predict gene functions and infer the molecular phenotypic consequences of genes of interest validated by previous studies without the need for an actual KO experiment.

\subsection{GenKI is robust and scalable}
To assess the robustness of GenKI inference, we collected scRNA-seq data \cite{zeisel2018molecular} from mouse neurons with Rett syndrome (RTT), a severe neurodevelopmental disorder. Mutations in Mecp2, a transcriptional repressor required to maintain normal neuronal functions, are known to cause RTT \cite{nan1997mecp2, lyst2015rett}. This data set contains two replicates with 2,054 and 2,156 neurons, respectively. We independently analyzed these two replicates with GenKI, in which we virtually knocked out the same KO gene Mecp2. Given the high similarity of these two biological replicates, GenKI would be robust if it generated roughly equivalent gene ranks across them. Indeed, we found high consistency between the rankings of the two reported rank lists (Spearman’s correlation coefficient $\rho$ = 0.82).
Finally, we evaluated the computation efficiency of GenKI. Supplementary Figure \ref{genki_s9} shows the results of the analysis, comparing the total running time with respect to different sizes of input scRNA-seq data sets. The running time for GenKI consists of scGRN construction, training, and inference. We simulated four random data sets at different scales for this comparison. Without using GPUs, GenKI exhibited a 2.8- to 4.9-fold faster running speed than scTenifoldKnk tested on equivalent hardware. GenKI is expected to run even faster by enabling the fast GPU implementation optimized by PyTorch Geometric \cite{fey2019fast}.

\section{Discussion}
In this study, we showcased the functionality and performance of GenKI in virtual KO experiments. We first evaluated the inference performance of GenKI using simulated data sets (SERGIO and BEELINE). Next, we used scRNA-seq data sets generated in real KO experiments to show that GenKI could predict gene functions by identifying and annotating KO-responsive genes. The functional predictions were found to be consistent with original studies in which WT and KO scRNA-seq data sets were generated.

Our main contribution in this work is to provide a neural network-based virtual KO analytical tool, which encodes the gene expression matrix to a latent space given its underlying scGRN. To the best of our knowledge, GenKI is the first virtual KO tool using a graph-based generative model to infer KO-responsive genes and their shared functions. Several computational tools have been developed for similar purposes to predict the effects of genetic perturbation using single-cell data. scGen \cite{lotfollahi2019scgen} and CPA \cite{lotfollahi2023predicting}, both running in a supervised manner, require massive training data labeled with various perturbations to train their autoencoder-based models. CellOracle \cite{kamimoto2023dissecting} can simulate gene expression in response to TFs perturbation by signal propagation through its inferred scGRN. However, this simulation is linear and does not quantify the level of perturbation at individual gene level. More importantly, it requires scATAC-seq data along with the corresponding scRNA-seq data to build the scGRN prior to making such an inference, which may limit its application. scTenifoldKnk \cite{osorio2022sctenifoldknk} is the only virtual KO tool with the identical input requirements as GenKI. Like GenKI, scTenifoldKnk only requires WT scRNA-seq data for its prediction analysis. It employs manifold alignment \cite{ma2011manifold} to project WT and virtual KO scGRNs to a joint low-dimensional space and calculate the differences between them. Given the minimalistic design, GenKI shares with scTenifoldKnk several key advantages such as being species agnostic—that is, they both work with scRNA-seq data from humans and animal models alike. By applying these tools directly to human data instead of surrogate animals, researchers may avoid pitfalls caused by overextending their conclusions from animal models to humans. Additionally, both GenKI and scTenifoldKnk allow any gene to be knocked out, regardless as to whether the KO genes are functionally vital or not. Knocking out a vital gene tends to cause fatal consequences and is, therefore, impractical to generate animal models for its KO.

GenKI outperforms scTenifoldKnk in the following aspects. First, scTenifoldKnk only utilizes the WT scGRN, while GenKI takes into account both the WT gene expression profile and scGRN. Second, the VGAE model, which consists of two message passing layers, collects information up to the second-order neighborhood of the network. In contrast, manifold alignment adopted in scTenifoldKnk only maintains the similarity of directly connected neighbors of the network, which results in different levels of inference power. In addition, GenKI shows better scalability, being able to process tens of thousands of cells within a reasonable time. Once the GenKI model is trained, the model can be reused for virtual KO of any genes in the data. While in order to do the same, scTenifoldKnk must re-solve the manifold alignment problem for each KO gene by eigen decomposition, which is considered computationally intensive and time-consuming. Last, GenKI avoids a pitfall in numerical computation in scTenifoldKnk. scTenifoldKnk performs a virtual KO experiment by removing the edges of a KO gene in the scGRN, which results in an asymmetric Laplacian matrix containing negative values. This potentially leads to eigenvectors of the Laplacian matrix with imaginary parts when solved by eigen decomposition. scTenifoldKnk practically adds 1 to all entries in obtained scGRNs to guarantee that all the entries are positive and only uses the real parts of obtained eigenvectors. GenKI’s architecture allows it to bypass this problem because it employs neural networks to solve the optimization problem, which has been shown to be numerically more stable than eigen decomposition \cite{nguyen2022deep}.

We addressed the question that end users may often have, i.e., “Are KO-responsive genes more likely to be differentially expressed?” DE analysis, followed by gene function enrichment analysis, are often used to identify the perturbed gene expression programs in order to understand the function of the KO gene. The problem is that the perturbation effect of the KO gene may propagate on the underlying network but may not direct reflected as observable and measurable changes in gene expression. GenKI, on the other hand, works on scGRNs directly to leverage unobservable network-level information—GenKI identifies perturbed genes through modelling underlying networks. Therefore, in contrast to DE analysis that can only detect perturbed genes with significant expression level changes, GenKI is likely to detect perturbed genes even there are less or no significant expression level changes. Perturbed genes without expression level changes are not uncommon. For instance, given a gene that is under control of multiple regulators, even if one of its regulators is knocked out, the remaining regulators may still be functioning to compensate and stabilize the given gene’s expression. Additionally, with the default setting, GenKI produced fewer significant genes than a typical DE analysis, which may improve the interpretability of gene function. In conclusion, GenKI is not an alternative to DE analysis, but rather a complementary technique that produces more targeted results.

The limitations of GenKI are mostly inherited from it being virtual. GenKI cannot be used to predict the regulatory direction of KO-responsive genes, which is important in learning cell responses to external stimuli \cite{anders2010differential}. If future refinements enable directional predictions, GenKI may improve with its potential ability to simulate the effect of overexpression. Also, GenKI, like scTenifoldKnk, currently performs a virtual KO experiment by removing all the edges of a KO gene in the WT scGRN. This action might be naïve given the complexity of a biological system. A virtual KO scGRN could be better modeled by simulating the virtual KO effect in a more probabilistic manner. Alternatively, there are many available priors involved in many different types of KO; hence a Bayesian treatment may facilitate the KO inference. GenKI is also inapplicable to bulk RNA-seq data, as genes in such data lose their variability in terms of gene expressions, which results difficulty in scGRN construction using PC regression and assigning expression values to node attributes in a graph. Recent advances in cell pseudo-temporal ordering enable us to map the underlying scGRNs throughout time \cite{reid2016pseudotime, xu2022scintime} and eventually learn temporal KO effects including cell-cell communication \cite{yang2023sctenifoldxct} in a dynamic manner. GenKI can be improved by incorporating a dynamic inference module to investigate such effects on cell or organ development.

\def\dis{\mathrm{dis}}
\def\MLP{\mathrm{MLP}}
\def\total{\mathrm{total}}
\def\fader{\mathrm{Fader}}
\def\NLC{\mathrm{NLC}}
\newcommand{\coloneqq}{\mathrel{\vcenter{:}}=}

\chapter[CrossmodalNet: INTERPRETABLE MODELING OF TIME-RESOLVED SINGLE-CELL GENE-PROTEIN EXPRESSION]{CrossmodalNet: INTERPRETABLE MODELING OF TIME-RESOLVED SINGLE-CELL GENE-PROTEIN EXPRESSION\footnote{Reprinted with permission from Y. Yang et al., "Interpretable modeling of time-resolved single-cell gene–protein expression with CrossmodalNet," Briefings in Bioinformatics, vol. 24, no. 6, pp. bbad342, 2023.}}\label{chap:crossmodal}

\section{Introduction}
Compared to traditional single-cell techniques \cite{osorio2020sctenifoldnet, xu2022scintime, osorio2022sctenifoldknk, yang2023sctenifoldxct, jindal2018discovery, yang2023gene} that measure only one aspect of cellular activity, the ability of multimodal \cite{ling2023graph, han2022g, han2022geometric} approaches has the potential to significantly improve our understanding of cellular behavior and function, thereby shedding light on a vast array of biological questions. Cellular Indexing of Transcriptomes and Epitopes by Sequencing (CITE-seq) is a cutting-edge sequencing method that allows simultaneous measurement of gene and surface protein expression at the single-cell level \cite{stoeckius2017simultaneous}.
CITE-seq, however, faces some challenges. First, CITE-seq experiments are costly and require specialized equipment and trained personnel. Second, the number of available antibodies limits the number of surface proteins that CITE-seq can measure. This is problematic when attempting to analyze complex cell populations. Antibody cross-reactivity and non-specific binding may also result in false CITE-seq discoveries \cite{restani2002cross}. Moreover, RNA and surface proteins are sometimes intrinsically with low correlation \cite{yuan2022clustering}.

Machine learning methods have been developed to learn the relationship between genes and proteins and translate between single-cell measurements of these two modalities. Seurat 4 \cite{hao2021integrated} and totalVI \cite{gayoso2021joint} solve this mapping problem in  reference-mapping approach. By learning and integrating the relationship between RNA and protein expression in a large reference data set, Seurat 4 and totalVI predict surface protein abundance from a query scRNA-seq data and assign cell types. However, these processes can be computationally expensive, and lack of large reference data makes it difficult to gain biological insights. 
Recent work scIPENN \cite{lakkis2022multi} is a multi-use framework for CITE-seq and scRNA-seq integration with surface protein abundance prediction and imputation. However, scIPENN’s RNN blocks could cause the gradient vanishing problem and potentially hinder the training process \cite{pascanu2013difficulty}. More importantly, its inability to interpret what and how genes significantly regulate protein expression levels over time may limit its application in understanding essential cellular system questions.

Here we propose CrossmodalNet, an interpretable machine learning model with customized adaptive loss that learns to translate between modalities of genes and proteins using CITE-seq data while encoding temporal information, and can accurately predict protein expression using only scRNA-seq data at chosen time points. We evaluate the performance of our model using three publicly available time-resolved CITE-seq data sets including hematopoietic stem and progenitor cells (HSPCs) and human peripheral blood mononuclear cells (PBMCs). By combining the interpretability of linear models with the flexibility of non-linear models, we show that our model decomposes transcriptional information of cells into basal and temporal domain, with the latter forming an easy-to-interpret time embedding. Using the learned time embedding, we demonstrate that our model outperforms other benchmarking methods for protein prediction at both observed and unobserved time points. Moreover, we show our model is capable of elucidating noise-free causal gene-protein relationships that are typically investigated in large-scale genomic studies.

\section{Methods}
Let $X\in \mathbb{R}^{m\times g}$ and $Y\in \mathbb{R}^{m\times p}$ represent gene and protein expression profiles from CITE-seq, respectively, with $m$ cells, $g$ genes, and $p$ proteins. We are given a training data set $\mathcal{D}=\{(x_1,y_1,t_1), \dots, (x_m,y_m,t_m)\}$, and our objective is to learn $y_i$ given $x_i$ at discrete time $t_i$ for each cell $i$ using data $\mathcal{D}$.

\begin{figure}
\centering
\includegraphics[width=0.70\columnwidth]{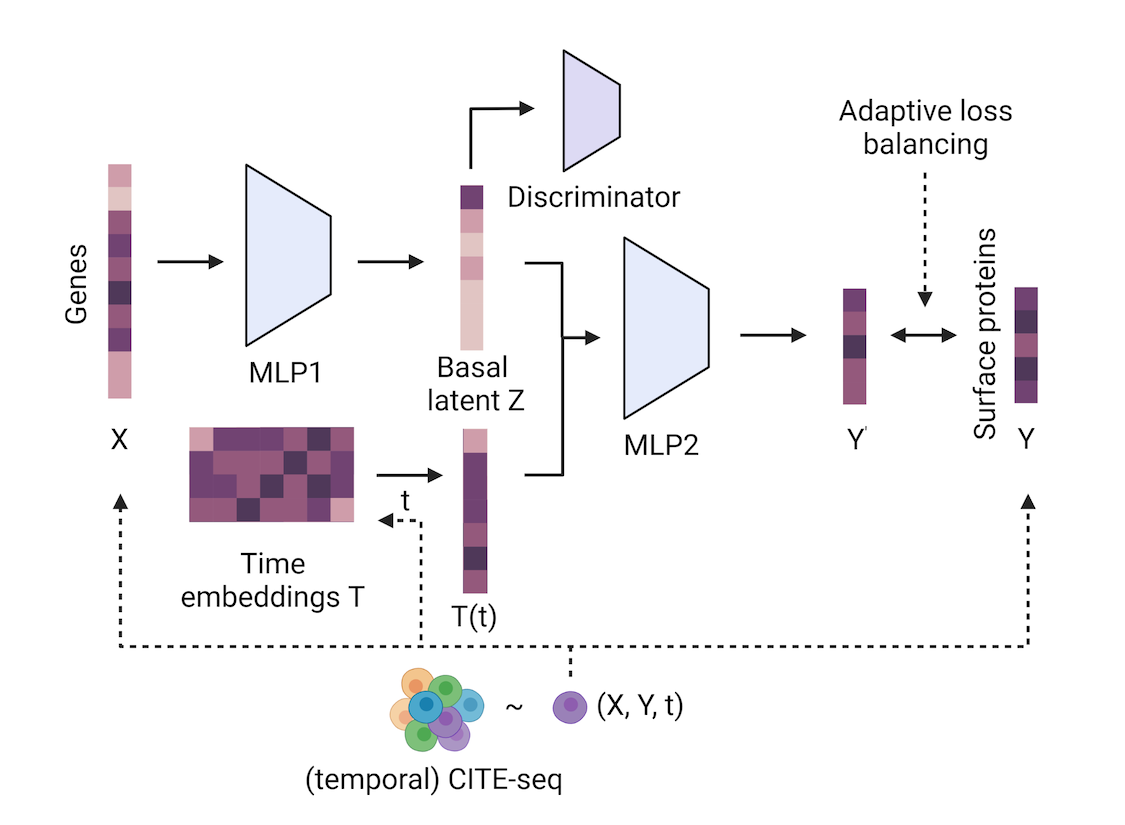}
\caption{The framework of CrossmodalNet predicting protein expressions given gene expression profiles using temporal CITE-seq data.}
\label{fig1}
\end{figure}

We adopt a framework similar to Fader Networks \cite{lample2017fader} that facilitates the incorporation of interpretable covariates. This framework enables a nonlinear mapping between gene and protein expression, and we utilize a multi-task training strategy to optimize the model's performance (Figure~\ref{fig1}).  Recent studies such as CPA \cite{lotfollahi2023predicting} and its subsequent research, MultiCPA \cite{inecik2022multicpa}, leverage the Fader Networks for predictions of drug responses and genetic perturbations. Below, we provide further details about our approach.

\subsection{Fader Networks}
We begin by initializing a learnable time embedding, denoted as $T\in \mathbb{R}^{c\times d}$, where $c$ represents the number of unique time classes, and $d$ represents the latent dimension of the model. The $d$-dimensional time representation for the $i$-th sample is denoted as $T_i \coloneqq T(t_i)$. 

To define the architecture of our model, we introduce the Linear-BatchNorm block $C_k$, which generates $k$ output features. Our model consists of two Multi-Layer Perceptrons (MLPs). The first MLP (MLP1) consists of $C_d - ReLU - Dropout - C_d$ and the second MLP (MLP2) consists of $C_d - ReLU - C_{out}$, without BatchNorm \cite{santurkar2018does} in the last output layer. 

\subsubsection{Discriminator objective}
We introduce a discriminator that calculates the probability of a time point $t_i$ given the input $x_i$. The objective function of the discriminator is defined as
\begin{align*}
\mathcal{L}_{\dis}(\theta_{\dis};\theta_{\MLP1}) = -\frac{1}{m}\sum_{i = 1}^m \log P_{\theta_{\dis}}\bigl(t_i|F_{\theta_{\MLP1}}(x_i)\bigr),
\end{align*}
where $\theta_{\dis}$ and $\theta_{\MLP1}$ are parameters of discriminator and MLP1, respectively. A well-trained discriminator will enable a cell’s basal latent state disentangled from the time.

\subsubsection{Adversarial objective}
Denote the the basal latent state as $z_i = F_{\theta_{\MLP1}}(x_i)$, we next aggregate  $z_i$ of cell $i$ and its time representation $T_i$ into a unified space, and then map the sum to the protein expression. The Fader loss, given the discriminator parameters $\theta_{\dis}$, is:
\begin{align*}
&\mathcal{L}_{\fader}(\theta_{\MLP1}, \theta_{\MLP2} ; \theta_{\dis}) \\
= &\frac{1}{m}\sum_{i = 1}^m \begin{aligned}[t]
\biggl(&\|F_{\theta_{\MLP2}}\bigl(F_{\theta_{\MLP1}}(x_i),T_i\bigr) - y_i\|^2_2\\
& + \lambda_{\dis}\cdot\log P_{\theta_{\dis}}\bigl(t_i; F_{\theta_{\MLP1}}(x_i)\bigr)\biggr),  
\end{aligned}
\end{align*}
where $\theta_{\MLP2}$ is parameters of MLP2 and $\lambda_{\dis}$ is a regularize parameter. Minimizing the training loss requires optimizing both the mean squared error (MSE) reconstruction loss, which is used to reconstruct $y_i$, and the cross entropy loss, which is used to predict $t_i$. We denote the reconstructed protein expression by
\[\widehat y_i = F_{\theta_{\MLP2}}\bigl(F_{\theta_{\MLP1}}(x_i),T_i\bigr).\]

\subsection{Multitask training}
We propose a de novo loss function called the negative log-correlation (NLC) loss, as illustrated in Figure S1 in the supplementary material. The NLC loss effectively governs the correlation between the predicted values $\widehat y_i$ and the actual values $y_i$, and it can be utilized for backpropagation. The formulation for the NLC loss is given by:
\[\mathcal{L}_{\NLC, i} = -\log\left(\max\left(\frac{(\text{Corr}(\widehat y_i, y_i) + 1)}{2}, \epsilon\right)\right),\]
where $\epsilon$ is a small value to stable the computation. Denote the NLC loss by
\[\mathcal{L}_{\NLC} = \sum_{i=1}^m \mathcal{L}_{\NLC, i},\]
and the total loss of the entire model is defined as:
\[\mathcal{L}_{\total} = w_1 \mathcal{L}_{\NLC} + w_2\mathcal{L}_{\fader}(\theta_{\MLP1}, \theta_{\MLP2} ; \theta_{\dis}),\]
where $w_1$ and $w_2$ are adaptive parameters balancing the two losses.

We employ GradNorm method proposed by \cite{chen2018gradnorm} to optimize the loss function, which would improve model performance and reduce overfitting when compared to single-task models. Denote the parameters of the first layer in $\theta_{\MLP1}$ as $\omega$. To this end, we define 
\[G_{\NLC} = \|\nabla_{\omega} w_1 \mathcal{L}_{\NLC}\|_2^2,\text{ and }G_{\fader} = \|\nabla_{\omega} w_2 \mathcal{L}_{\fader}\|_2^2,\]
where $G_{\NLC}$ and $G_{\fader}$ represent the $L_2$ norms of the gradients of the weighted single-task loss with respect to $\omega$. We introduce the average gradient norm across both tasks, denoted as $\bar G = \frac{1}{2}(G_{\NLC} + G_{\fader})$, as a key component of the GradNorm approach, where the goal of GradNorm is to match the scale of $G_{\NLC}$ and $G_{\fader}$. 

Let $\mathcal{L}_{\NLC}(t)$ and $\mathcal{L}_{\fader}(t)$ be the loss function values at the $t$-th iteration. During the training process at the $t$-th iteration, we calculate the NLC loss ratios, defined as 
\[\widetilde{\mathcal{L}}_{\NLC}(t) = \mathcal{L}_{\NLC}(t) / \mathcal{L}_{\NLC}(0),\]
where $\mathcal{L}_{\NLC}(0)$ represents the loss value at the initialization stage. A similar definition is applied for $\widetilde{\mathcal{L}}_{\fader}(t)$ and $\mathcal{L}_{\fader}(0)$. Furthermore, we compute the average loss function value at time $t$ as $\bar L(t) = \frac{1}{2}(\widetilde{\mathcal{L}}_{\NLC}(t) + \widetilde{\mathcal{L}}_{\fader}(t))$. The relative inverse training rates for each loss is denoted by $r_{\NLC}(t) = \tilde L_{\NLC}(t) / \bar L(t)$, and $r_{\fader}(t)$ employing a similar formulation. For a detailed illustration of the training process with CrossmodalNet, refer to Algorithm~\ref{alg}.

\begin{algorithm}
   \caption{Training with CrossmodalNet}\label{alg}
   \begin{algorithmic}[1]  
      \REQUIRE data set $\mathcal{D}$, time embedding $\{T_i\}_{i =1 }^m$
      \STATE Initialize $w_1 = w_2 = 1$
      \STATE Initialize network weights $\theta_{\dis}$, $\theta_{\MLP1}$ and $\theta_{\MLP2}$
      \STATE Compute the loss function value $\mathcal{L}_{\NLC}(0)$ and $\mathcal{L}_{\fader}(0)$.
      \FOR{$t = 1$ {\bfseries to} max\_train\_steps}
         \IF{$t \bmod \text{learning\_step} = 0$}
            \STATE Calculate discriminator loss $\mathcal{L}_{\dis}$
            \STATE Compute gradients $\nabla_{\theta} \mathcal{L}_{\dis}$ with respect to $\theta_{\dis}$
            \STATE Update $\theta_{\dis}$ using $\nabla_{\theta} \mathcal{L}_{\dis}$
         \ELSE
            \STATE Compute $\mathcal{L}_{\fader}(t)$, $\mathcal{L}_{\NLC}(t)$ and $\mathcal{L}_{\total}(t)$
            \STATE Compute standard gradients $\nabla_{\theta} \mathcal{L}_{\total}(t)$ with respect to $\theta_{\MLP1}$ and $\theta_{\MLP2}$
            \STATE Compute $G_{\NLC}$, $G_{\fader}$ and $\bar{G}$.
            \STATE Compute
            \begin{align*}
               \mathcal{L}_G &= \|G_{\NLC} - \bar G\times r_{\NLC}^{\alpha}\|_1 + \|G_{\fader} - \bar G\times r_{\fader}^{\alpha}\|_1
            \end{align*}
            \STATE Compute GradNorm gradients $\nabla_{w_i} \mathcal{L}_G$ for $i = 1, 2$
            \STATE Update $w_1$ and $w_2$ using $\nabla_{w_i} \mathcal{L}_G$
            \STATE Update $\theta_{\MLP1}$ and $\theta_{\MLP2}$ using $\nabla_{\theta} \mathcal{L}_{\total}$
         \ENDIF
      \ENDFOR
   \end{algorithmic}
\end{algorithm}

\subsection{Hyperparameter tuning and implementation}
We conduct a random hyperparameter search with 100 trials using Ray Tune v2.0.0 \cite{liaw2018tune}. The distribution of values for the hyperparameter search is provided in Table S1 in the supplementary material. 

To implement this process, we initially divide the cells into three data sets: training (80\%), validation (5\%), and testing (15\%). We perform random sampling for in-distribution predictions to ensure that the proportions of cells at different time points are equal in each set. For out-of-distribution predictions, the cells at a specific time point are reserved for the testing set, while the remaining cells are designated for the training and validation sets. During the training process, we set the maximum iteration number to 500 and implement early stopping after the Pearson correlation coefficient of the validation set reaches its maximum for 10 consecutive iterations. The Adam optimizer is used for all training tasks.

\subsection{Causal gene-protein relationship inference}
Due to the nonlinearity of CrossmodalNet, we employ saliency maps \cite{simonyan2013deep} to discern the importance of input features for the output. In the context of causal gene-protein relationship analysis, we compute the saliency of a protein $j$ with respect to genes by aggregating all cells:
\[\gamma_j = {\sum_{i = 1}^m \frac{\partial y_i(j)}{\partial {x_i}}} \in \mathbb{R}^{g}.\]
Since our latent inference only contains basal information, we anticipate the saliency analysis will reveal more noise-free gene-protein relationships.

\section{Experiment setup}
\subsection{Data}
We use three real CITE-seq data sets for model training and evaluation.
\begin{itemize}
    \item HSPC data set: This data set was collected from over 70,000 mobilized peripheral CD34+ HSPCs isolated from four healthy human donors across three time points from Kaggle Open Problems in Single-Cell Analysis \cite{kaggle2022} guided by \cite{velten2017human} where 140 surface proteins were measured.
    \item Myeloid data set: This data set was collected from over 47,000 peripheral blood mononuclear cells of patients with advanced biliary tract cancer (BTC) across three time points following anti-PD-1 treatment where 99 surface proteins were measured \cite{keenan2022circulating}. Four CD14+ monocyte sub-populations by responsive BTC patients are used.
    \item PBMC data set: This data set containing a total of 161,764 cells was collected from eight volunteers enrolled in an HIV vaccine trial at three pre- and post-vaccination time points, where 228 surface proteins were measured \cite{hao2021integrated}.
\end{itemize}

\subsection{Preprocessing}
We perform RNA library-size normalization and log1p transformation on scRNA-seq data using NormalizeData function from Seurat v4.0.2 package \cite{hao2021integrated}. We perform dsb transformation for surface protein data using DSBNormalizeProtein function from dsb package v1.0.2 \cite{mule2022normalizing}. Both normalization methods are performed after data split. All default settings are retained and used.

\subsection{Model evaluation}
We evaluate our model against several baseline methods including linear regression, ridge regression, lightGBM \cite{ke2017lightgbm}, MLP, Seurat 4, totalVI, and scIPENN. We utilize scikit-learn v1.2 to build the linear and ridge regression models with default hyperparameters. The LightGBM model is built using LightGBM v3.3.5, and we use randomized cross-validation to select the optimal hyperparameters. To benchmark our method against multimodal reference mapping methods, we follow the documentation of Seurat 4 and totalVI (sci-tools v1.0.2). Pytorch v2.0 \cite{paszke2019pytorch} and pytorch-lightning v2.0.2 \cite{falcon2019pytorch} are used to create the MLP models. Similar to how we tune our model, we utilize Ray to determine the best model structure and hyperparameters.  We evaluate each model's performance using MSE and Pearson correlation coefficient.

\section{Results}
\subsection{CrossmodalNet accurately translates between gene-protein modalities}
Table~\ref{in-table} compares the performance of CrossmodalNet and other methods under the in-distribution setting. Our result indicates that CrossmodalNet is capable of achieving the highest Pearson correlation coefficient of all methods. In addition, even though linear models show good predictive performance, they are not superior to lightGBM and MLP models. In particular, MLPs trained with MSE loss achieve the relatively low MSE, which is expected. It should be noted, however, these MSE models do not produce the highest Pearson correlation coefficient, indicating that a single MSE loss may not be the optimal choice for this task. Notably, the multimodal reference-based mapping methods employed in this study, namely Seurat 4 and totalVI, generate normalized protein expression values based on their built-in normalization approaches. Therefore, their MSE values cannot be directly compared to other methods that predict dsb-normalized protein abundance. 

\begin{table}[htbp]
   \caption{In-distribution predictions.}\label{in-table}
   \centering
   \begin{tabular}{ccccccc}
      \hline
      Data set & \multicolumn{2}{c}{HSPC} & \multicolumn{2}{c}{Myeloid} & \multicolumn{2}{c}{PBMC}\\
      \hline
      Method & Corr & MSE & Corr & MSE & Corr & MSE\\
      \hline
      CrossmodalNet & \textbf{0.593} & 0.259 & \textbf{0.819} & 0.249 & \textbf{0.448} & 1.772\\
      MLP+MSE & 0.561 & \textbf{0.140} & 0.792 & 0.160 & 0.385 & 1.784\\
      MLP+NLC & 0.490 & 0.316 & 0.560 & 0.670 & 0.390 & \textbf{1.682}\\
      lightGBM & 0.575 & 0.259 & 0.791 & \textbf{0.151} & 0.388 & 1.902\\
      Linear reg. & 0.559 & 0.269 & 0.745 & 0.301 & 0.386 & 1.908\\
      Ridge reg. & 0.560 & 0.268 & 0.772 & 0.304 & 0.387 & 1.907\\
      scIPENN & 0.379 & 0.456 & 0.429 & 0.489 & 0.083 & 2.241\\
      totalVI & 0.413 & - & 0.153 & - & 0.253 & -\\
      Seurat 4 & 0.441 & - & 0.341 & - & 0.429 & -\\
      \hline
   \end{tabular}
\end{table}

\subsection{CrossmodalNet generalizes to unobserved time point}
To demonstrate the generalization of the CrossmodalNet model, we hold out cells at an intermediate time point and train with cells preceding and following the time point. Specifically, we hold out the HSPC from day 3 and train day 2 and 4; the monocytes from week 2 and train week 1 and 3; the PBMC from day 3 and train day 0 and 7, respectively. After obtaining the learnable time embedding, the unseen time representation is inferred through a linear interpolation between two learned vectors of the time embedding. During testing, we manually concatenate this inferred time representation vector with basal latent representations of cells given by the trained model to obtain predictions.

In Table~\ref{out-table}, we compare the performance of CrossmodalNet and other methods under this setting. Our results show that CrossmodalNet outperforms other methods in most scenarios, indicating its high generalizability. Interestingly, our model trained with the myeloid and PBMC data exhibit relatively inferior performance relative to in-distribution predictions, whereas the model trained with HSPC data do not compromise. This might reflect a more prominent temporal batch effect in the myeloid and PBMC data, which cannot be modeled linearly. In addition, we see that the Pearson correlation coefficients of the MLP models trained with NLC loss are greater than those with MSE loss. This observation suggests that our NLC loss may aid neural networks at a certain level of generalization. Compared to other methods, the reference-mapping methods that project query cells to the most similar cell in the reference do not generalize well, indicating that they could mask the presence of de novo cell states that are outside the scope of the reference data.

\begin{table}[htbp]
   \caption{Out-distribution predictions.}\label{out-table}
   \centering
   \begin{tabular}{ccccccc}
      \hline
      Data set & \multicolumn{2}{c}{HSPC} & \multicolumn{2}{c}{Myeloid} & \multicolumn{2}{c}{PBMC}\\
      \hline
      Method & Corr & MSE & Corr & MSE & Corr & MSE\\
      \hline
      CrossmodalNet & \textbf{0.564} & 0.268 & \textbf{0.246} & 0.382 & 0.393 & 1.889\\
      MLP+MSE & 0.511 & \textbf{0.134} & 0.093 & \textbf{0.202} & 0.402 & 1.832 \\
      MLP+NLC & 0.549 & 0.260 & 0.150 & 0.372 & \textbf{0.413} & \textbf{1.727}\\
      lightGBM & 0.554 & 0.267 & 0.098 & 0.829 & 0.392 & 1.928\\
      Linear reg. & 0.534 & 0.282 & 0.102 & 0.910 & 0.391 & 1.934\\
      Ridge reg. & 0.535 & 0.281 & 0.105 & 0.862 & 0.391 & 1.932\\
      scIPENN & 0.374 & 0.437 & 0.238 & 0.408 & 0.074 & 2.249\\
      totalVI & 0.339 & - & 0.104 & - & 0.284 & -\\
      Seurat 4 & 0.428 & - & 0.181 & - & 0.376 & -\\
      \hline
   \end{tabular}
\end{table}

\subsection{CrossmodalNet infers causal gene-protein relationships}
The study of the relationship between genes and surface proteins is essential for comprehending the molecular basis of numerous biological processes and diseases and has significant implications for the development of new diagnostic tools, treatments, and preventative measures. Since our basal latent inference excludes temporal information, we expect our saliency analysis would uncover gene-protein relationships with less bias from other factors, such as time. In the following, we demonstrate that our model elucidates these relationships that are commonly investigated in large-scale genomic studies.

The development of HSPCs is tightly regulated by changes in gene and protein expression, but there is currently limited understanding of how these two measurements co-vary in HSPCs when they develop into more mature blood cells. The first CITE-seq data set contains more than 70,000 CD34+ HSPCs based in \cite{velten2017human}, which suggests that discrete cell populations are established only when differentiation has progressed to the level of restricted progenitors associated with the upregulation of surface protein CD38.

\begin{figure}
\centering
    \begin{subfigure}[b]{0.5\columnwidth}
     \includegraphics[width=0.85\linewidth]{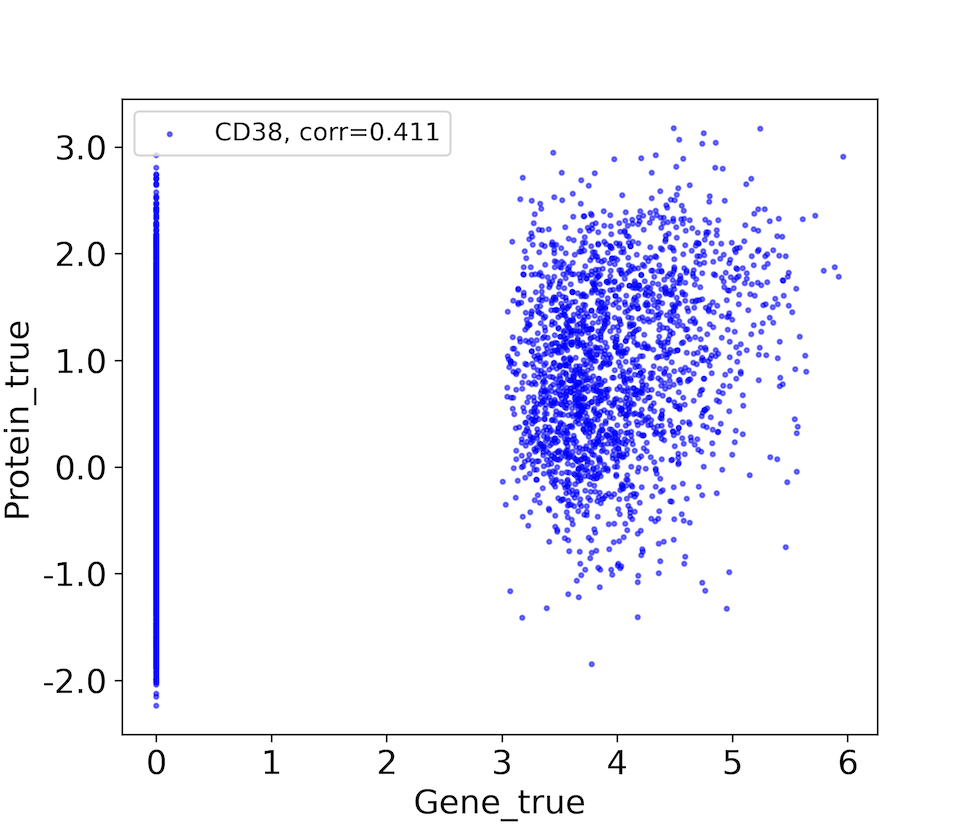}
        \caption{Scatter plot of CD38 gene and \\
        protein expression.}
        \label{CD38_raw}
    \end{subfigure}%
    \begin{subfigure}[b]{0.5\columnwidth}
        \includegraphics[width=0.85\linewidth]{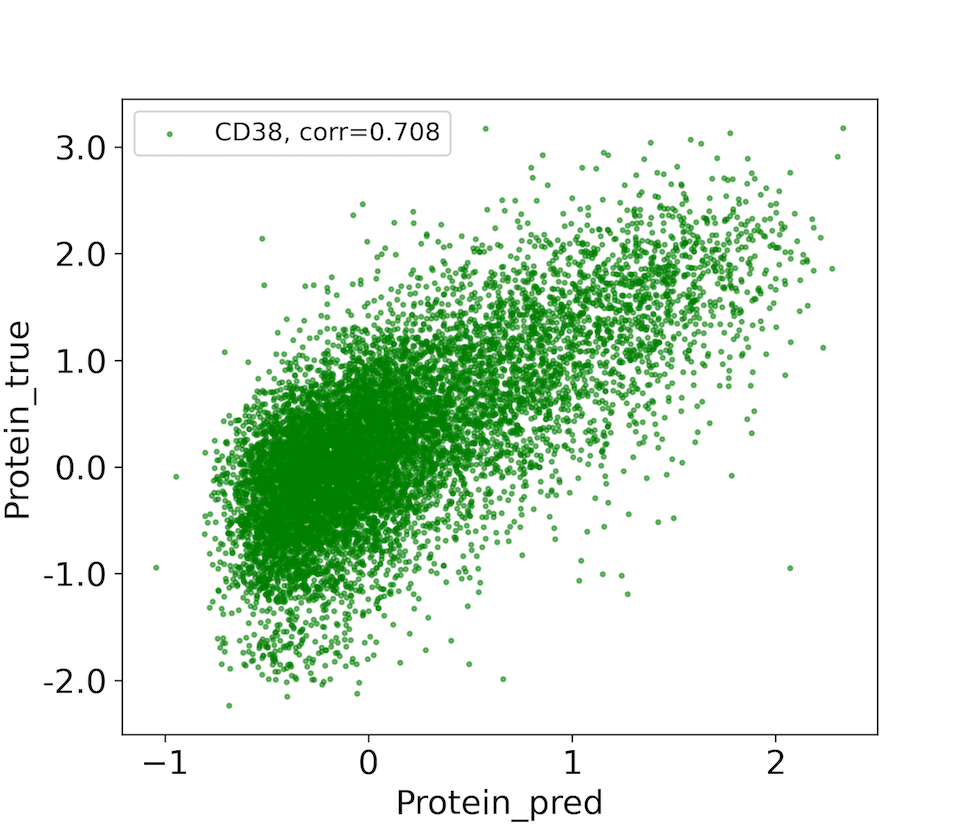}
        \caption{Scatter plot of CD38 protein expression and prediction.}
        \label{CD38_predict}
    \end{subfigure}

    \begin{subfigure}[b]{0.5\columnwidth}
        \includegraphics[width=0.85\linewidth]{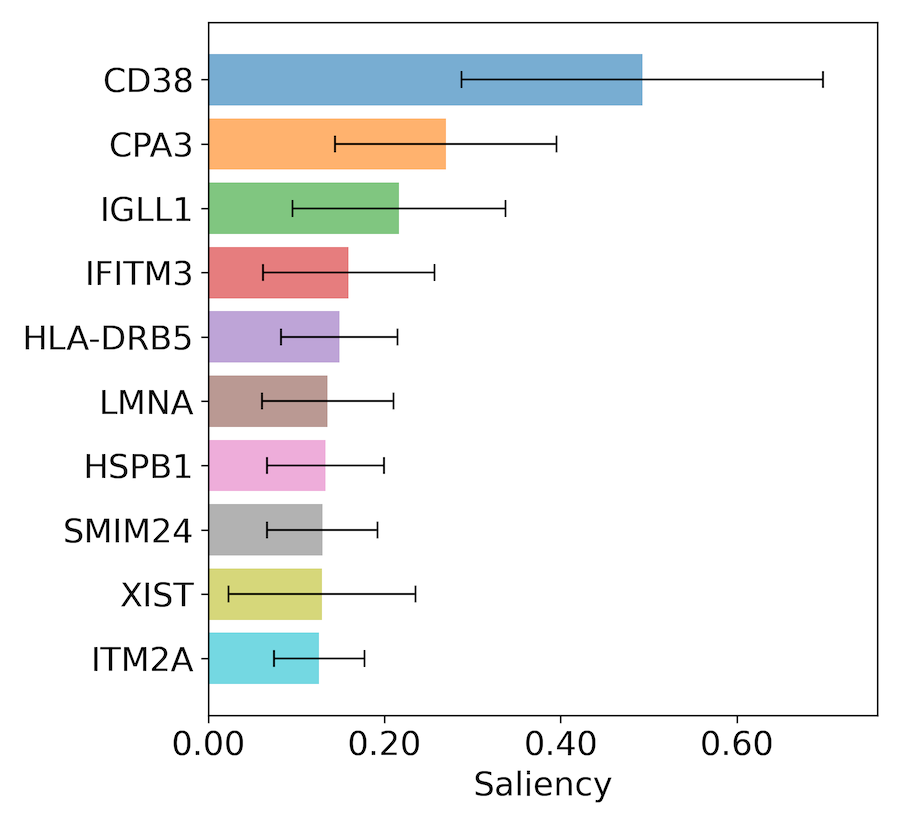}
        \caption{Top ten genes ranked by \\
        CD38 saliency.}
        \label{CD38_saliency}
    \end{subfigure}%
    \begin{subfigure}[b]{0.5\columnwidth}
        \includegraphics[width=0.85\linewidth]{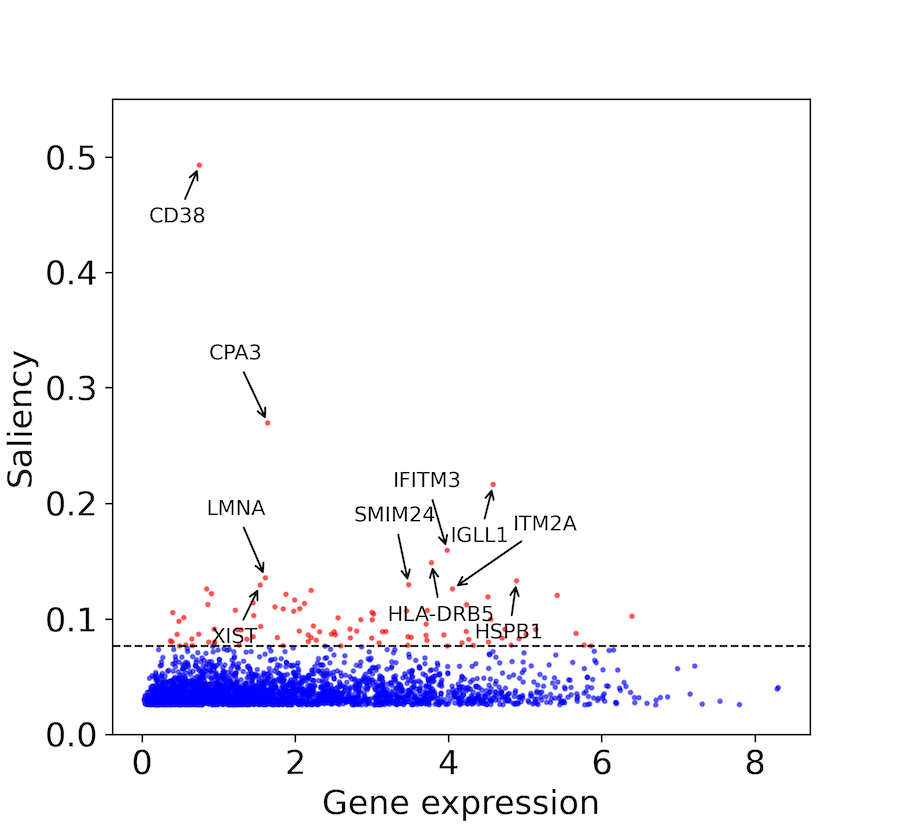}
        \caption{Scatter plot of gene expression and CD38 saliency.}
        \label{CD38_gene_saliency}
    \end{subfigure}
    
    \caption{Gene-protein relationship analysis for protein CD38.}
    \label{fig_CD38}
\end{figure}

Consistent with observations from CITE-seq studies \cite{stoeckius2017simultaneous}, our findings indicate that predicted protein levels of CD38, similar to the true protein values (0.708, Figure \ref{CD38_predict}), differ from the gene expression level of the corresponding encoding gene, with our predicted values exhibiting slightly greater correlation (0.411, Figure \ref{CD38_raw}, and 0.481, Figure S2a). Figure S3a and S3b visualize feature plots of CD38 in gene and predicted protein domain. 


We visualize the learned high dimensional time embedding in PCA where three time vectors are found to be almost evenly separated from each other (Figure S4a). This may explain why our linear interpolation inference generalizes well to this data set. Next, we compute the saliency of features to determine which features (genes) our model pays attention to for predicting the expression level of CD38. Figure \ref{CD38_saliency} depicts the saliency ranking of the top ten most prominent genes. Gene \textit{CD38} is at the top of the list despite its low expression level, which causes it to be obscured by other highly expressed genes (Figure \ref{CD38_gene_saliency}), indicating that our method successfully recognizes the gene \textit{CD38}'s significant contribution to the protein CD38’s expression. Gene Set Enrichment Analysis (GSEA) with the KEGG pathway database using top 100 saliency genes ranks \textit{Hematopoietic cell lineage} at the top (Figure \ref{GSEA_CD38}), which corresponds to the potential central role of CD38 in cell differentiation presented in the study of HSPCs \cite{velten2017human}. Together, our results show our model accurately models the intrinsic gene-protein relationships across time and shed light on the molecular mechanism underlying the functioning of CD38.

\begin{figure}
    \begin{center}
    \begin{subfigure}[b]{0.5\columnwidth}
        \includegraphics[width=0.85\linewidth]{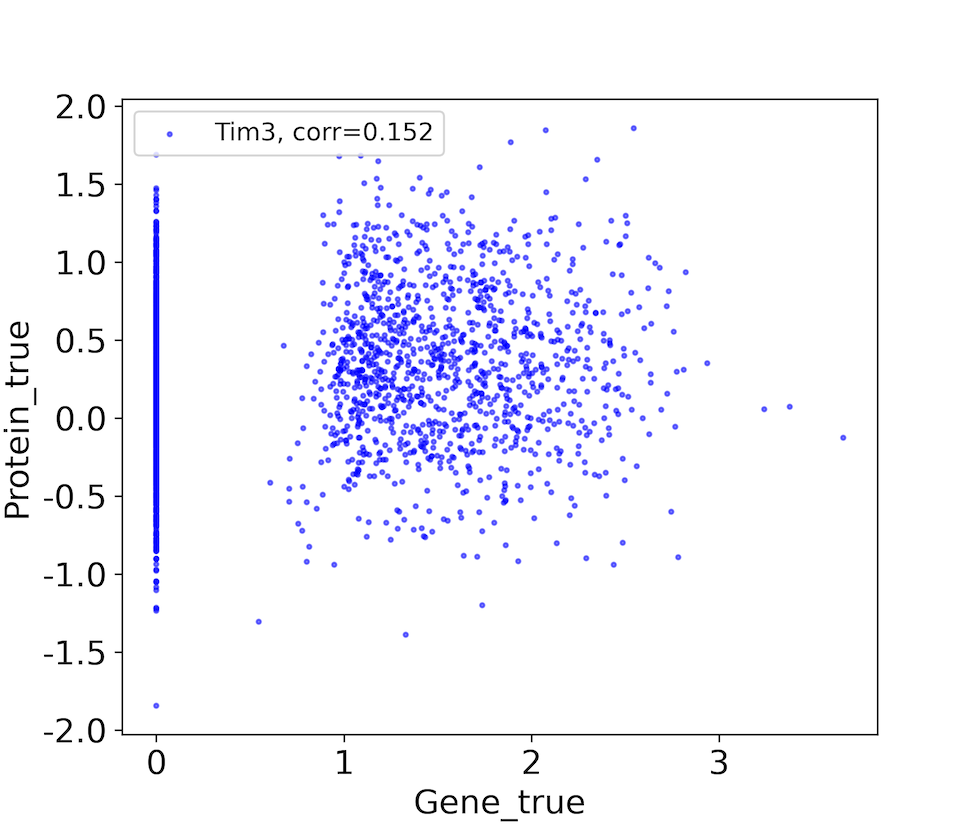}
        \caption{Scatter plot of Tim3 gene and \\
        protein expression.}
        \label{Tim3_raw}
    \end{subfigure}%
    \begin{subfigure}[b]{0.5\columnwidth}
        \includegraphics[width=0.85\linewidth]{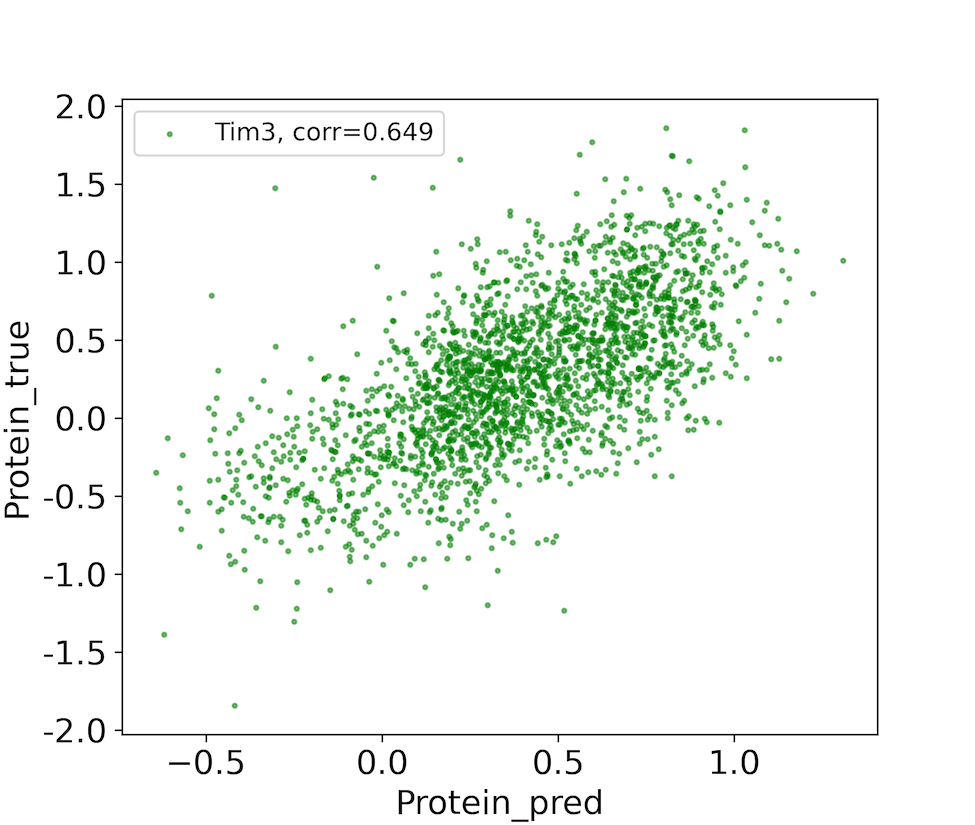}
        \caption{Scatter plot of Tim3 protein expression and prediction.}
        \label{Tim3_predict}
    \end{subfigure}

    \begin{subfigure}[b]{0.5\columnwidth}
        \includegraphics[width=0.85\linewidth]{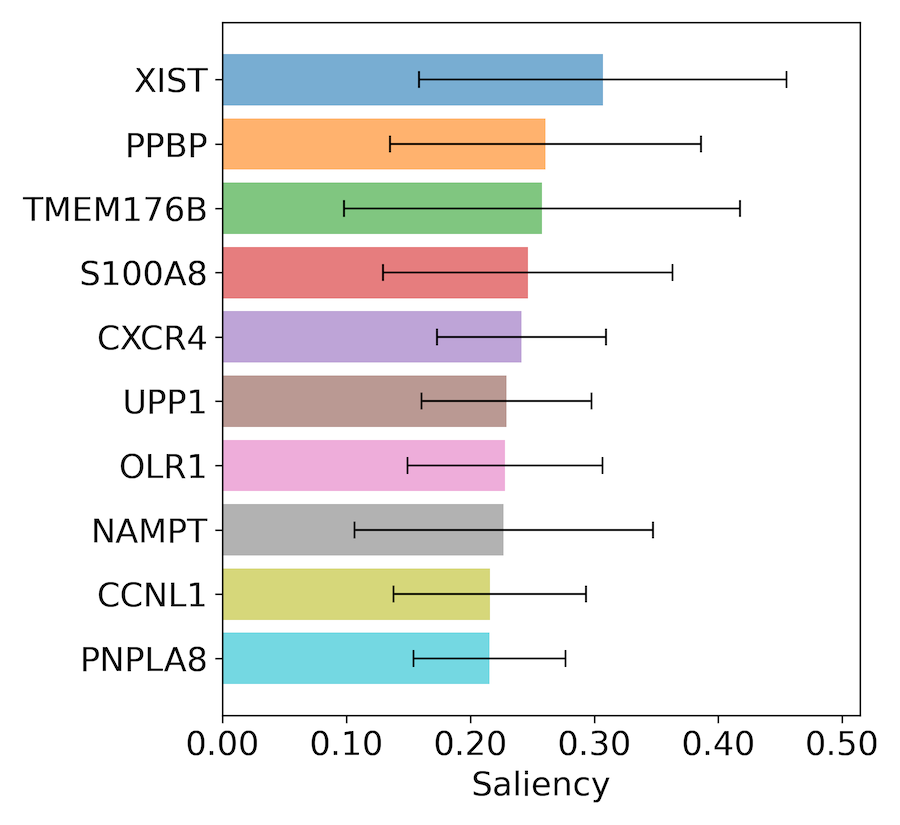}
        \caption{Top ten genes ranked by \\
        Tim3 saliency.}
        \label{Tim3_saliency}
    \end{subfigure}%
    \begin{subfigure}[b]{0.5\columnwidth}
        \includegraphics[width=0.85\linewidth]{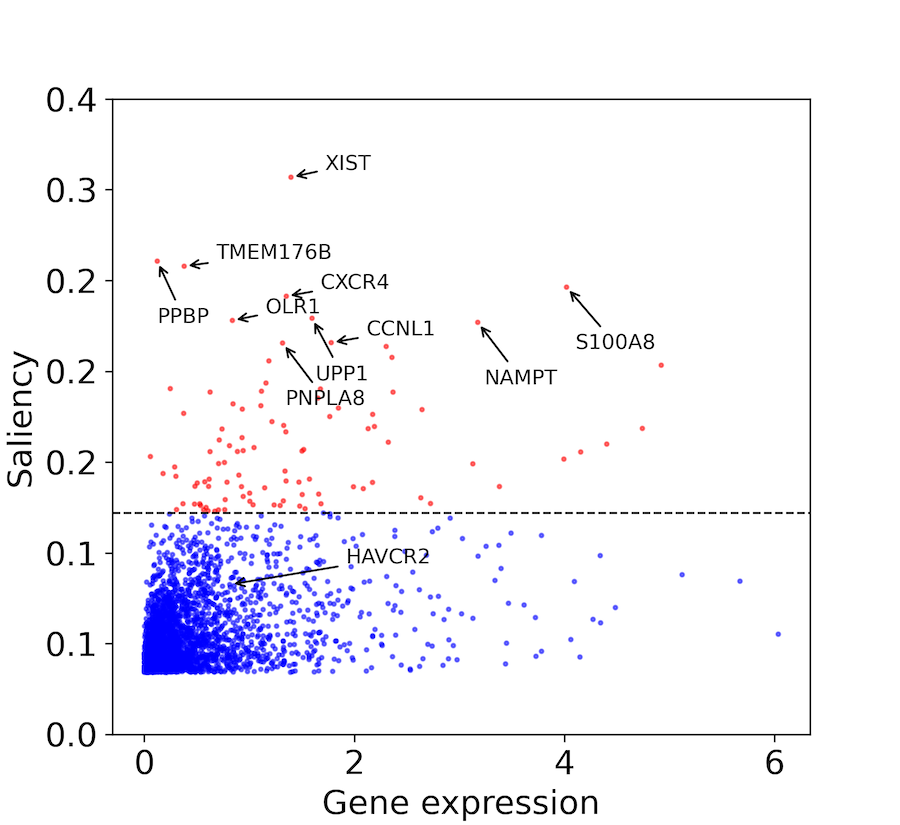}
        \caption{Scatter plot of gene expression and Tim3 saliency.}
        \label{Tim3_gene_saliency}
    \end{subfigure}
    
    \caption{Gene-protein relationship analysis for protein Tim3.}
    \label{fig_Tim3}
    \end{center}
\end{figure}

Myeloid cells contribute to immunotherapy resistance; however, their role in response to checkpoint inhibition (CPI) in anti-PD-1 refractory cancers is unclear. Next we explore the second CITE-seq data set from \cite{keenan2022circulating}, where the researchers conclude that CD14+ monocytes linked to anti-PD-1 resistance in human biliary cancer cause T cell paralysis. T cells are dysfunctional when co-cultured with the monocytes that express high levels of Tim3.

We first demonstrate that the expression of surface protein Tim3 is significantly distinct from its coding gene \textit{HAVCR2} (0.152, Figure \ref{Tim3_raw}). Our model's prediction (0.649, Figure \ref{Tim3_predict}) coincides with the contrast (0.168, Figure S2b) and their feature plots are shown in Figure S3c and S3d. 
Figure S4b depicts the learned time embedding, and the proximity of weeks 2 and 3 indicates that an interpretable embedding has been learned to reveal their inherent similarity.
The saliency ranking of the ten most prominent Tim3 genes is depicted in Figure \ref{Tim3_saliency}. Due to the intrinsically low correlations between the protein Tim3 and gene \textit{HAVCR2}, the gene has not been ranked highly (Figure \ref{Tim3_gene_saliency}) and has had less of an impact on our prediction of Tim3. GSEA ranks \textit{phagosome} as the most significant pathway (Figure \ref{GSEA_Tim3}), correlating with the finding that PD-1 signaling can polarize macrophages to an M2 phenotype, cause defects in phagocytosis, and impair antitumor immunity \cite{keenan2022circulating}.

The upregulation of the HLA-DR (Human Leukocyte Antigen-DR) protein on monocytes allows for efficient antigen presentation to T cells, thereby initiating an immune response against HIV. Monitoring HLA-DR expression on monocytes after HIV vaccination offers insight into the activation and functionality of the immune response. The third data set \cite{hao2021integrated}, which is used to construct a multimodal atlas of human PBMCs, reveals a robust innate immune response to HIV vaccination in monocyte populations where HLA-DR is highly expressed.

\begin{figure}
    \begin{center}
    \begin{subfigure}[b]{0.5\columnwidth}
        \includegraphics[width=0.85\linewidth]{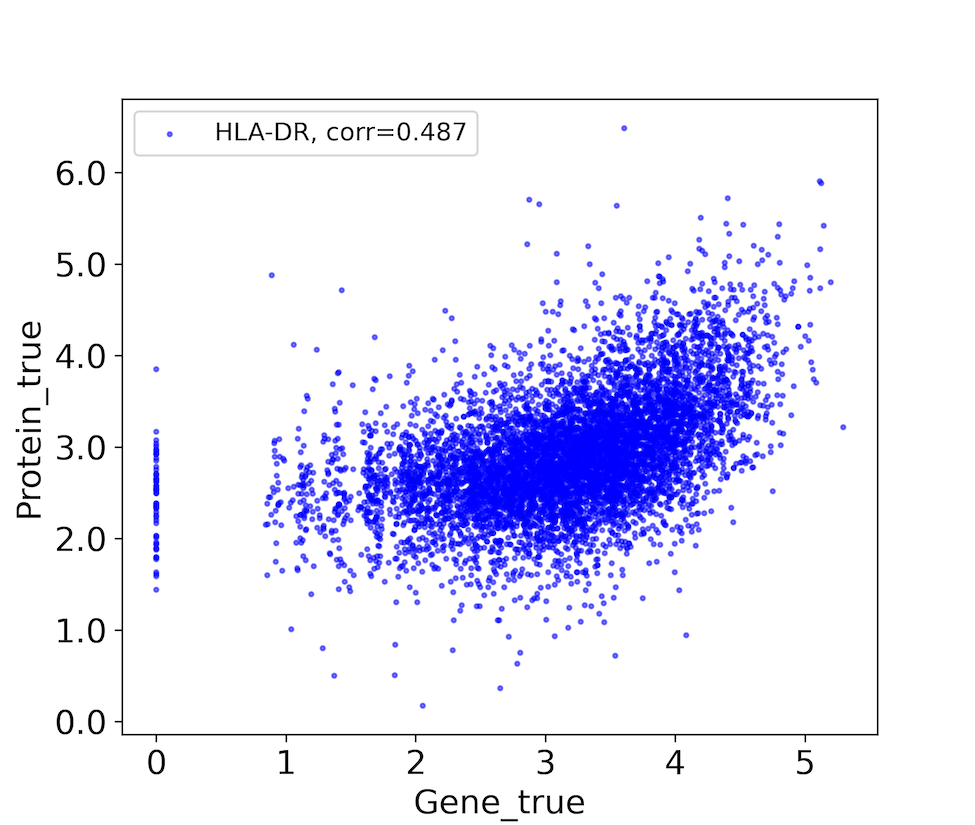}
        \caption{Scatter plot of HLA-DR gene and \\
         protein expression.}
        \label{HLA_raw}
    \end{subfigure}%
    \begin{subfigure}[b]{0.5\columnwidth}
        \includegraphics[width=0.85\linewidth]{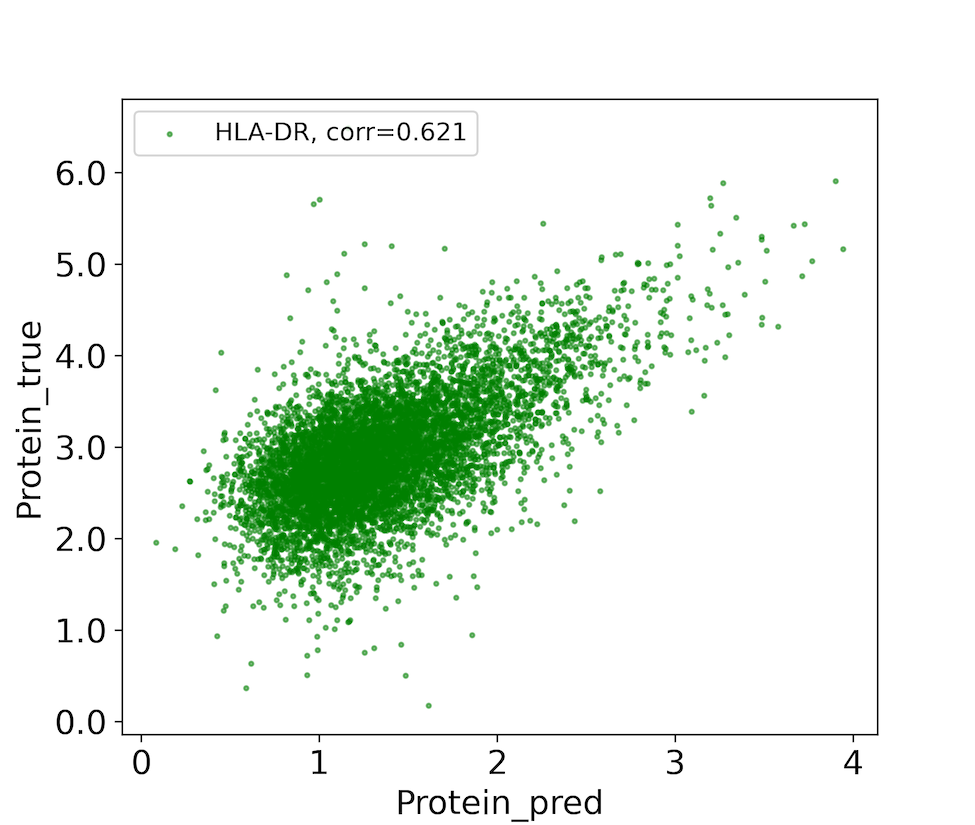}
        \caption{Scatter plot of HLA-DR protein expression and prediction.}
        \label{HLA_predict}
    \end{subfigure}

    \begin{subfigure}[b]{0.5\columnwidth}
        \includegraphics[width=0.85\linewidth]{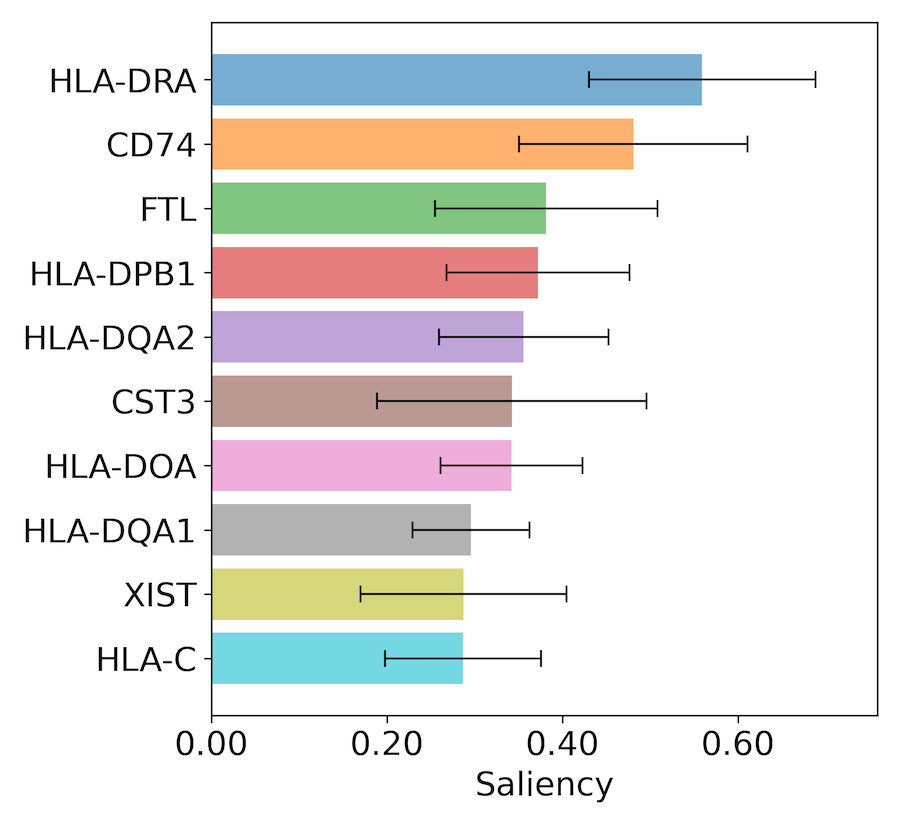}
        \caption{Top ten genes ranked by \\
         HLA-DR saliency.}
        \label{HLA_saliency}
    \end{subfigure}%
    \begin{subfigure}[b]{0.5\columnwidth}
        \includegraphics[width=0.85\linewidth]{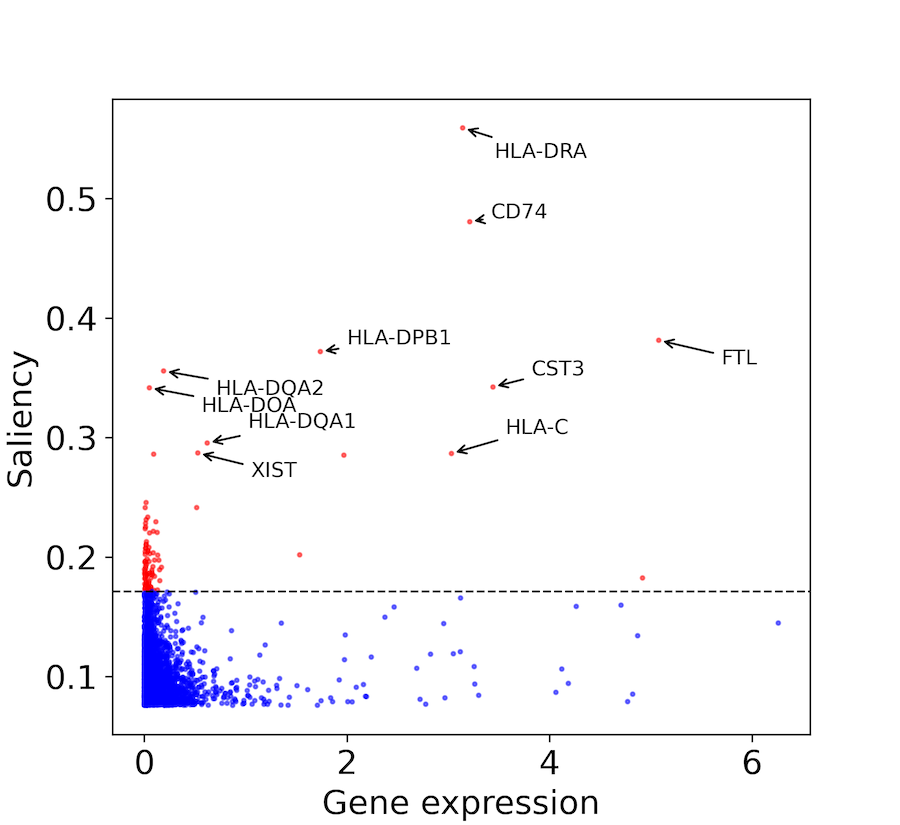}
        \caption{Scatter plot of gene expression and HLA-DR saliency.}
        \label{HLA_gene_saliency}
    \end{subfigure}
    \caption{Gene-protein relationship analysis for protein HLA-DR.}
    \label{fig_HLA}
    \end{center}
\end{figure}

\begin{figure}
    \vskip 0.15in
    \begin{center}
    \begin{subfigure}[b]{0.5\columnwidth}
        \includegraphics[width=0.85\linewidth]{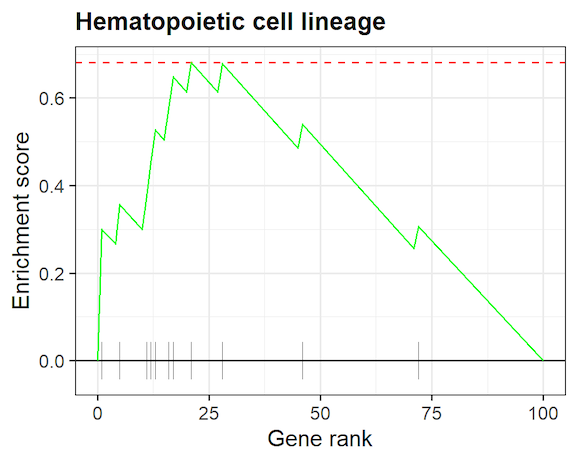}
        \caption{Hematopoietic pathway associated with the protein CD38.}
        \label{GSEA_CD38}
    \end{subfigure}%
    \hfill
    \begin{subfigure}[b]{0.5\columnwidth}
        \includegraphics[width=0.85\linewidth]{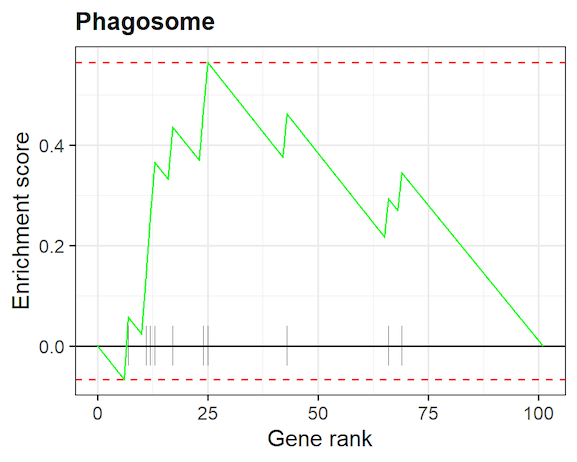}
        \caption{Phagosome pathway associated with the protein Tim3.}
        \label{GSEA_Tim3}
    \end{subfigure}    
    \hfill
    \begin{subfigure}[b]{0.5\columnwidth}
        \includegraphics[width=0.85\linewidth]{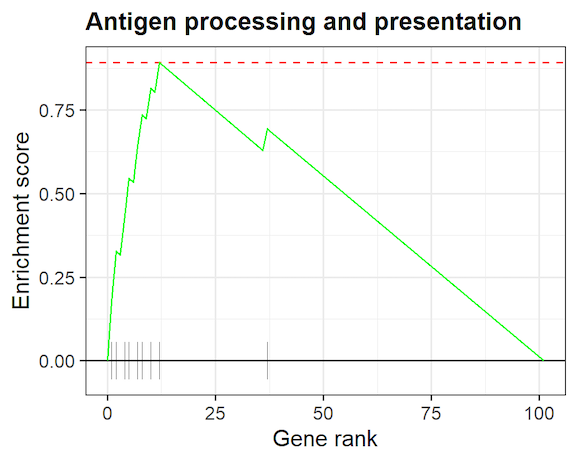}
        \caption{Antigen processing and presentation pathway associated with the protein HLA-DR.}
        \label{GSEA_HLA}
    \end{subfigure}    
    \caption{GSEA plots depict the top enriched pathways associated with CD38, Tim3, and HLA-DR proteins. Gene rank represents the position of each gene based on its saliency.}
    \label{GSEA}
    \end{center}
\vskip -0.2in
\end{figure}

The inherent correlation between surface protein HLA-DR and its coding gene \textit{HLA-DRA} (0.487, Figure \ref{HLA_raw}) is higher than the previous instances. We observe a stronger correlation between the gene and protein predictions of our model (0.543, Figure S2c) as it learns (0.621, Figure \ref{HLA_predict}).Feature plots of the gene and predicted protein exhibit more similar patterns in Figure S3e and S3f.
Figure S4c depicts the learned time embedding, and the greater cosine similarity between post-vaccination days 3 and 7 demonstrates that our model recognizes the vaccination effect, distinguishing them from day 0 prior to the vaccination.
Figure \ref{HLA_saliency} depicts the saliency ranking of the ten most prominent Tim3 genes including six genes encoding HLA molecules (HLA-DRA, HLA-DPB1, HLA-DQA2, HLA-DOA, HLA-DQA1, HLA-C), some of which are not highly expressed (Figure \ref{HLA_gene_saliency}). GSEA identifies \textit{Antigen processing and presentation} as the most significant pathway (Figure \ref{GSEA_HLA}), highlighting the essential antigen-presenting role that monocyte serves in initiating the immune response following vaccination. 

\subsection{CrossmodalNet is scalable}
We simulate four CITE-seq data sets of varying sizes as input and evaluate the scalability of our model by comparing the total training time with other baseline methods. Without using GPUs, our model exhibits a 7.4- to 14.3-fold faster running speed than the average of baseline methods tested on equivalent hardware (Intel Xeon 6248R at 3.0 GHz with 24 GB RAM requested, Table S2). Our model is expected to run faster by optimizing GPU implementation. The scalability of our model would allow large-scale experiments involving thousands of cells, in accordance with modern norms.
 
\section{Discussion}
We present CrossmodalNet as a highly interpretable and scalable model that can be generalized for the prediction of proteomics data from transcriptomics data. 
Our experiments show that our model with the customized adaptive loss outperforms benchmarking methods including linear and nonlinear models, as evidenced by higher Pearson correlation in most benchmarking scenarios. In practice, once scRNA-seq data of a cell system is available, our model can accurately estimate the patterns of surface proteins. Thus, our model provides an in-silico alternative to CITE-seq experiments and may facilitate the generation of hypotheses and the design of experiments.

Our method for decomposing temporal gene expression into basal and time embeddings of cells is an advancement step towards understanding the mechanisms that govern gene-protein regulation and cell state transitions. The time embedding offers insights into the underlying mechanisms of cell system development and is useful for predicting protein expression in a time-specific manner. In addition, our inference for basal embedding uncovered by our model is biologically interpretable, and we demonstrate that causal gene-protein relationships that provides a fundamental understanding of how genetic information is translated into functional proteins can be deduced from this inference. This quantitative understanding is essential for identifying cellular development and can be expanded to detect disease-causing genes, develop new drugs, and understand complex cellular processes.

Despite CrossmodalNet has been systematically evaluated and demonstrated to be an accurate, interpretable, and scalable model, it has limitations. For instance, the current multiomics experiments only measured a handful of discrete time points. By incorporating larger time-resolved multiomics data with more time points, we expect to generate a pre-trained model that can serve as a generalized tool for subsequent single-cell analyses. Our model is also inapplicable to bulk sequencing data, as it requires heterogeneity at single-cell resolution. There are many available transcriptomic priors, hence a Bayesian treatment may enhance the performance and interpretability of our model. Future work could be directed to investigate strategies for incorporating gene-protein prior knowledge into model training, and to extend this work to model scATAC-seq data \cite{buenrostro2015single}. As scATAC-seq and scRNA-seq are naturally causally related, we expect to discover more robust causal relations underlying the central dogma of molecular biology. Recent developments in spatial CITE-seq \cite{liu2023high} make it possible to measure the abundance of genes and proteins while preserving cell localization. CrossmodalNet could be upgraded to incorporate such data for more precise examination of cells or organs.
\chapter{SUMMARY AND CONCLUSIONS \label{cha:Summary}}

In this doctoral dissertation, using single-cell data, I presented an in-depth review and study to learn genotype-phenotype relationships. I designed and implemented three computational tools, scTenifoldXct, GenKI, and CrossmodalNet, using modern systems biology and machine learning techniques to decode the complex map of interactions governing cellular behavior.

scTenifoldXct \cite{yang2023sctenifoldxct} is a semi-supervised tool for predicting cell-cell interactions from scRNA-seq data. I combined intra- and inter-cellular gene networks to detect cell-cell interaction, providing robust statistical confidence for detecting results. The tool reveals patterns of cell-cell communication and highlights the role of particular genes in these interactions. In the case of inflammatory skin diseases, for instance, scTenifoldXct identified ligand-receptor pairs with the potential to drive disease pathogenesis. This tool also performs differential interaction analysis by comparing cell-cell interaction patterns between tissue samples, which makes it a potent measure for detecting subtle differential interactions. 

I then presented GenKI \cite{yang2023gene}, a variational graph autoencoder-based virtual gene KO tool. GenKI learns simultaneously the latent representations of scRNA-seq data of WT samples and the underlying scGRN that contribute to observed phenotypes. This novel method enables the inference of gene knockouts without the need for actual knockout data, thereby providing a powerful predictive tool for gene function. In a case study involving Trem2-KO in microglia cells and Nkx2-1-KO in AT1 cells, GenKI was able to identify significantly enriched genes, shedding light on the functional implications of these genes.

CrossmodalNet \cite{yang2023interpretable}, the third tool, is a machine learning model that utilizes multimodal CITE-seq data to discover the relationship between gene and protein expression. It accurately predicts protein expression using only scRNA-seq data at specific time points, making it a valuable tool for comprehending the gene-protein translation process. The tool reveals significant insights into the gene-protein translation process, demonstrating that specific genes have a substantial impact on protein expression levels. This was evident in the case study involving human peripheral blood mononuclear cells, where CrossmodalNet effectively predicted protein expression at various time points and revealed significant phenotype-leading genes.

In conclusion, this doctoral dissertation provides novel computational tools and methods for analyzing single-cell data, thereby making an essential contribution to the field of cellular biology. These tools can aid in the discovery of the molecular drivers of cellular processes, paving the way for the development of novel therapeutic strategies and enhancing our ability to combat a vast array of diseases.

\let\oldbibitem\bibitem
\renewcommand{\bibitem}{\setlength{\itemsep}{0pt}\oldbibitem}
\bibliographystyle{ieeetr}

\phantomsection
\addcontentsline{toc}{chapter}{REFERENCES}

\renewcommand{\bibname}{{\normalsize\rm REFERENCES}}


%
%
%
%
%

\begin{appendices}
\titleformat{\chapter}{\centering\normalsize}{APPENDIX \thechapter}{0em}{\vskip .5\baselineskip\centering}
\renewcommand{\appendixname}{APPENDIX}


\phantomsection

\chapter{scTenifoldXct}


\section{Supplementary Tables}

\begin{table}[htbp]
\centering
\begin{tabular}{|p{1.6cm}|p{1.6cm}|p{3.6cm}|p{2.98cm}|p{4.06cm}|}
\hline
\textbf{Number of cells} & \textbf{Number of genes} & \textbf{Time for GRN construction (s)} & \textbf{Time for training (s)} & \textbf{Peak memory required (MB)} \\
\hline
100 & 1,000 & 11.71 $\pm$ 0.09 & 22.80 $\pm$ 0.64 & 719.39 $\pm$ 2.65 \\
\hline
100 & 3,000 & 43.42 $\pm$ 2.79 & 167.15 $\pm$ 9.67 & 2,551.31 $\pm$ 11.27 \\
\hline
500 & 3,000 & 365.56 $\pm$ 5.35 & 191.55 $\pm$ 6.11 & 2,743.01 $\pm$ 2.78 \\
\hline
500 & 5,000 & 1045.38 $\pm$ 66.62 & 478.80 $\pm$ 10.98 & 6,684.62 $\pm$ 10.53 \\
\hline
1,000 & 5,000 & 2595.48 $\pm$ 125.87 & 540.06 $\pm$ 30.11 & 6,856.17 $\pm$ 45.04 \\
\hline
\end{tabular}
\caption{Running time and memory consumption with respect to data set sizes (n = 10).}
All evaluations were implemented on the public platform Google Colab, which operates under Ubuntu 18.04 64 bits and is composed of a four-core Intel Xeon CPU processor at 2.30 GHz with 27.3 GB available random-access memory (RAM).
\label{xct_t5s}
\end{table}

\begin{table}[htbp]
\centering
\begin{tabular}{|p{4.1cm}|p{4.5cm}|p{1.5cm}|}
\hline
\multicolumn{1}{|p{4.1cm}|}{\textbf{LR pairs}} & 
\multicolumn{1}{|p{4.5cm}|}{\textbf{LR distance on manifold}} & 
\multicolumn{1}{|p{1.5cm}|}{\textbf{\textit{p-value}}} \\ 
\hline
\textit{CCL2-CCR7} & 
0.0010 & 
0.002 \\ 
\hline
\textbf{\textit{CCL19-CCR7}} & 
0.0010 & 
0.003 \\ 
\hline
\textit{CCL2-CXCR4} & 
0.0010 & 
0.003 \\ 
\hline
\textit{VCAM1-ITGB2} & 
0.0010 & 
0.003 \\ 
\hline
\textit{CCL19-CXCR4} & 
0.0011 & 
0.004 \\ 
\hline
\textit{CXCL12-CCR7} & 
0.0011 & 
0.004 \\ 
\hline
\textbf{\textit{CXCL12-CXCR4}} & 
0.0012 & 
0.004 \\ 
\hline
\textit{CXCL14-CCR7} & 
0.0013 & 
0.005 \\ 
\hline
\textit{CXCL14-CXCR4} & 
0.0014 & 
0.005 \\ 
\hline
\textit{CCL26-CCR7} & 
0.0017 & 
0.007 \\ 
\hline
\textit{CCL26-CXCR4} & 
0.0019 & 
0.009 \\ 
\hline
\textit{CXCL2-CCR7} & 
0.0021 & 
0.009 \\ 
\hline
\textit{CXCL1-CCR7} & 
0.0022 & 
0.010 \\ 
\hline
\textit{CXCL2-CXCR4} & 
0.0023 & 
0.010 \\ 
\hline
\textit{CTGF-ITGB2} & 
0.0025 & 
0.011 \\ 
\hline
\textit{CXCL1-CXCR4} & 
0.0025 & 
0.011 \\ 
\hline
\textbf{\textit{CCL2-TNF}} & 
0.0026 & 
0.012 \\ 
\hline
\textit{CYR61-ITGB2} & 
0.0036 & 
0.016 \\ 
\hline
\textit{DCN-VEGFA} & 
0.0043 & 
0.019 \\ 
\hline
\textit{CXCL14-CCR6} & 
0.0047 & 
0.021 \\ 
\hline
\textit{CCL2-VEGFA} & 
0.0047 & 
0.022 \\ 
\hline
\textit{CCL19-CCR6} & 
0.0049 & 
0.023 \\ 
\hline
\textit{CXCL12-CCR6} & 
0.0049 & 
0.023 \\ 
\hline
\textit{CCL2-CCR6} & 
0.0051 & 
0.024 \\ 
\hline
\textit{FGF7-VEGFA} & 
0.0054 & 
0.027 \\ 
\hline
\textit{CCL26-CCR6} & 
0.0071 & 
0.040 \\ 
\hline
\textit{CXCL1-CCR6} & 
0.0072 & 
0.041 \\ 
\hline
\textit{S100A8-ITGB2} & 
0.0073 & 
0.041 \\ 
\hline
\textit{CXCL2-CCR6} & 
0.0075 & 
0.044 \\ 
\hline
\textit{BMP4-VEGFA} & 
0.0076 & 
0.045 \\ 
\hline
\end{tabular}
\caption{scTenifoldXct identified LR pairs between fibroblasts and dendritic cells in lesional skin.}
\label{xct_t1s}
\end{table}
\begin{table}[htbp]
\centering
\begin{tabular}{|p{4.1cm}|p{4.5cm}|p{1.5cm}|}
\hline
\textbf{LR pairs} & \textbf{LR distance on manifold} & \textbf{\textit{p-value}} \\
\hline
\textit{CXCL8-CXCR4} & 0.0007 & 0.001 \\
\hline
\textit{CCL22-CXCR4} & 0.0013 & 0.004 \\
\hline
\textit{CXCL8-CXCR6} & 0.0016 & 0.007 \\
\hline
\textbf{\textit{CCL17-CCR6}} & 0.0017 & 0.007 \\
\hline
\textit{CXCL8-CCR6} & 0.0019 & 0.009 \\
\hline
\textit{S100A8-CD69} & 0.0019 & 0.009 \\
\hline
\textbf{\textit{CCL22-CCR6}} & 0.0020 & 0.010 \\
\hline
\textit{CCL17-CXCR6} & 0.0021 & 0.010 \\
\hline
\textit{CCL22-CXCR6} & 0.0021 & 0.011 \\
\hline
\textbf{\textit{CCL17-CXCR4}} & 0.0021 & 0.011 \\
\hline
\textit{HLA-DMA-CD74} & 0.0024 & 0.013 \\
\hline
\textit{S100A8-ITGB2} & 0.0026 & 0.015 \\
\hline
\textit{TNF-NR3C1} & 0.0027 & 0.016 \\
\hline
\textit{INHBA-TGFBR3} & 0.0030 & 0.019 \\
\hline
\textit{S100B-TNF} & 0.0036 & 0.023 \\
\hline
\textit{TNF-TNFRSF1B} & 0.0037 & 0.024 \\
\hline
\textit{S100A9-ITGB2} & 0.0042 & 0.027 \\
\hline
\textit{TNFRSF1B-TNF} & 0.0043 & 0.028 \\
\hline
\textit{TNF-ICOS} & 0.0043 & 0.028 \\
\hline
\textit{CCL17-CCR10} & 0.0044 & 0.029 \\
\hline
\textit{TNF-CD83} & 0.0046 & 0.030 \\
\hline
\textit{CCL4-CXCR6} & 0.0047 & 0.031 \\
\hline
\textit{CCL17-CCR4} & 0.0050 & 0.032 \\
\hline
\textit{CCL4-CCR6} & 0.0051 & 0.033 \\
\hline
\textit{CXCL8-CCR10} & 0.0053 & 0.034 \\
\hline
\textit{CCL22-CCR10} & 0.0054 & 0.035 \\
\hline
\textit{AREG-NTRK2} & 0.0059 & 0.038 \\
\hline
\textit{VEGFA-NTRK2} & 0.0064 & 0.042 \\
\hline
\textit{CCL22-CCR4} & 0.0064 & 0.042 \\
\hline
\textit{CXCL8-CCR4} & 0.0067 & 0.044 \\
\hline
\textit{CCL19-CXCR4} & 0.0067 & 0.044 \\
\hline
\textit{CCL19-CXCR6} & 0.0072 & 0.049 \\
\hline
\textit{INHBA-ACVR1B} & 0.0073 & 0.050 \\
\hline
\end{tabular}
\caption{scTenifoldXct identified LR pairs between dendritic cells and T cells in lesional skin.}
\label{xct_t2s}
\end{table}

\begin{table}[htbp]
\centering
\begin{tabular}{|p{5.2cm}|p{1.4cm}|p{1.4cm}|p{1.4cm}|p{3.0cm}|}
\hline
\textbf{Term} & \textbf{Overlap} & \textbf{\textit{P-value}} & \textbf{Adj. \textit{P-value}} & \textbf{Genes} \\
\hline
\textit{Beta-1 integrin cell surface interactions} & 4/66 & 3.5E-08 & 3.9E-06 & \textit{MDK, SPP1, ITGA6, ITGA5} \\
\hline
\textit{ECM-receptor interaction} & 4/84 & 9.3E-08 & 3.9E-06 & \textit{ITGB5, SPP1, ITGA6, ITGA5} \\
\hline
\textit{Integrin cell surface interactions} & 4/85 & 9.8E-08 & 3.9E-06 & \textit{ITGB5, SPP1, ITGA6, ITGA5} \\
\hline
\textit{Integrin family cell surface interactions} & 3/26 & 3.2E-07 & 9.4E-06 & \textit{ITGB5, ITGA6, ITGA5} \\
\hline
\textit{Focal adhesion} & 4/233 & 5.6E-06 & 1.3E-04 & \textit{ITGB5, SPP1, ITGA6, ITGA5} \\
\hline
\textit{Signal transduction} & 6/1020 & 6.4E-06 & 1.3E-04 & \textit{ITGB5, ACKR3, ITGA6, ACKR1, ITGA5, CXCL3} \\
\hline
\textit{Arrhythmogenic right ventricular cardiomyopathy (ARVC)} & 3/75 & 8.2E-06 & 1.4E-04 & \textit{ITGB5, ITGA6, ITGA5} \\
\hline
\textit{Dilated cardiomyopathy} & 3/100 & 1.9E-05 & 2.6E-04 & \textit{ITGB5, ITGA6, ITGA5} \\
\hline
\textit{Integrin-mediated cell adhesion} & 3/100 & 1.9E-05 & 2.6E-04 & \textit{ITGB5, ITGA6, ITGA5} \\
\hline
\textit{TNF-alpha effects on cytokine activity, cell motility, and apoptosis} & 3/135 & 4.8E-05 & 5.6E-04 & \textit{ACKR3, ITGA5, CXCL3} \\
\hline
\end{tabular}
\caption{GO enrichment analysis results of putative ligand and receptor genes of upregulated interactions between hepatocytes and endothelial cells in HCC (n = 10).}
\label{xct_t3s}
\end{table}

\begin{table}[htbp]
\centering
\begin{tabular}{|p{2.8cm}|p{2.6cm}|p{2.6cm}|p{2.6cm}|p{1.4cm}|p{1.33cm}|}
\hline
\textbf{LR pairs} & \textbf{LR distance on manifold (KO)} & \textbf{LR distance on manifold (WT)} & \textbf{Difference in LR distances} & \textbf{\textit{p-value}} & \textbf{Adj. \textit{p-value}} \\
\hline
\textbf{\textit{IL33-IL1RL1}} & 0.0202 & 0.0048 & 0.0154 & 0 & 0 \\
\hline
\textit{STC1-TNF} & 0.0124 & 0.0088 & 0.0036 & 1.4E-06 & 0.0009 \\
\hline
\textit{CSF1-TNF} & 0.0127 & 0.0091 & 0.0036 & 1.5E-06 & 0.0010 \\
\hline
\textit{CSF2-TNF} & 0.0111 & 0.0075 & 0.0036 & 1.9E-06 & 0.0012 \\
\hline
\textit{EDN1-EDNRB} & 0.0143 & 0.0108 & 0.0035 & 3.2E-06 & 0.0018 \\
\hline
\textit{EDN3-EDNRB} & 0.0145 & 0.0111 & 0.0034 & 6.1E-06 & 0.0028 \\
\hline
\textit{ANGPT1-TNF} & 0.0150 & 0.0116 & 0.0034 & 6.5E-06 & 0.0029 \\
\hline
\end{tabular}
\caption{Differential interactions between AT2 cells and Basophils in lung.}
LR pairs were ranked according to the distance differences between gene pairs on the aligned manifolds of two cell types in the latent space. L1 norm was used to show the distance differences with directions, which correspond to the change in the interaction strength between KO and WT tissues.
\label{xct_t4s}
\end{table}

\newpage

\section{Supplementary Figures}

\begin{figure}[h]
\centering
\includegraphics[scale=.45]{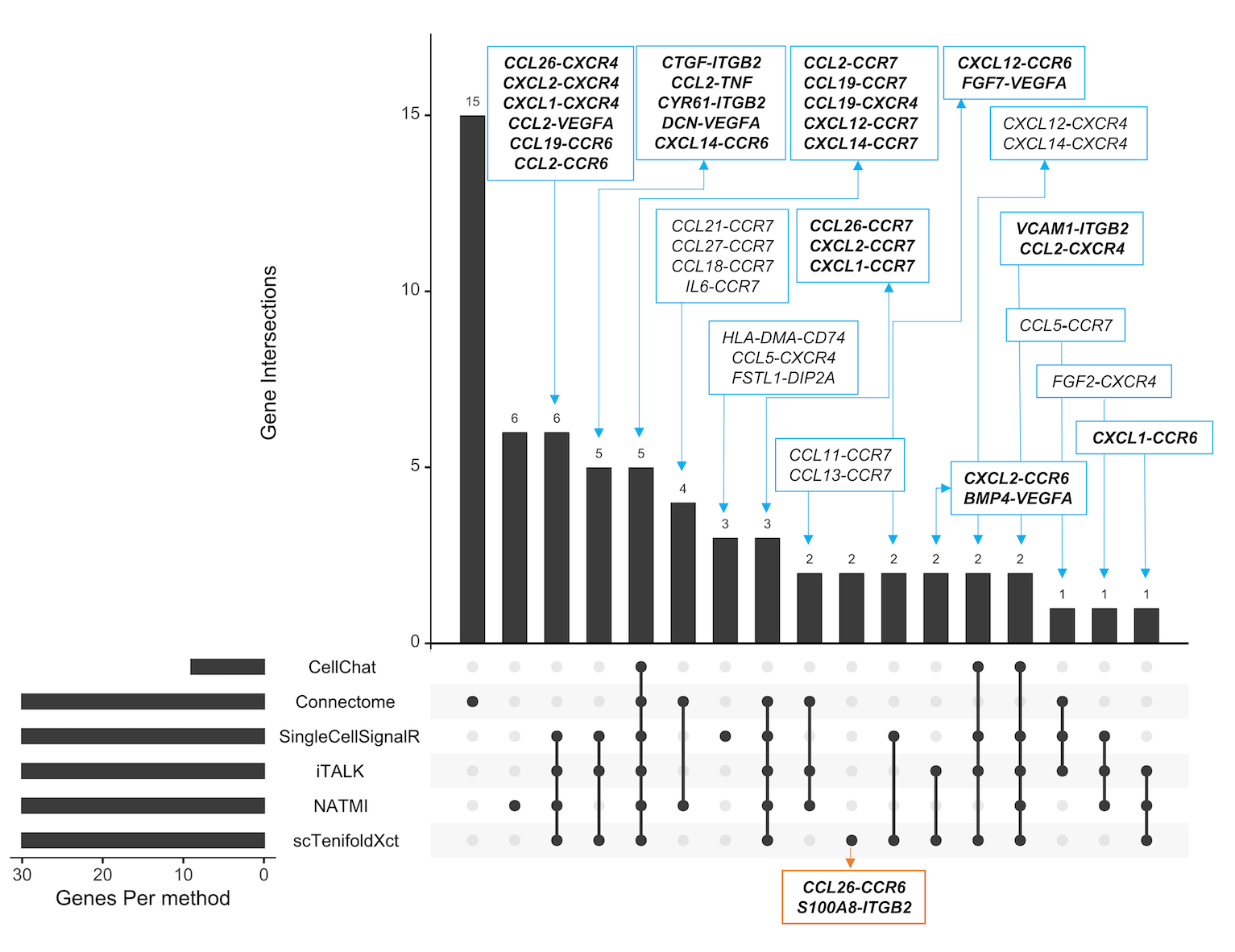}
\caption{Upset plot showing overlap of most significant interactions for six methods using the same resource.}
\raggedright
Six methods using the same OmniPath database were implemented at their own optimal thresholds for the inflammatory skin data set. scTenifoldXct overlapped more with consensus inferred by other methods and detected two exclusive LR pairs.
\label{xct_s1}
\end{figure}


\begin{figure}[h]
\centering
\includegraphics[scale=.45]{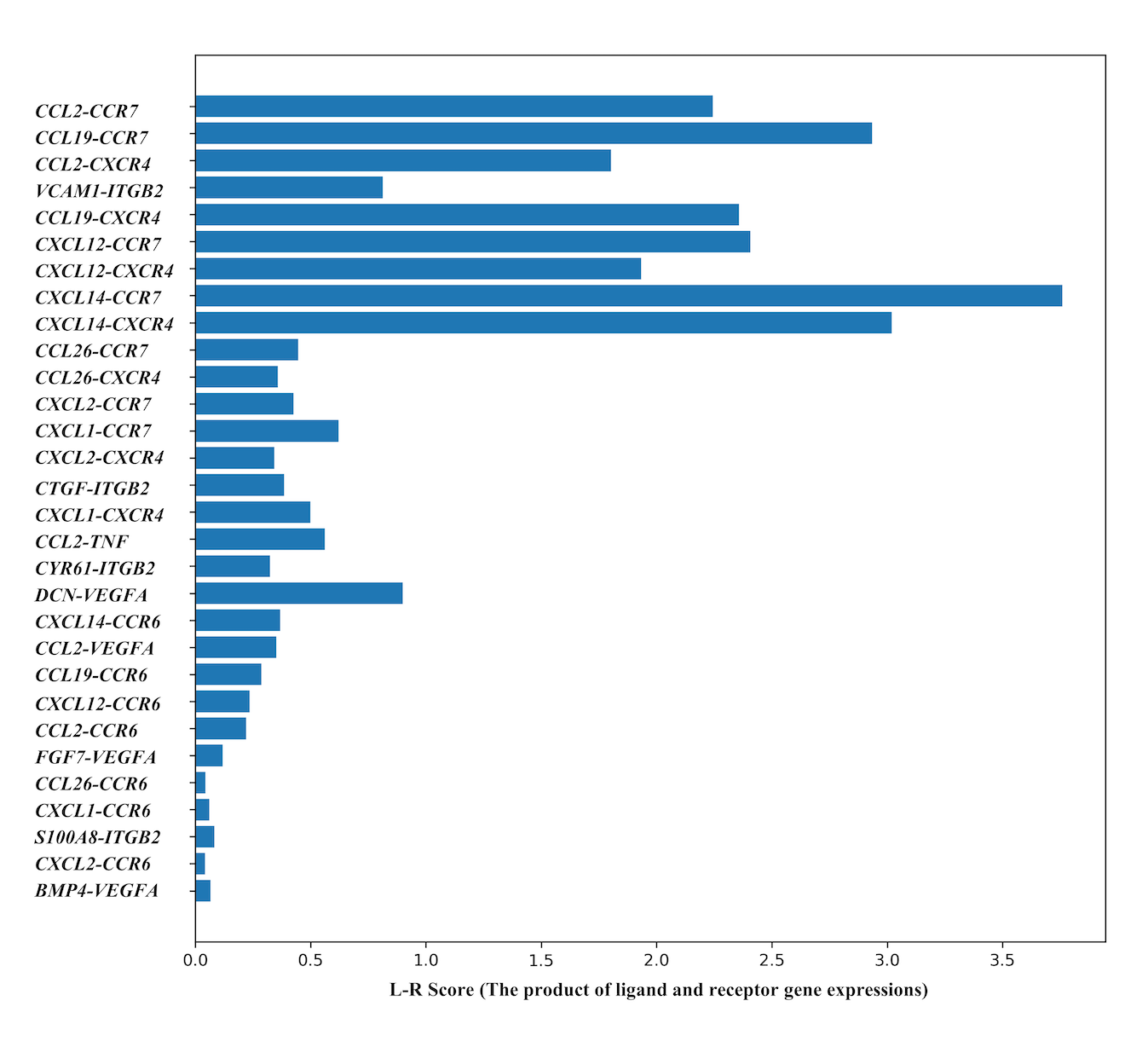}
\caption{Products of ligand and receptor expression of significant LR pairs detected by scTenifoldXct.}
\label{xct_s2}
\end{figure}


\begin{figure}[h]
\centering
\includegraphics[scale=.50]{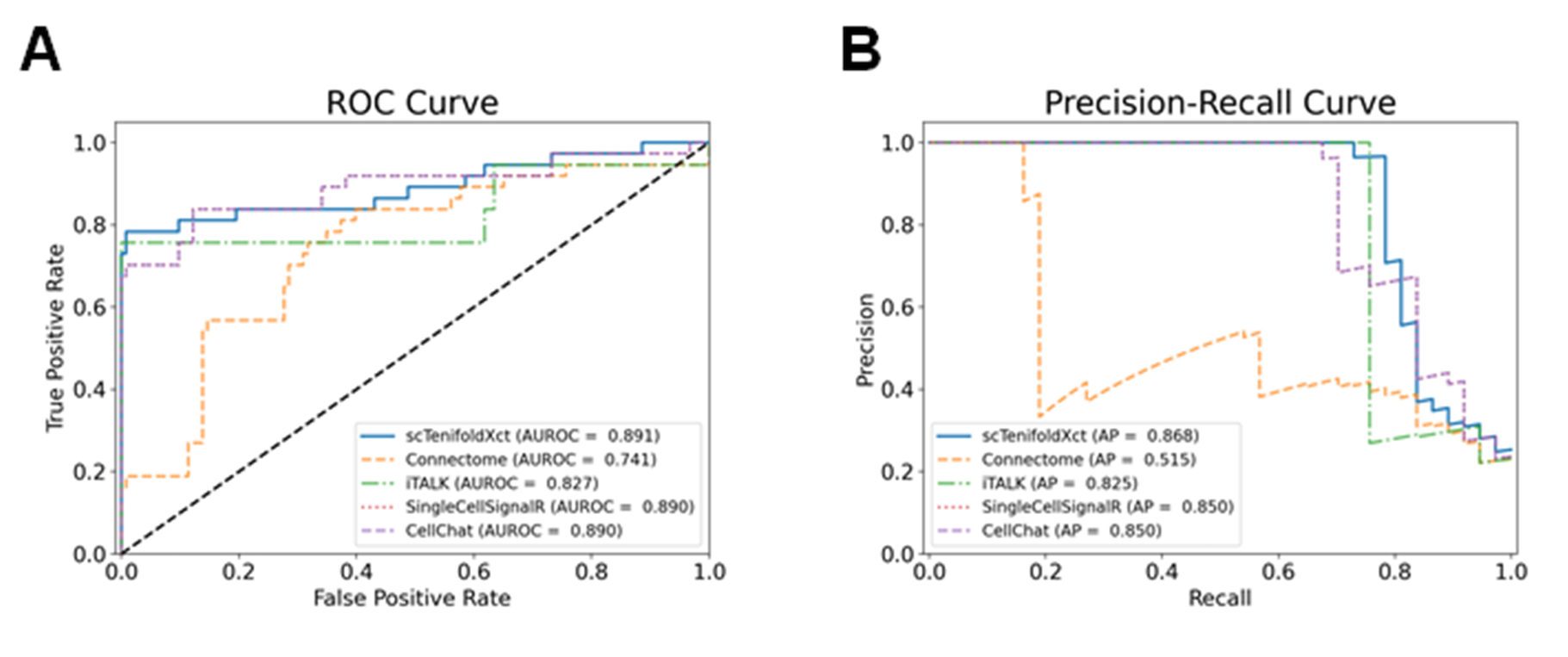}
\caption{The receiver operating characteristic (ROC) curves and the precision-recall curves presenting the comparison of the performance between scTenifoldXct and other methods.}
\raggedright
(\textbf{A}) The ROC curves for scTenifoldXct, SingleCellSignalR, Connectome, iTALK, and CellChat. NATMI was excluded from this evaluation because not all the LR pairs could be ranked. (\textbf{B}) The precision-recall curves of the methods. SingleCellSignalR and CellChat both constructed the LR score using a similar form of hill function, causing them to overlap in the ranking-based analyses.
\label{xct_s3}
\end{figure}


\begin{figure}[h]
\centering
\includegraphics[scale=.55]{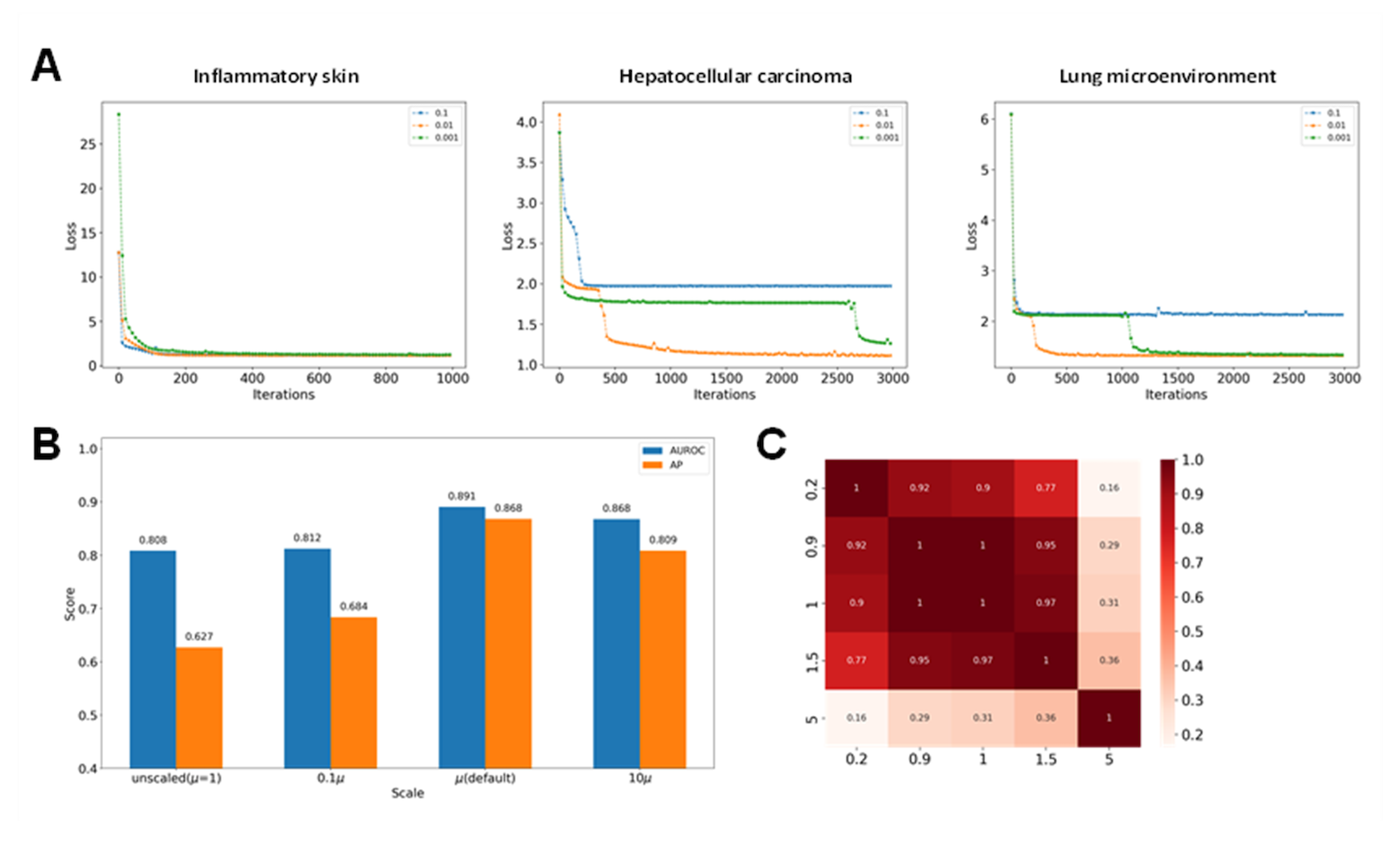}
\caption{Selection of hyperparameters for manifold alignment.}
\raggedright
(\textbf{A}) The learning curves at three different learning rates across iterations for three data sets used in the manuscript showing the single- and two-sample scTenifoldXct analysis. (\textbf{B}) Bar chart showing AUROCs and APs of single-sample scTenifoldXct analysis running with different scaling factor $\mu$ with the inflammatory skin data set. (\textbf{C}) Heatmap showing Pearson correlation coefficients between aligned distances across a range of scale factors in hyperparameter \textit{$\beta$} in two-sample scTenifoldXct analysis using the hepatocellular carcinoma data set. The numbers of significant LR pairs when the scale factor of \textit{$\beta$} $=$ 0.2, 1, 1.5, 5 and their intersection with the default setting of \textit{$\beta$} are 0/0, 10/10, 10/10, 9/3 respectively.
\label{xct_s4}
\end{figure}


\chapter{GenKI}

\section{Supplementary Tables}

\begin{table}[htbp]
\centering
\begin{tabular}{|p{1.6cm}|p{3.2cm}|p{3.2cm}|p{2.5cm}|}
\hline
\textbf{Data set} & \textbf{Cell type of interest (cell number)} & \textbf{Target KO gene} & \textbf{Source} \\
\hline
Microglia & Microglia (648) & Trem2 & GEO: GSE130627 \\
\hline
Lung & AT1 cells (624) & Nkx2-1 & GEO: GSE129628 \\
\hline
Intestine & Enterocytes (502) & Hnf4a \& Smad4 & GEO: GSE112946 \\
\hline
COVID-19 & Epithelial cells (108) & STAT1 & EGAS 00001004481 \\
\hline
Nervous system & Neurons (2,054 \& 2,156) & Mecp2 & SRA: SRP135960 \\
\hline
\end{tabular}
\caption{Summary of real scRNA-seq applications of GenKI analysis.}
\label{genki_t1s}
\end{table}

\begin{longtable}{|p{3cm}|p{3cm}|}
\hline
\textbf{Genes} & \textbf{KL Divergence} \\
\hline
\endhead
TREM2 & 1.84E5 \\
\hline
CTSD & 8.67 \\
\hline
APOE & 8.40 \\
\hline
CD74 & 8.38 \\
\hline
LYZ2 & 7.41 \\
\hline
CD9 & 5.94 \\
\hline
TYROBP & 5.35 \\
\hline
CD52 & 4.45 \\
\hline
CTSZ & 4.26 \\
\hline
CTSB & 4.11 \\
\hline
ACTB & 3.98 \\
\hline
H2-AA & 3.20 \\
\hline
CYBA & 2.79 \\
\hline
NDUFA4 & 2.76 \\
\hline
CX3CR1 & 2.65 \\
\hline
TTR & 2.47 \\
\hline
COX6C & 1.88 \\
\hline
ENPP2 & 1.61 \\
\hline
ADAP2OS & 1.53 \\
\hline
TIMP2 & 1.51 \\
\hline
\caption{Trem2-KO responsive genes in microglia cells.}
\label{genki_t2s}
\end{longtable}

\begin{longtable}{|p{4cm}|p{1.4cm}|p{1.4cm}|p{1.4cm}|p{6.2cm}|}
\hline
\textbf{Term} & \textbf{Overlap} & \textbf{p-value} & \textbf{Adj. p-value} & \textbf{Genes} \\
\hline
\endhead
Interleukin-2 signaling pathway & 7/847 & 1.1E-05 & 1.3E-03 & CX3CR1; CD52; CTSZ; TIMP2; CTSD; ACTB; CTSB \\
\hline
Other semaphorin interactions & 2/16 & 1.1E-04 & 4.0E-03 & TYROBP; TREM2 \\
\hline
Mitochondrial pathway of apoptosis & 3/97 & 1.2E-04 & 4.0E-03 & CTSZ; CTSD; CTSB \\
\hline
MHC class $I\!I$ antigen presentation & 3/103 & 1.4E-04 & 4.0E-03 & CD74; CTSD; CTSB \\
\hline
Lysosome & 3/121 & 2.3E-04 & 5.0E-03 & CTSZ; CTSD; CTSB \\
\hline
Adaptive immune system & 5/606 & 2.7E-04 & 5.0E-03 & CD74; TYROBP; CYBA; CTSD; CTSB \\
\hline
Alzheimer's disease & 3/169 & 6.1E-04 & 9.8E-03 & NDUFA4; APOE; COX6C \\
\hline
Semaphorin interactions & 2/66 & 2.0E-03 & 2.7E-02 & TYROBP; TREM2 \\
\hline
FSH regulation of apoptosis & 3/263 & 2.2E-03 & 2.7E-02 & CX3CR1; APOE; ACTB \\
\hline
Immune system & 5/998 & 2.5E-03 & 2.8E-04 & CD74; TYROBP; CYBA; CTSD; CTSB \\
\hline
\caption{Pathway enrichment analysis results of Trem2-KO responsive genes in microglia cells (n = 10).}
\label{genki_t3s}
\end{longtable}

\begin{longtable}{|p{4cm}|p{1.4cm}|p{1.4cm}|p{1.4cm}|p{6.2cm}|}
\hline
\textbf{Term} & \textbf{Overlap} & \textbf{p-value} & \textbf{Adj. p-value} & \textbf{Genes} \\
\hline
\endhead
regulation of hippocampal neuron apoptotic process & 3/5 & 8.5E-09 & 5.5E-06 & CX3CR1; TYROBP; TREM2 \\
\hline
macrophage activation involved in immune response & 3/13 & 2.4E-07 & 3.8E-05 & CX3CR1; TYROBP; TREM2 \\
\hline
neutrophil degranulation & 7/481 & 2.6E-07 & 3.8E-05 & TYROBP; TTR; CTSZ; TIMP2; CYBA; CTSD; CTSB \\
\hline
neutrophil activation involved in immune response & 7/485 & 2.8E-07 & 3.8E-05 & TYROBP; TTR; CTSZ; TIMP2; CYBA; CTSD; CTSB \\
\hline
neutrophil mediated immunity & 7/488 & 2.9E-07 & 3.8E-05 & TYROBP; TTR; CTSZ; TIMP2; CYBA; CTSD; CTSB \\
\hline
microglial cell activation & 3/22 & 1.3E-06 & 1.4E-04 & CX3CR1; TYROBP; TREM2 \\
\hline
positive regulation of transport & 4/91 & 1.8E-06 & 1.7E-04 & CYBA; TREM2; APOE; ACTB \\
\hline
regulation of interleukin-6 production & 4/110 & 3.9E-06 & 3.2E-04 & CD74; TYROBP; CYBA; TREM2 \\
\hline
regulation of microglial cell mediated cytotoxicity & 2/5 & 9.4E-06 & 6.3E-04 & CX3CR1; TYROBP \\
\hline
innate immune response & 5/302 & 9.8E-06 & 6.3E-04 & CX3CR1; CD74; TYROBP; CYBA; TREM2 \\
\hline
positive regulation of amyloid-beta clearance & 2/7 & 1.9E-05 & 1.1E-03 & TREM2; APOE \\
\hline
regulation of microglial cell migration & 2/7 & 1.9E-05 & 1.1E-03 & CX3CR1; TREM2 \\
\hline
negative regulation of long-term synaptic potentiation & 2/8 & 2.6E-05 & 1.3E-03 & TYROBP; APOE \\
\hline
lipoprotein metabolic process & 2/9 & 3.4E-05 & 1.5E-03 & APOE; CTSD \\
\hline
synapse pruning & 2/9 & 3.4E-05 & 1.5E-03 & CX3CR1; TREM2 \\
\hline
regulation of toll-like receptor 2 signaling pathway & 2/11 & 5.2E-05 & 2.0E-03 & CYBA; TREM2 \\
\hline
lipoprotein catabolic process & 2/11 & 5.2E-05 & 2.0E-03 & APOE; CTSD \\
\hline
positive regulation of interleukin-6 production & 3/76 & 5.7E-05 & 2.1E-03 & CD74; TYROBP; CYBA \\
\hline
negative regulation of amyloid fibril formation & 2/12 & 6.2E-05 & 2.1E-03 & TREM2; APOE \\
\hline
negative regulation of cell activation & 2/14 & 8.6E-05 & 2.8E-03 & TREM2; APOE \\
\hline
\caption{GO enrichment analysis results of Trem2-KO responsive genes in microglia cells (n = 20).}
\label{genki_t4s}
\end{longtable}

\begin{longtable}{|l|c|l|c|l|c|}
\hline
\textbf{Genes} & \textbf{KL Divergence} & \textbf{Genes} & \textbf{KL Divergence} & \textbf{Genes} & \textbf{KL Divergence} \\
\hline
\endhead
NKX2-1 & 49.89 & LGALS3 & 2.87E-05 & STMN1 & 2.50E-05 \\
\hline
IGFBP2 & 3.75E-05 & WFDC2 & 2.86E-05 & EMB & 2.49E-05 \\
\hline
CD74 & 3.73E-05 & TUBA1A & 2.84E-05 & PXDC1 & 2.46E-05 \\
\hline
SFTPD & 3.65E-05 & LMO7 & 2.84E-05 & CCND2 & 2.46E-05 \\
\hline
FOS & 3.64E-05 & MAL & 2.81E-05 & H2-AA & 2.45E-05 \\
\hline
JUN & 3.60E-05 & ANXA1 & 2.81E-05 & EPCAM & 2.44E-05 \\
\hline
CYR61 & 3.35E-05 & F3 & 2.80E-05 & ID2 & 2.43E-05 \\
\hline
DAG1 & 3.34E-05 & SFTPC & 2.79E-05 & CD24A & 2.37E-05 \\
\hline
EGR1 & 3.32E-05 & TPPP3 & 2.76E-05 & GAS6 & 2.36E-05 \\
\hline
NDNF & 3.31E-05 & KRT18 & 2.75E-05 & ACTN4 & 2.35E-05 \\
\hline
JUNB & 3.26E-05 & RPS2 & 2.73E-05 & H2-AB1 & 2.35E-05 \\
\hline
NAPSA & 3.26E-05 & ANO1 & 2.70E-05 & PLOD2 & 2.33E-05 \\
\hline
AQP1 & 3.25E-05 & RNASE4 & 2.70E-05 & RAB25 & 2.31E-05 \\
\hline
CTSH & 3.25E-05 & PRNP & 2.69E-05 & CLDN7 & 2.27E-05 \\
\hline
NPC2 & 3.19E-05 & SPOCK2 & 2.69E-05 & AGER & 2.27E-05 \\
\hline
SFTA2 & 3.13E-05 & S100A11 & 2.68E-05 & ARL6IP1 & 2.24E-05 \\
\hline
TIMP3 & 3.12E-05 & DDC & 2.65E-05 & MBIP & 2.20E-05 \\
\hline
RGCC & 3.10E-05 & LAMC2 & 2.65E-05 & TMEM108 & 2.11E-05 \\
\hline
DUSP1 & 3.05E-05 & ATP1B1 & 2.64E-05 & H2-EB1 & 2.08E-05 \\
\hline
IER2 & 3.03E-05 & MFGE8 & 2.64E-05 & PAICS & 2.05E-05 \\
\hline
SCGB1A1 & 3.03E-05 & CLU & 2.63E-05 & PDLIM1 & 2.00E-05 \\
\hline
CLIC5 & 3.02E-05 & PHLDA1 & 2.62E-05 & POPDC3 & 1.99E-05 \\
\hline
S100A6 & 2.95E-05 & PTMA & 2.61E-05 &  &  \\
\hline
ANXA2 & 2.93E-05 & IGFBP7 & 2.61E-05 &  &  \\
\hline
TMSB4X & 2.93E-05 & MUC1 & 2.61E-05 &  &  \\
\hline
CLDN3 & 2.93E-05 & S100A10 & 2.60E-05 &  &  \\
\hline
KRT8 & 2.91E-05 & HOPX & 2.58E-05 &  &  \\
\hline
RTKN2 & 2.90E-05 & AKAP5 & 2.55E-05 &  &  \\
\hline
KRT19 & 2.90E-05 & CYSTM1 & 2.52E-05 &  &  \\
\hline
NNAT & 2.88E-05 & FJX1 & 2.51E-05 &  &  \\
\hline
\caption{Nkx2-1-KO responsive genes in AT1 cells.}
\label{genki_t5s}
\end{longtable}

\begin{longtable}{|p{4cm}|p{1.4cm}|p{1.4cm}|p{1.4cm}|p{6.2cm}|}
\hline
\textbf{Term} & \textbf{Overlap} & \textbf{p-value} & \textbf{Adj. p-value} & \textbf{Genes} \\
\hline
\endhead
positive regulation of macromolecule metabolic process & 12/384 & 5.2E-08 & 6.1E-05 & CD74; EGR1; CLDN3; RGCC; ANXA2; ID2; NKX2-1; MBIP; CTSH; GAS6; CLU; F3 \\
\hline
chemical homeostasis within a tissue & 3/7 & 2.3E-06 & 7.2E-04 & NAPSA; SFTPD; CTSH \\
\hline
surfactant homeostasis & 3/7 & 2.3E-06 & 7.2E-04 & NAPSA; SFTPD; CTSH \\
\hline
regulation of cell population proliferation & 14/764 & 2.5E-06 & 7.2E-04 & JUN; RTKN2; LAMC2; CLU; CLDN3; CCND2; RGCC; RAB25; EPCAM; ID2; CTSH; IGFBP7; JUNB; S100A11 \\
\hline
positive regulation of cell population proliferation & 11/474 & 3.6E-06 & 8.2E-04 & CD74; CCND2; RAB25; EPCAM; ID2; RTKN2; S100A6; CTSH; LAMC2; GAS6; AQP1 \\
\hline
positive regulation of gene expression & 11/482 & 4.3E-06 & 8.2E-04 & CD74; EGR1; CLDN3; RGCC; ID2; NKX2-1; MBIP; CTSH; GAS6; CLU; F3 \\
positive regulation of fibroblast proliferation & 4/28 & 4.9E-06 & 8.3E-04 & CD74; S100A6; GAS6; AQP1 \\
\hline
regulation of cell migration & 10/408 & 6.5E-06 & 9.4E-04 & CD74; ANXA1; CLDN3; TMSB4X; DAG1; NKX2-1; CTSH; LAMC2; ACTN4; F3 \\
\hline
cellular protein metabolic process & 10/417 & 7.8E-06 & 1.0E-03 & NAPSA; SFTPC; SFTPD; IGFBP2; CTSH; PLOD2; IGFBP7; RPS2; GAS6; MFGE8 \\
\hline
regulation of apoptotic process & 13/742 & 9.5E-06 & 1.0E-03 & CD74; JUN; ANXA1; ARL6IP1; ACTN4; CLU; AQP1; KRT18; CCND2; CTSH; GAS6; PHLDA1; PTMA \\
\hline
negative regulation of cytokine production & 7/182 & 9.8E-06 & 1.0E-03 & PRNP; ANXA1; RGCC; TMSB4X; SFTPD; GAS6; AGER \\
\hline
negative regulation of programmed cell death & 9/381 & 2.6E-05 & 2.5E-03 & CD74; ANXA1; KRT18; CCND2; ARL6IP1; CTSH; GAS6; PTMA; AQP1 \\
\hline
negative regulation of apoptotic process & 10/485 & 2.9E-05 & 2.6E-03 & CD74; ANXA1; KRT18; CCND2; ARL6IP1; CTSH; NDNF; GAS6; PTMA; AQP1 \\
\hline
regulation of fibroblast proliferation & 4/46 & 3.8E-05 & 3.1E-03 & CD74; S100A6; GAS6; AQP1 \\
\hline
positive regulation of neuron death & 4/47 & 4.1E-05 & 3.2E-03 & EGR1; PRNP; FOS; CLU \\
\hline
positive regulation of cell migration & 7/269 & 1.2E-04 & 7.9E-03 & CLDN3; RAB25; CTSH; LAMC2; ACTN4;F3;S100A11 \\
\hline
cellular response to reactive oxygen species & 4/63 & 1.3E-04 & 7.9E-03 & JUN; ANXA1; FOS; AQP1 \\
\hline
positive regulation of cell death & 4/66 & 1.5E-04 & 8.0E-03 & EGR1; PRNP; FOS; CLU \\
\hline
negative regulation of intrinsic apoptotic signaling pathway in response to DNA damage & 3/26 & 1.6E-04 & 8.0E-03 & CD74; MUC1; CLU \\
\hline
positive regulation of G1/S transition of mitotic cell cycle & 3/26 & 1.6E-04 & 8.0E-03 & CCND2; ANXA1; RGCC \\
\hline
\caption{GO enrichment analysis results of Nkx2-1-KO responsive genes in AT1 cells (n = 20).}
\label{genki_t6s}
\end{longtable}

\begin{longtable}{|p{3cm}|p{3cm}|}
\hline
\textbf{Genes} & \textbf{KL Divergence} \\
\hline
\endhead
HNF4A & 7.03E2 \\
\hline
SMAD4 & 10.57 \\
\hline
REG3B & 1.29E-02 \\
\hline
REG3G & 1.09E-02 \\
\hline
DMBT1 & 4.18E-03 \\
\hline
REG1 & 3.05E-03 \\
\hline
CCK & 1.91E-03 \\
\hline
GCG & 1.43E-03 \\
\hline
GSTA1 & 1.05E-03 \\
\hline
UGT2B34 & 3.34E-04 \\
\hline
GCLM & 3.33E-04 \\
\hline
SFXN1 & 3.02E-04 \\
\hline
LAMP2 & 2.90E-04 \\
\hline
TSC22D1 & 2.58E-04 \\
\hline
\caption{Hnf4a \& Smad4-KO responsive genes in enterocytes.}
\label{genki_t7s}
\end{longtable}

\begin{longtable}{|p{4cm}|p{1.4cm}|p{1.4cm}|p{1.4cm}|p{4.2cm}|}
\hline
\textbf{Term} & \textbf{Overlap} & \textbf{p-value} & \textbf{Adj. p-value} & \textbf{Genes} \\
\hline
\endhead
glutathione metabolic process & 2/43 & 4.0E-04 & 3.8E-02 & GSTA1; GCLM \\
\hline
defense response to Gram-positive bacterium & 2/70 & 1.0E-03 & 3.8E-02 & DMBT1; REG3G \\
\hline
carbohydrate homeostasis & 2/70 & 1.1E-03 & 3.8E-02 & HNF4A; GCG \\
\hline
regulation of peptide hormone secretion & 2/74 & 1.2E-03 & 3.8E-02 & HNF4A; GCG \\
\hline
glucose homeostasis & 2/86 & 1.6E-03 & 3.8E-02 & HNF4A; GCG \\
\hline
epithelial cell differentiation & 2/101 & 2.2E-03 & 3.8E-02 & DMBT1; GSTA1 \\
\hline
regulation of insulin secretion & 2/104 & 2.3E-03 & 3.8E-02 & HNF4A; GCG \\
\hline
sulfur compound biosynthetic process & 2/113 & 2.8E-03 & 3.8E-02 & GSTA1; GCLM \\
\hline
epithelium development & 2/122 & 3.2E-03 & 3.8E-02 & DMBT1; GSTA1 \\
\hline
regulation of protein secretion & 2/125 & 3.3E-03 & 3.8E-02 & HNF4A; GCG \\
\hline
negative regulation of cell growth & 2/126 & 3.4E-03 & 3.8E-02 & SMAD4; HNF4A \\
\hline
negative regulation of growth & 2/126 & 3.4E-03 & 3.8E-02 & SMAD4; HNF4A \\
\hline
positive regulation of cell proliferation involved in heart morphogenesis & 1/5 & 3.5E-03 & 3.8E-02 & SMAD4 \\
\hline
lysosomal protein catabolic process & 1/5 & 3.5E-03 & 3.8E-02 & LAMP2 \\
\hline
negative regulation of keratinocyte differentiation & 1/5 & 3.5E-03 & 3.8E-02 & REG3G \\
\hline
positive regulation of histone H3-K9 acetylation & 1/5 & 3.5E-03 & 3.8E-02 & SMAD4 \\
\hline
cardiac endothelial cell differentiation & 1/5 & 3.5E-03 & 3.8E-02 & SMAD4 \\
\hline
endocardial cell differentiation & 1/5 & 3.5E-03 & 3.8E-02 & SMAD4 \\
\hline
negative regulation of cardiac muscle cell differentiation & 1/6 & 4.2E-03 & 3.8E-02 & SMAD4 \\
\hline
induction of bacterial agglutination & 1/6 & 4.2E-03 & 3.8E-02 & DMBT1 \\
\hline
\caption{GO enrichment analysis results of Hnf4a \& Smad4-KO responsive genes in enterocytes (n = 20).}
\label{genki_t8s}
\end{longtable}

\begin{longtable}{|p{2.2cm}|p{1.9cm}|p{1.9cm}|p{1.9cm}|p{1.8cm}|p{1.9cm}|p{1.8cm}|}
\hline
\textbf{DE method} & \textbf{Wilcoxon} & \textbf{Wilcoxon (logFC>1)} & \textbf{t-test} & \textbf{negbinom} & \textbf{MAST} & \textbf{DESeq2} \\
\hline
Input data & Normalized data & Normalized data & Normalized data & Raw counts & Normalized data & Raw counts \\
\hline
Model & Non-parametric test & Non-parametric test & Parametric test & Negative binomial model & Generalized linear model & Negative binomial model \\
\hline
DE genes & 1129 & 528 & 1019 & 878 & 1019 & 1137 \\
\hline
Up genes & 273 & 175 & 293 & 167 & 293 & 240 \\
\hline
Down genes & 856 & 353 & 726 & 711 & 726 & 897 \\
\hline
DE intersect & 71 (17) & 49 (15) & 70 (7) & 69 (24) & 74 (21) & 81 (11) \\
\hline
Up\&DE intersect & 42 & 29 & 41 & 28 & 44 & 35 \\
\hline
Down\&DE intersect & 29 & 20 & 29 & 41 & 30 & 46 \\
\hline
\caption{DE analyses of the Nkx2-1 KO experiment resulting from different DE methods.}
\label{genki_t9s}
\end{longtable}

\begin{longtable}{|p{3cm}|p{3cm}|}
\hline
\textbf{Genes} & \textbf{KL divergence} \\
\hline
STAT1 & 2.94E3 \\ \hline
S100A8 & 8.72 \\ \hline
HLA-DRA & 4.29 \\ \hline
S100A9 & 2.90 \\ \hline
FTL & 1.98 \\ \hline
WFDC2 & 1.59 \\ \hline
ANXA1 & 1.56 \\ \hline
LCN2 & 1.33 \\ \hline
CSTB & 1.21 \\ \hline
PPDPF & 1.13 \\ \hline
VIM & 1.04 \\ \hline
SLPI & 0.87 \\ \hline
HLA-DRB1 & 0.83 \\ \hline
LYPD2 & 0.81 \\ \hline
S100P & 0.80 \\ \hline
GPX1 & 0.66 \\ \hline
CTSB & 0.61 \\ \hline
PI3 & 0.59 \\ \hline
KRT19 & 0.55 \\ \hline
HSPB1 & 0.53 \\ \hline
FABP5 & 0.52 \\ \hline
AGR2 & 0.41 \\ \hline
LGALS1 & 0.37 \\ \hline
IL1RN & 0.35 \\ \hline
CTSD & 0.34 \\ \hline
HLA-DPA1 & 0.31 \\ \hline
HSPA5 & 0.27 \\ \hline
TIMP1 & 0.26 \\ \hline
\caption{STAT1-KO responsive genes in epithelial cells.}
\label{genki_t10s}
\end{longtable}

\begin{longtable}{|p{4cm}|p{1.4cm}|p{1.4cm}|p{1.4cm}|p{6.2cm}|}
\hline
\textbf{Term} & \textbf{Overlap} & \textbf{p-value} & \textbf{Adj. p-value} & \textbf{Genes} \\
\hline
\endhead
neutrophil degranulation & 10/481 & 5.2E-10 & 1.1E-07 & CSTB; SLPI; FABP5; LCN2; S100P; CTSD; S100A9; S100A8; CTSB; FTL \\
\hline
neutrophil activation involved in immune response & 10/485 & 5.7E-10 & 1.1E-07 & CSTB; SLPI; FABP5; LCN2; S100P; CTSD; S100A9; S100A8; CTSB; FTL \\
\hline
neutrophil mediated immunity & 10/488 & 6.0E-10 & 1.1E-07 & CSTB; SLPI; FABP5; LCN2; S100P; CTSD; S100A9; S100A8; CTSB; FTL \\
\hline
cytokine-mediated signaling pathway & 9/621 & 1.0E-07 & 1.4E-05 & IL1RN; ANXA1; STAT1; LCN2; HLA-DRA; TIMP1; VIM; HLA-DRB1; HLA-DPA1 \\
\hline
defense response to bacterium & 6/176 & 1.4E-07 & 1.5E-05 & SLPI; LCN2; PI3; S100A9; WFDC2; S100A8 \\
\hline
interferon-gamma-mediated signaling pathway & 4/68 & 2.4E-06 & 2.1E-04 & STAT1; HLA-DRA; HLA-DRB1; HLA-DPA1 \\
\hline
cellular response to cytokine stimulus & 7/482 & 3.5E-06 & 2.7E-04 & IL1RN; ANXA1; STAT1; LCN2; TIMP1; VIM; HLA-DPA1 \\
\hline
antigen processing and presentation of exogenous peptide antigen via MHC class II & 4/98 & 1.0E-05 & 6.6E-04 & HLA-DRA; CTSD; HLA-DRB1; HLA-DPA1 \\
\hline
antigen processing and presentation of peptide antigen via MHC class II & 4/100 & 1.1E-05 & 6.6E-04 & HLA-DRA; CTSD; HLA-DRB1; HLA-DPA1 \\
\hline
response to lipid & 4/114 & 1.8E-05 & 7.8E-04 & ANXA1; SLPI; S100A9; S100A8 \\
\hline
\caption{GO enrichment analysis results of STAT1-KO responsive genes in epithelial cells (n = 10).}
\label{genki_t11s}
\end{longtable}

\newpage
\section{Supplementary Figures}

\begin{figure}[h]
\centering
\includegraphics[scale=.70]{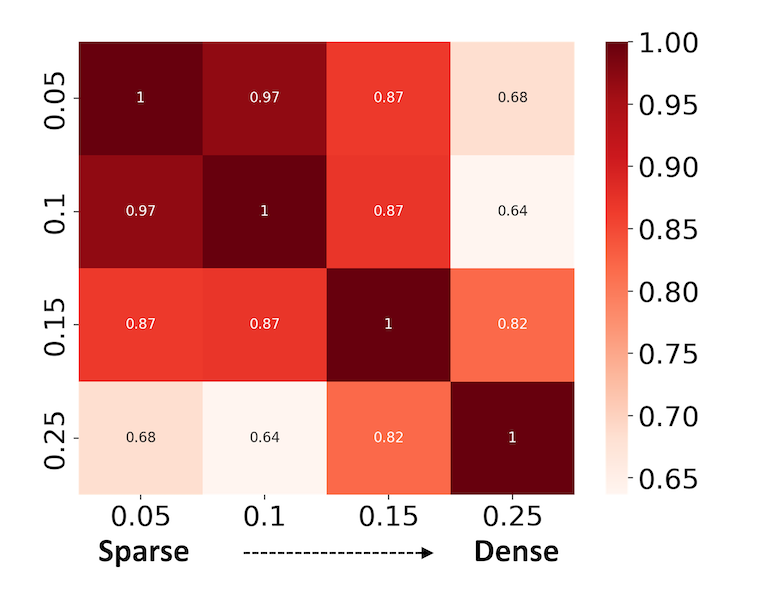}
\caption{Heatmap of Spearman correlation coefficients between scores of KL divergence given by GenKI across four different cutoffs.}
\label{genki_s1}
\end{figure}

\begin{figure}[h]
\centering
\includegraphics[scale=.55]{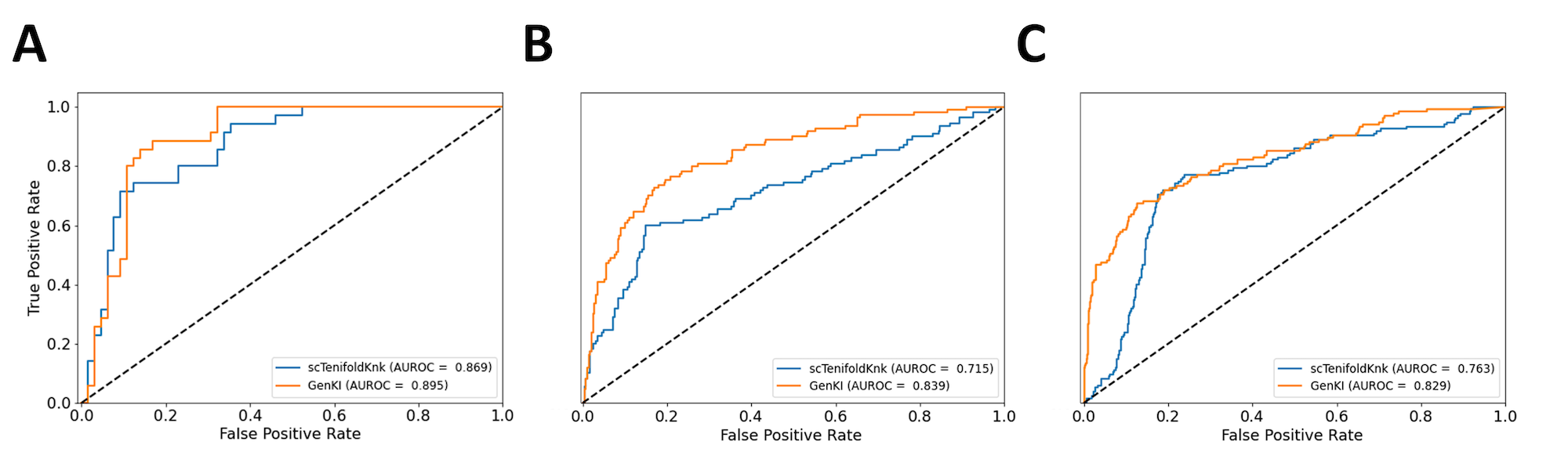}
\caption{ROC curves of virtual KO experiments using three simulated SERGIO data sets.}
\raggedright
(A) KO gene \#74 (35 edges) in SERGIO simulated data set of 100 genes by 2,700 cells. (B) KO gene \#47 (110 edges) in SERGIO simulated data set of 400 genes by 2,700 cells. (C) KO gene \#302 (135 edges) in SERGIO simulated data set of 1,200 genes by 2,700 cells.
\label{genki_s2}
\end{figure}

\begin{figure}[h]
\centering
\includegraphics[scale=.70]{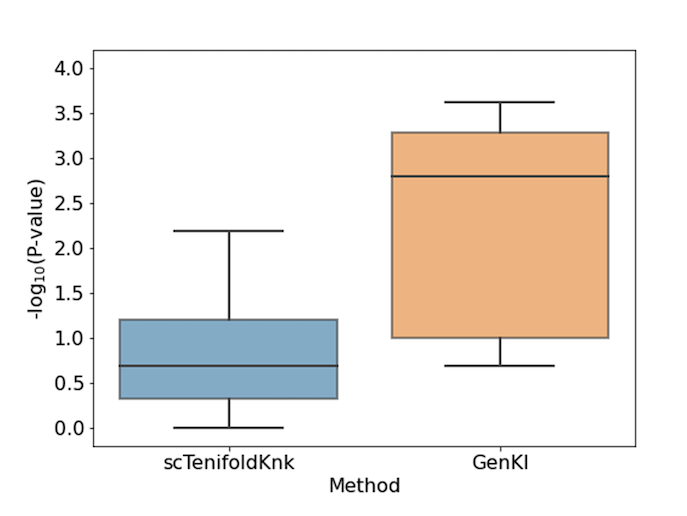}
\caption{Comparing p-value given by GenKI and scTenifoldKnk to genes that are close or distance neighbors of a KO gene in the BEELINE data set.}
\label{genki_s3}
\end{figure}

\begin{figure}[h]
\centering
\includegraphics[scale=.70]{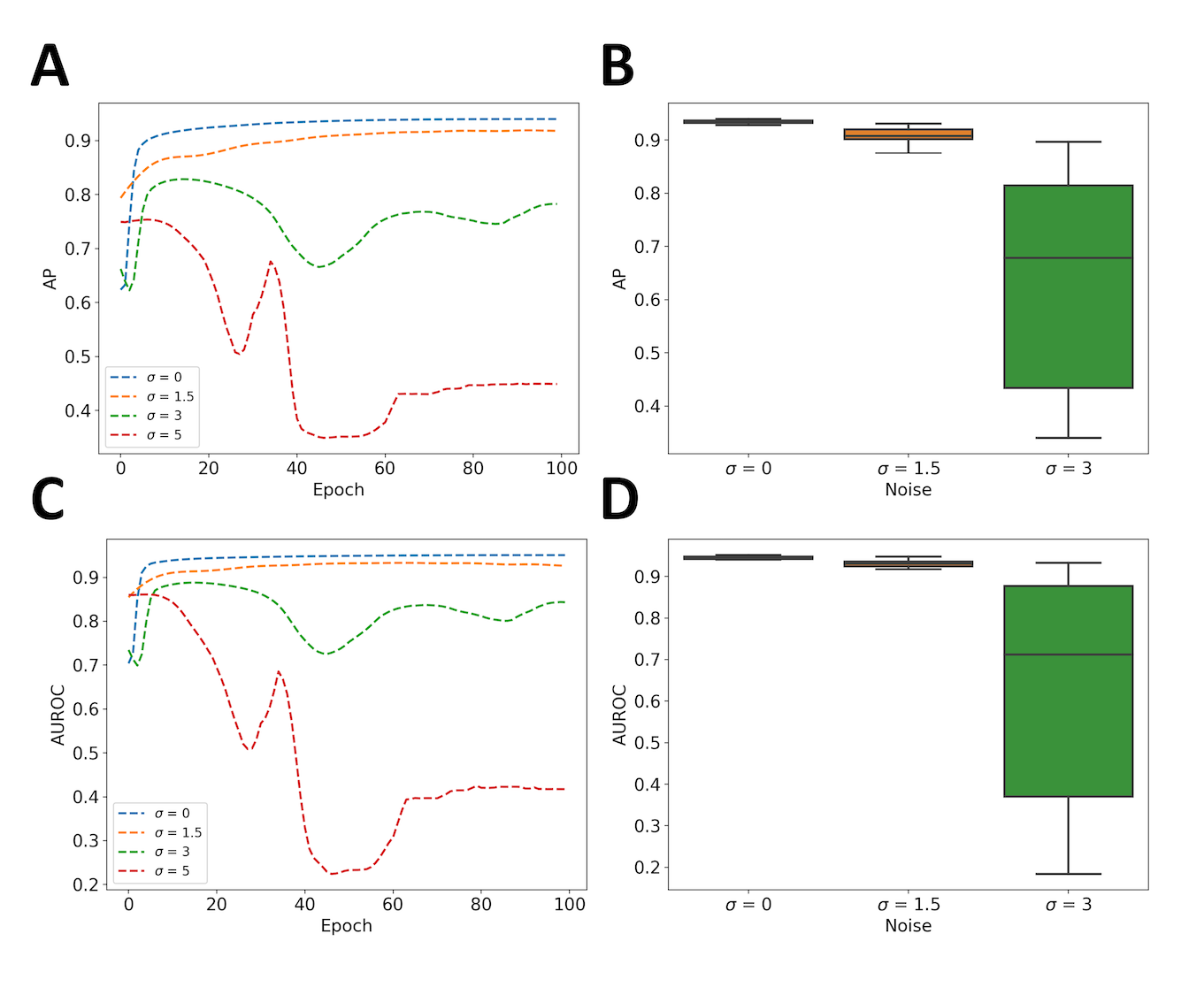}
\caption{Performance of GenKI in reconstructing scGRNs of microglia data set.}
\raggedright
(A) The levels of AP at different training epochs. (B) The APs between data with different levels of random noise in the gene expression profiles. Larger $\sigma$ values result in higher levels of random noise (see “Materials and Methods” for details). (C) The levels of AUROC at different training epochs. (D) The AUROCs between data with different levels of random noise in the gene expression profiles.
\label{genki_s4}
\end{figure}

\begin{figure}[h]
\centering
\includegraphics[scale=.70]{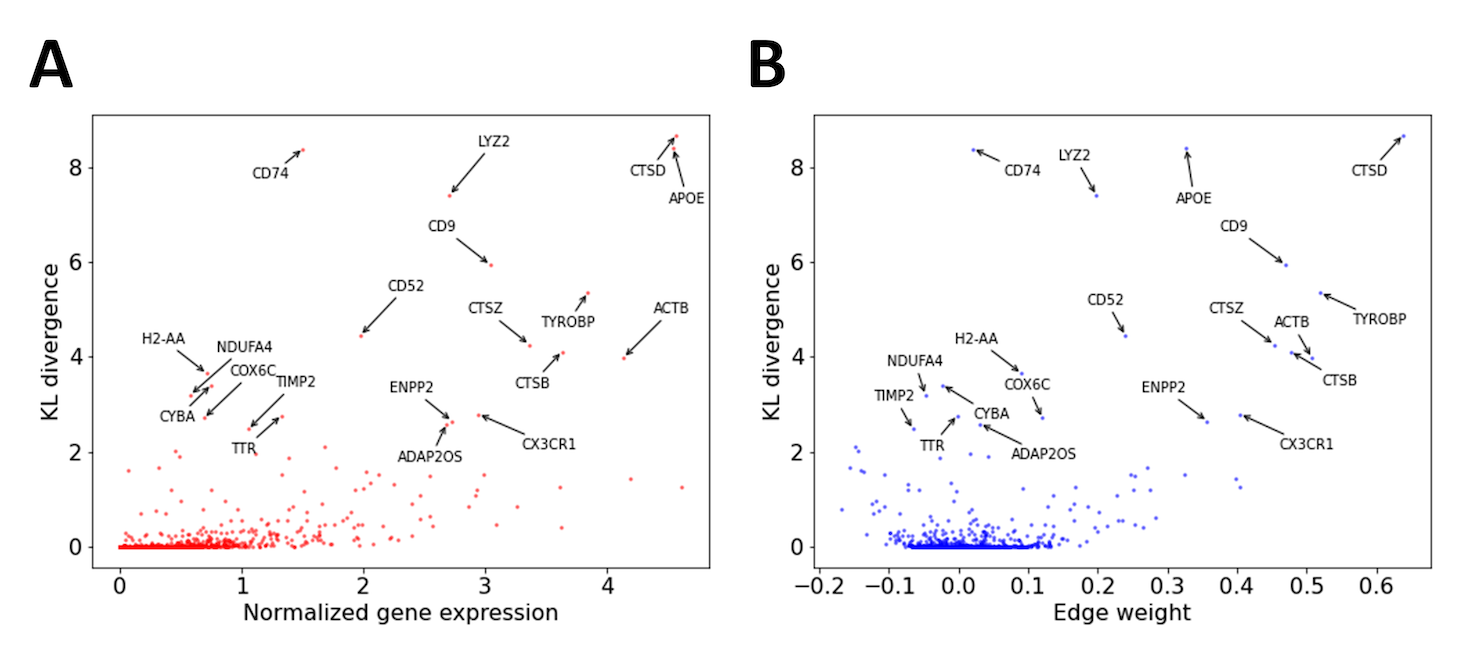}
\caption{KO responsive genes cannot be simply inferred from gene expression or edge weight with the KO gene.}
\raggedright
(A) Scatter plot revealing the relationship between normalized gene expression and KL divergence. (B) Scatter plot revealing the relationship between edge weight associated with Trem2 and KL divergence. The KO-responsive genes inferred by GenKI are annotated and Trem2 as an outlier is ignored for visualization purpose.
\label{genki_s5}
\end{figure}

\begin{figure}[h]
\centering
\includegraphics[scale=.55]{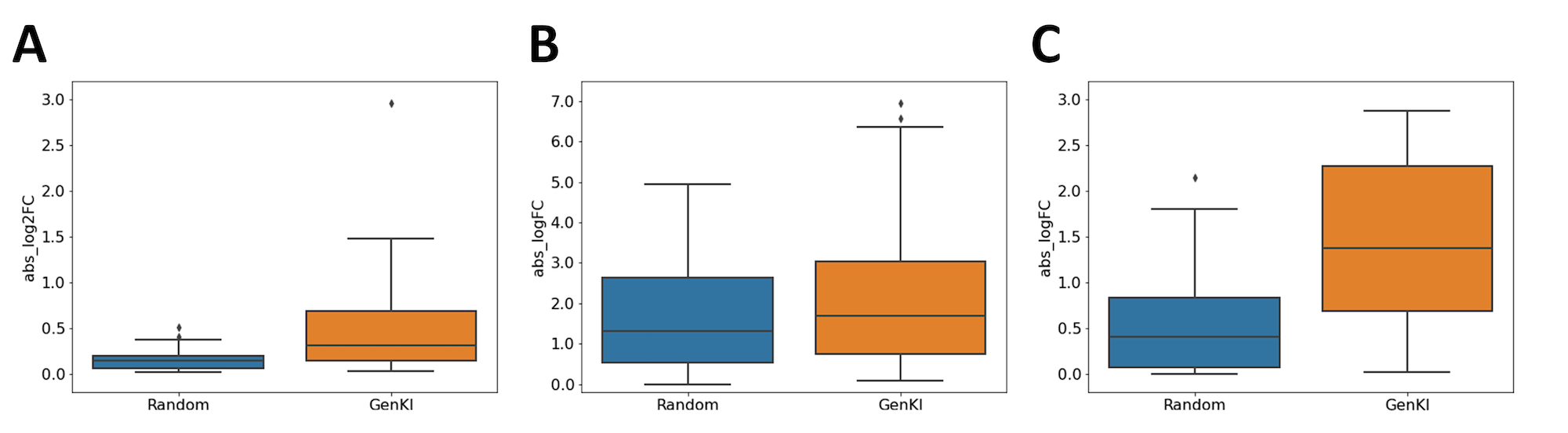}
\caption{Comparison of absolute values of fold-change (FC=WT/KO) between randomly selected non-perturbed genes and KO-responsive genes given by GenKI.}
\raggedright
(A) Microglia data set (B) Lung data set (C) Intestine data set. P-values of the one-sided t-test are 0.0168, 0.0177 and 0.0175, respectively.
\label{genki_s6}
\end{figure}

\begin{figure}[h]
\centering
\includegraphics[scale=.55]{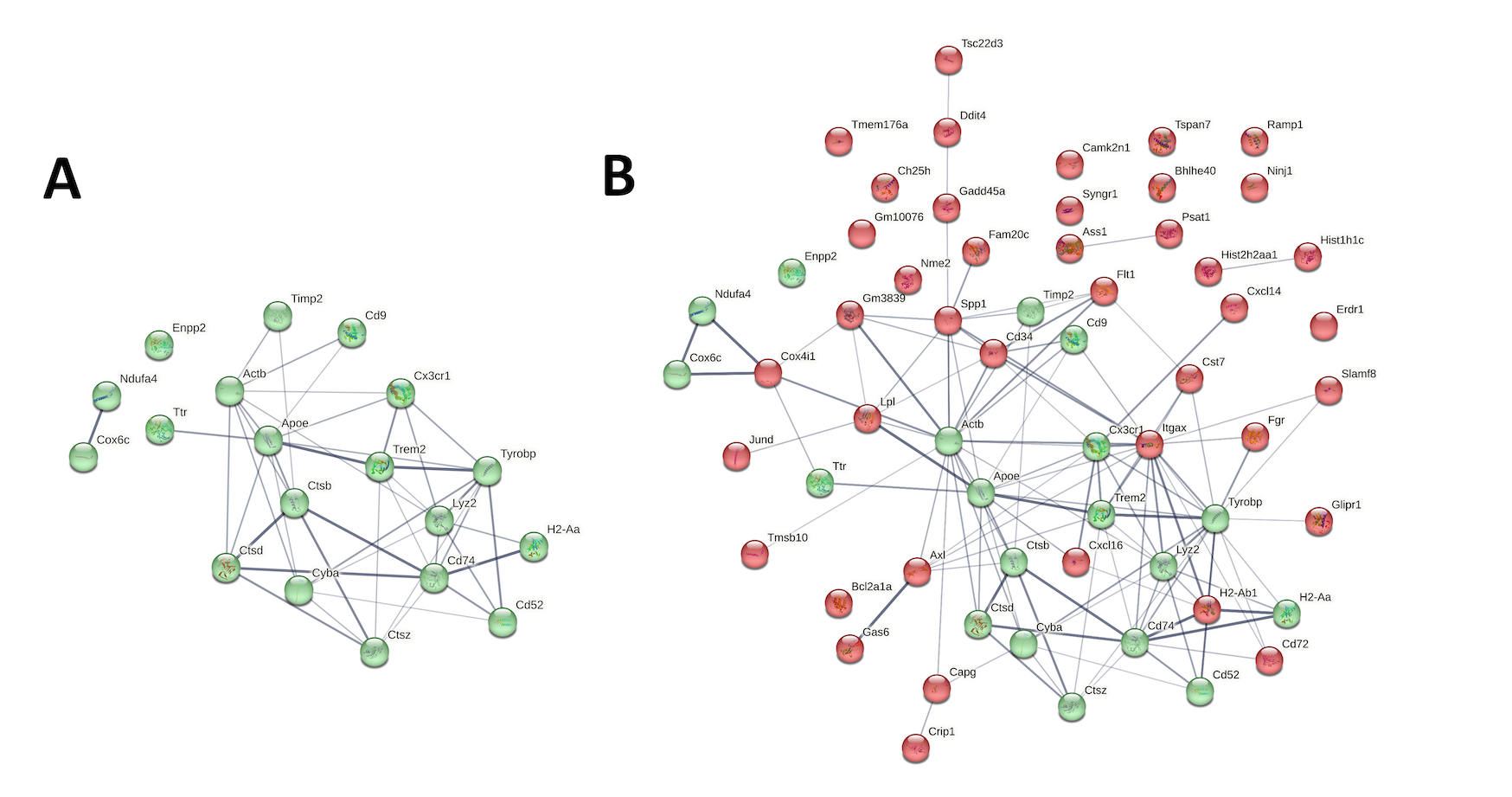}
\caption{STRING networks of Trem2-KO responsive genes before and after including DE genes.}
\raggedright
(A) Trem2-KO responsive genes (in green) without DE genes (B) Trem2-KO responsive genes with top ranked DE genes (in red).
\label{genki_s7}
\end{figure}

\begin{figure}[h]
\centering
\includegraphics[scale=.70]{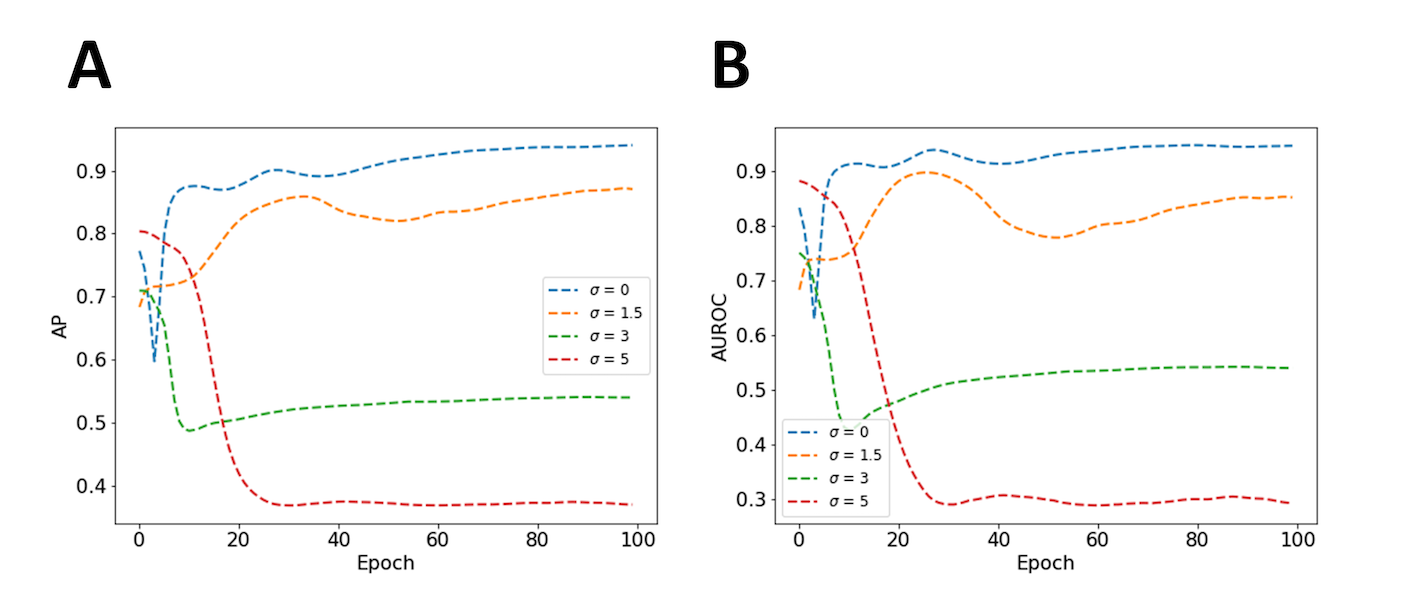}
\caption{Performance of GenKI in reconstructing scGRNs of COVID-19 data set.}
\raggedright
(A) The levels of AP at different training epochs. (B) The levels of AUROC at different training epochs. Larger $\sigma$ values result in higher levels of random noise.
\label{genki_s8}
\end{figure}

\begin{figure}[h]
\centering
\includegraphics[scale=.55]{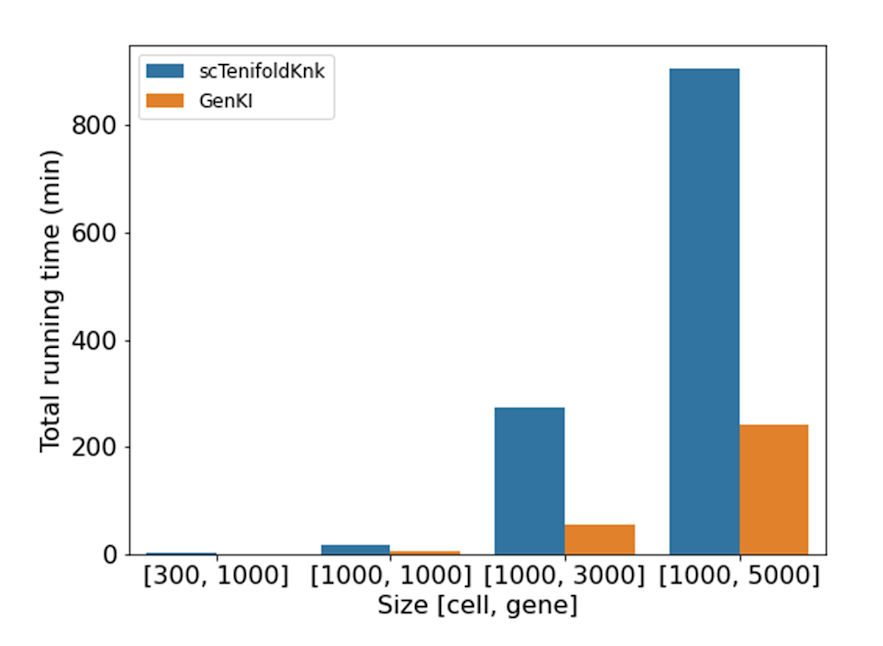}
\caption{Total running time of GenKI and scTenifoldKnk with respect to data set sizes.}
\raggedright
All evaluations were implemented on the equivalent hardware composed of a 16-core Intel Core i7-11700 CPU processor at 2.50 GHz with 15.7 GB available random-access memory (RAM). PyTorch profiler was used for GenKI and the system.time function for scTenifoldKnk.
\label{genki_s9}
\end{figure}

\pagebreak{}

\chapter{CrossmodalNet}

\section{Supplementary Tables}

\begin{table}[ht]
\begin{center}
\begin{small}
\begin{sc}
\begin{tabular}{lcccr}
\toprule
Module & Hyperparameter & Default value & Search space \\
\midrule
MLP General & Batch normalization & True & choice([True, False]) \\
MLP General & Learning rate & 0.001 & qloguniform(1e-4, 1e-1, 5e-5) \\
MLP General & Weight decay & 5e-6 & $10^{randint(-3, -7)}$ \\
MLP1    & dropout & 0.05 & choice([0, 0.05, 0.15, 0.3]) \\
MLP1 & Latent dimension & 512 & choice([256, 512])\\
Discriminator & Hidden dimension & 128 & choice([128, 32]) \\
Discriminator & Regularization & 0.5 & quniform(0, 2, 0.1) \\
Discriminator & Gradient penalty & 0.4 & quniform(0, 2, 0.1) \\
Discriminator & Learning rate & 0.0087 & qloguniform(1e-4, 1e-1, 5e-5) \\
Discriminator & Weight decay & 5e-5 & $10^{randint(-3, -7)}$ \\
Discriminator & Learning step & 3 & choice([3, 5, 10]) \\
Gradnorm   & Weight learning rate & 0.01475 & qloguniform(1e-4, 1e-1, 5e-5) \\
Gradnorm   & Alpha & 0.5 & quniform(0, 3, 0.1) \\
\bottomrule
\end{tabular}
\end{sc}
\end{small}
\caption{Random search spaces for hyperparameter tuning.}
\label{hyperparameter-table}
\end{center}
\end{table}

\begin{table}[ht]
\begin{center}
\begin{small}
\begin{sc}
\begin{tabular}{lcccc}
\toprule
Model & \begin{tabular}{@{}c@{}}D1 \\ (1,000, 3,000)\end{tabular} & \begin{tabular}{@{}c@{}}D2 \\ (3,000, 5,000)\end{tabular} & \begin{tabular}{@{}c@{}}D3 \\ (10,000, 8,000)\end{tabular} & \begin{tabular}{@{}c@{}}D4 \\ (50,000, 15,000) \end{tabular} \\
\midrule
CrossmodalNet    & 22.17 & 147.66 & 791.84 & 6539.13\\
MLP & 89.96 & 302.67 & 1604.53 & 7180.55\\
lightGBM    & 638.01 & 4364.12 & 29606.51 & 138094.40\\
Linear reg.   & 48.53 & 850.14 & 23845.28 & 136474.90\\
Ridge reg.    & 11.16 & 97.32 & 1099.67 & 11759.31\\
scIPENN      & 42.28 & 154.48 & 626.36 & 6516.53\\
\bottomrule
\end{tabular}
\end{sc}
\end{small}
\caption{Running time (s) on synthetic data sets (samples, features).}
\label{scale-table}
\end{center}
\end{table}

\newpage

\section{Supplementary Figures}

\begin{figure}[ht]
\begin{center}
\includegraphics[width=0.45\columnwidth]{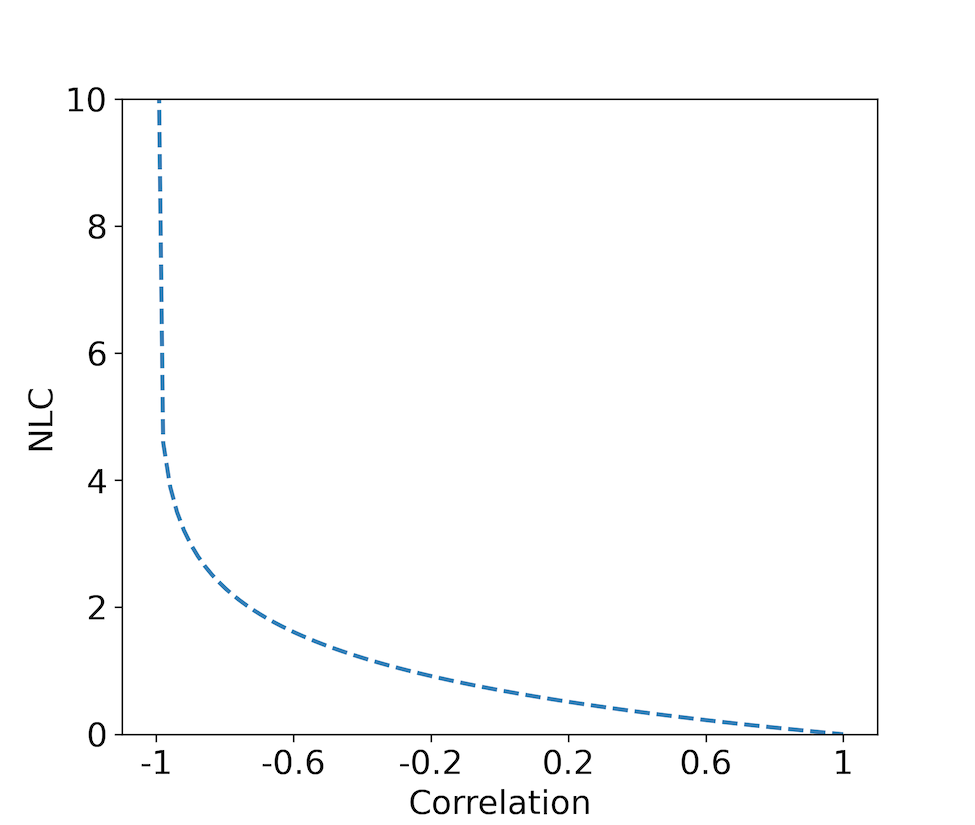}
\caption{NLC loss curve across correlations.}
\label{NLC_loss}
\end{center}
\end{figure}

\begin{figure}
    \begin{center}
    \begin{subfigure}[b]{0.33\columnwidth}
    \begin{center}       
        \includegraphics[width=0.85\linewidth]{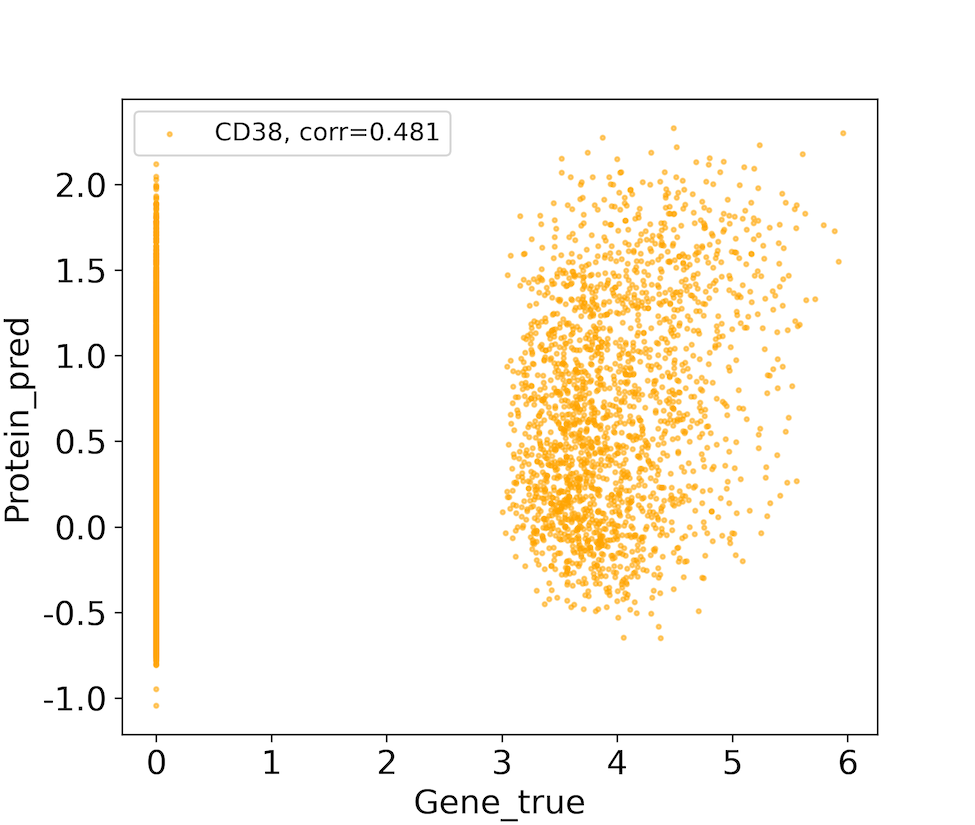}
        \caption{Scatter plot of CD38.}
        \label{CD38_sup}
    \end{center}
    \end{subfigure}%
    \hfill
    \begin{subfigure}[b]{0.33\columnwidth}
    \begin{center}
        \includegraphics[width=0.85\linewidth]{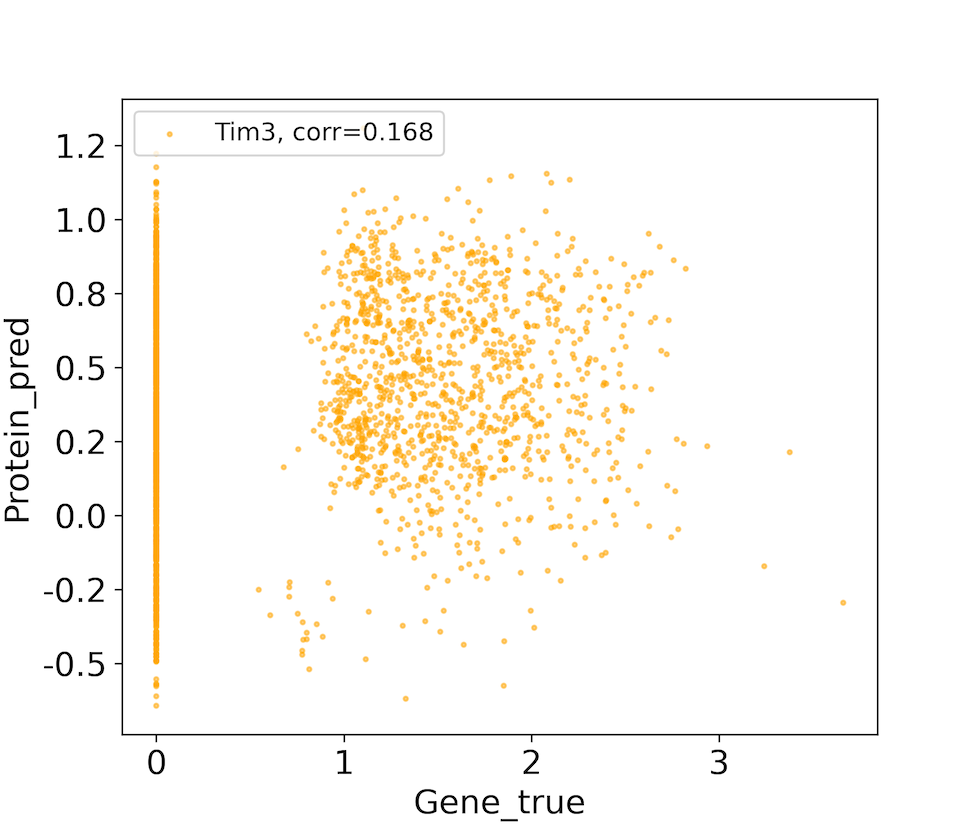}
        \caption{Scatter plot of Tim3.}
        \label{Tim3_sup}
    \end{center}
    \end{subfigure}    
    \hfill
    \begin{subfigure}[b]{0.33\columnwidth}
    \begin{center}
        \includegraphics[width=0.85\linewidth]{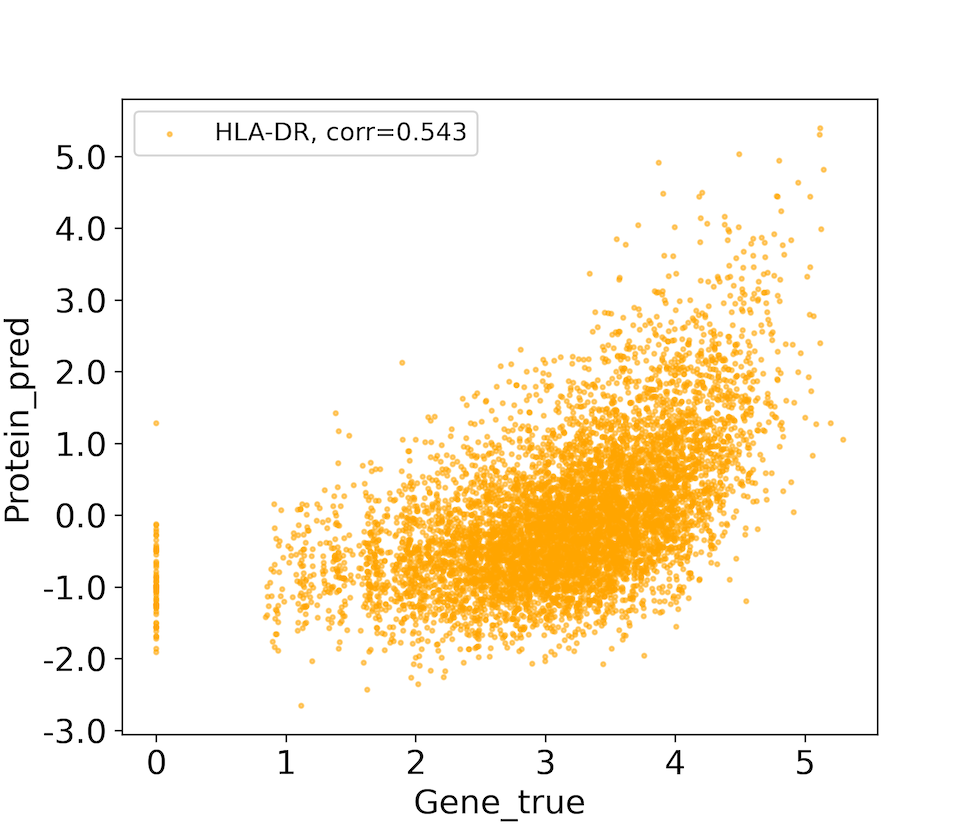}
        \caption{Scatter plot of HLA-DR.}
        \label{HLA_sup}
    \end{center}
    \end{subfigure}    
    \caption{Scatter plots of gene and predicted protein expression.}
    \label{t_emb}
    \end{center}
\end{figure}

\begin{figure}
    \begin{center}
    \begin{subfigure}[b]{0.35\columnwidth}
        \begin{center}
        \includegraphics[width=0.85\linewidth]{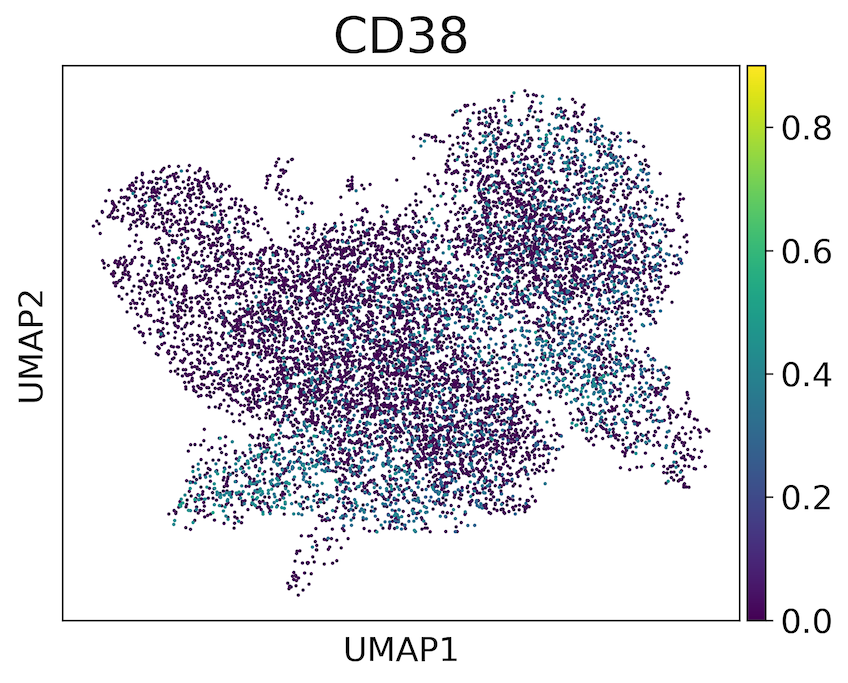}
        \caption{Feature plot of gene CD38.}
        \label{umap_CD38_gene}
        \end{center}
    \end{subfigure}%
    \hfill
    \begin{subfigure}[b]{0.35\columnwidth}
        \begin{center}
        \includegraphics[width=0.85\linewidth]{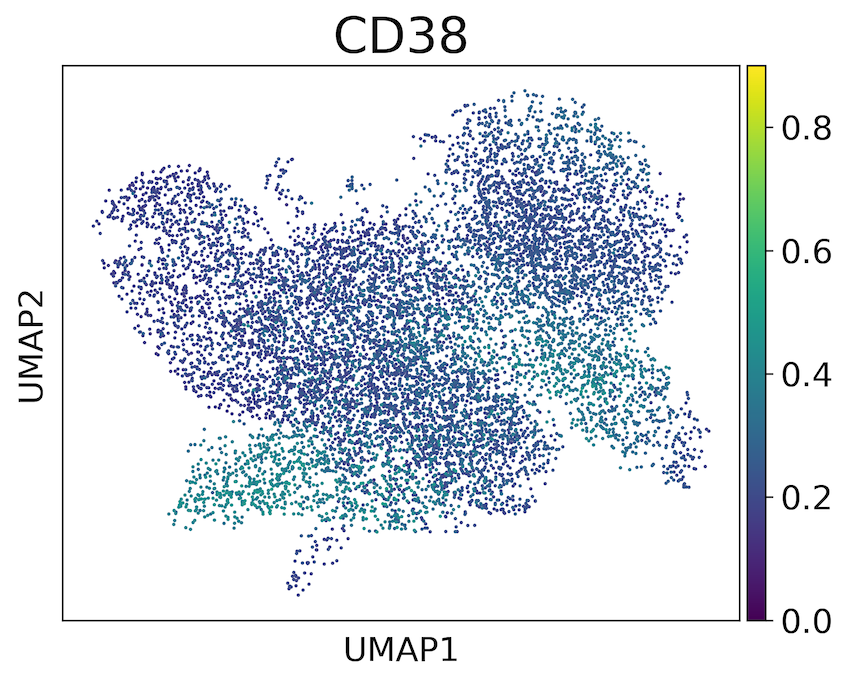}
        \caption{Feature plot of predicted protein CD38.}
        \label{umap_CD38_pred}
        \end{center}
    \end{subfigure}

    \begin{subfigure}[b]{0.35\columnwidth}
        \begin{center}
        \includegraphics[width=0.85\linewidth]{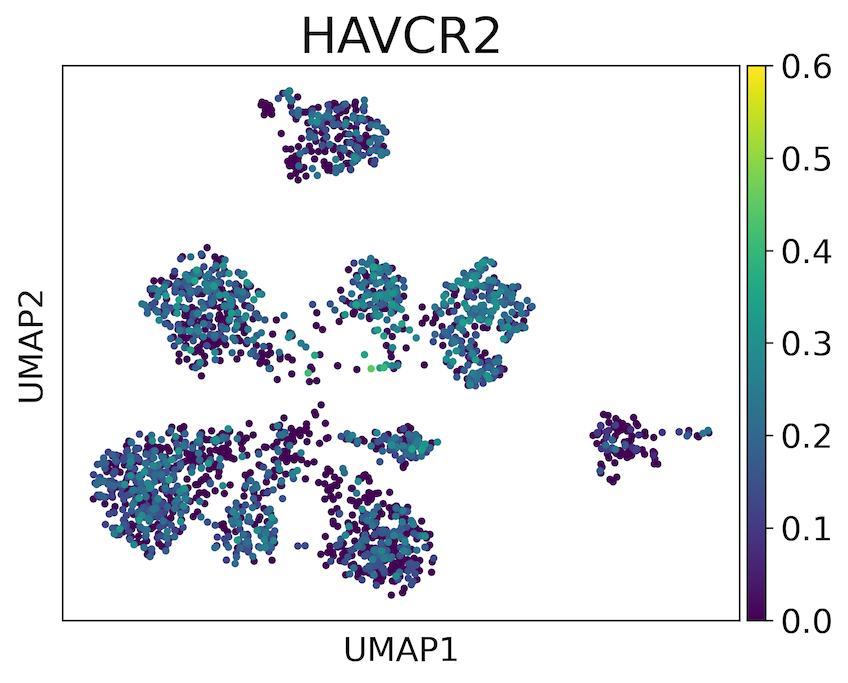}
        \caption{Feature plot of gene HAVCR2.}
        \label{umap_Tim3_gene}
        \end{center}
    \end{subfigure}%
    \hfill
    \begin{subfigure}[b]{0.35\columnwidth}
        \begin{center}
        \includegraphics[width=0.85\linewidth]{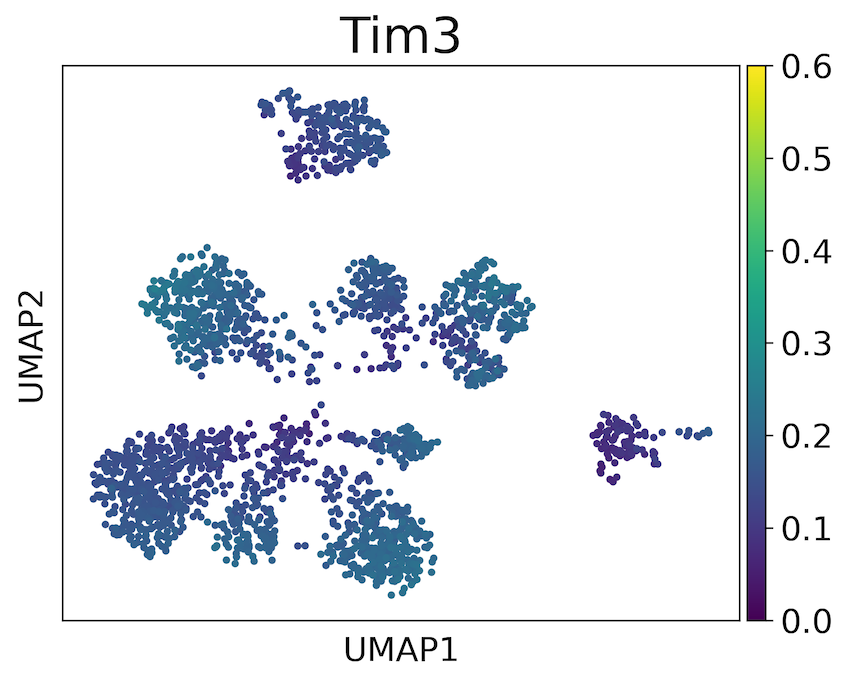}
        \caption{Feature plot of predicted protein Tim3.}
        \label{umap_Tim3_pred}
        \end{center}
    \end{subfigure}

    \begin{subfigure}[b]{0.35\columnwidth}
        \begin{center}
        \includegraphics[width=0.85\linewidth]{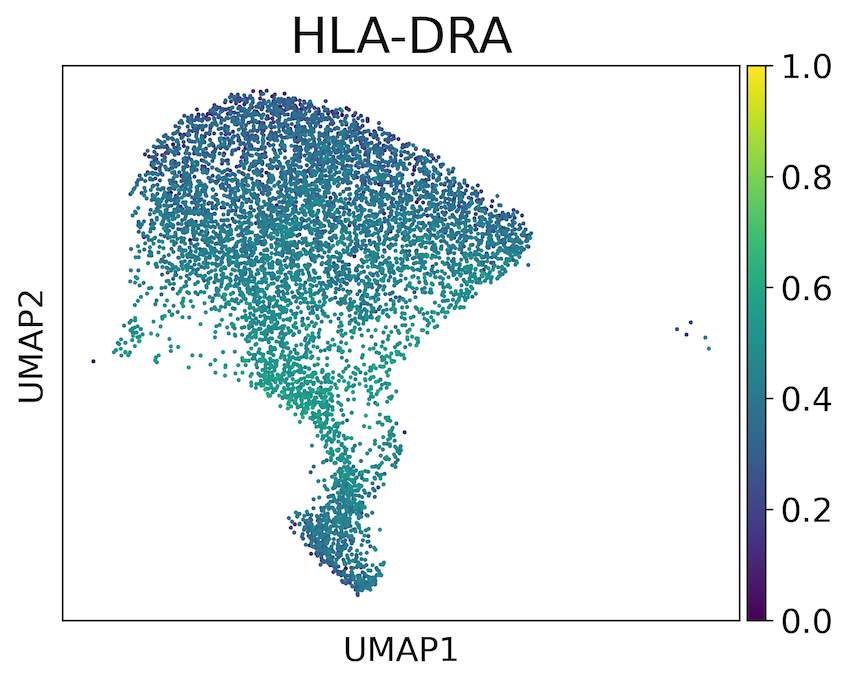}
        \caption{Feature plot of gene HLA-DRA.}
        \label{umap_HLA_gene}
        \end{center}
    \end{subfigure}%
    \hfill
    \begin{subfigure}[b]{0.35\columnwidth}
        \begin{center}
        \includegraphics[width=0.85\linewidth]{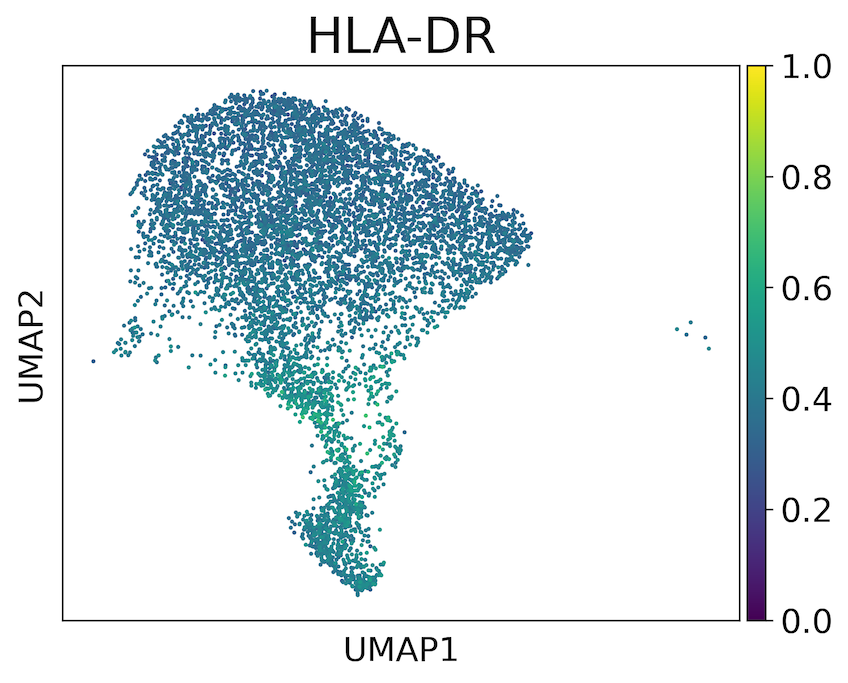}
        \caption{Feature plot of predicted protein HLA-DR.}
        \label{umap_HLA_pred}
        \end{center}
    \end{subfigure}
    
    \caption{Feature plots of markers. Left and right are normalized intensity-colored UMAP plots of gene expression and predicted protein expression in the testing data set, respectively.}
    \label{umap}
    \end{center}
\end{figure}

\begin{figure}
    \begin{center}
    \begin{subfigure}[b]{0.33\columnwidth}
    \begin{center}       
        \includegraphics[width=0.85\linewidth]{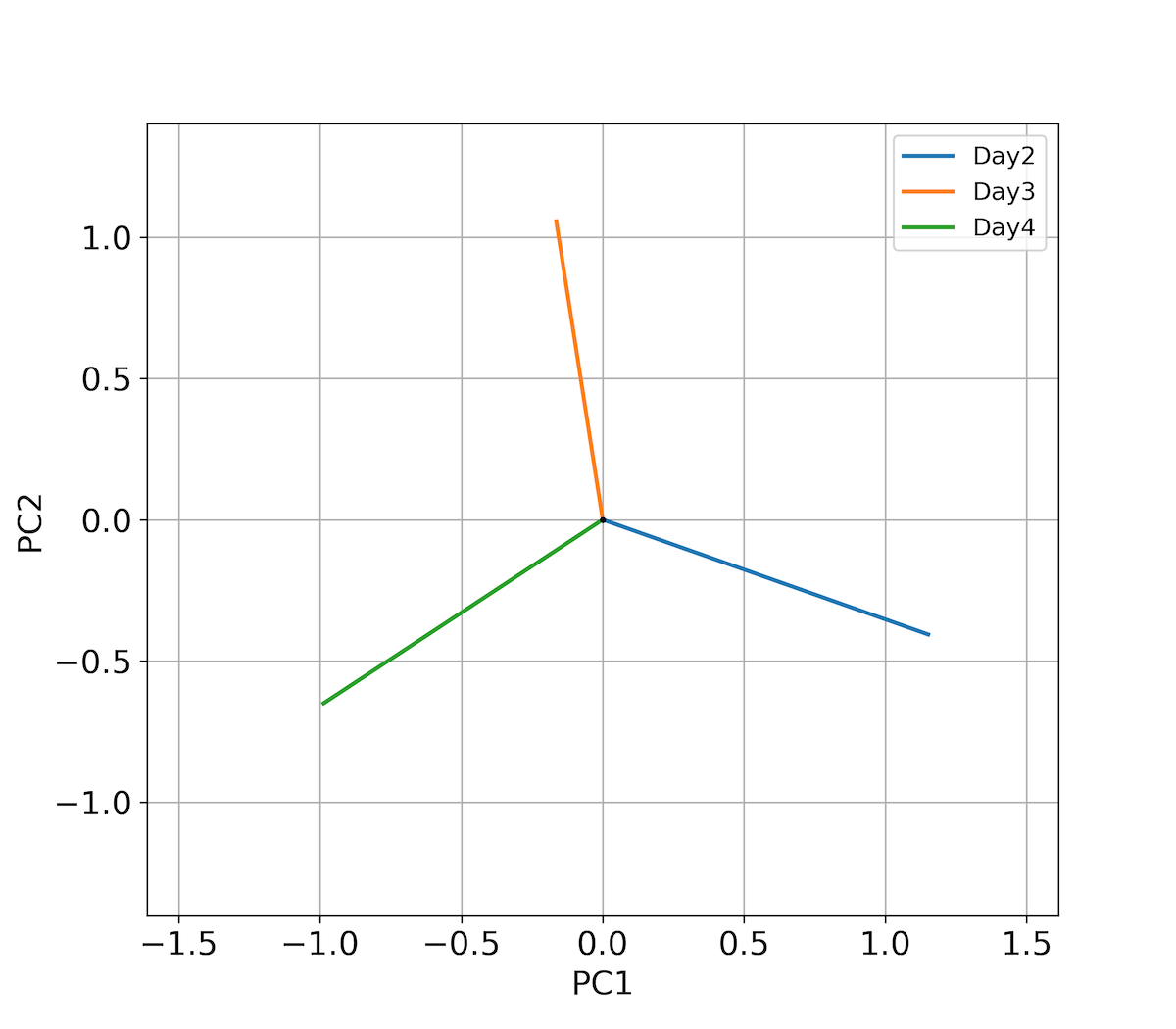}
        \caption{Latent time embedding trained on HSPC data.}
        \label{t_emb1}
    \end{center}
    \end{subfigure}%
    \hfill
    \begin{subfigure}[b]{0.33\columnwidth}
    \begin{center}
        \includegraphics[width=0.85\linewidth]{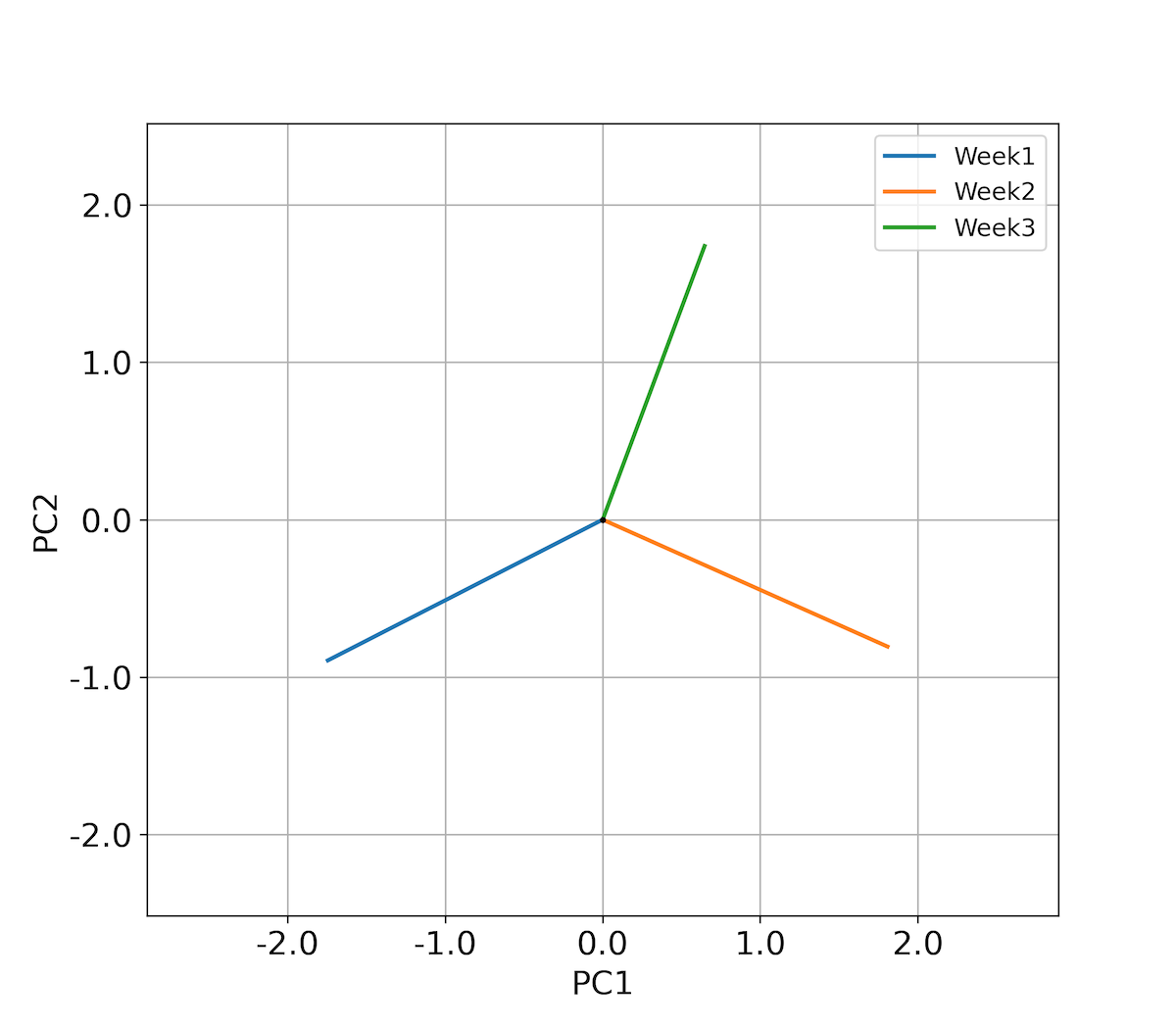}
        \caption{Latent time embedding trained on myeloid data.}
        \label{t_emb2}
    \end{center}
    \end{subfigure}    
    \hfill
    \begin{subfigure}[b]{0.33\columnwidth}
    \begin{center}
        \includegraphics[width=0.85\linewidth]{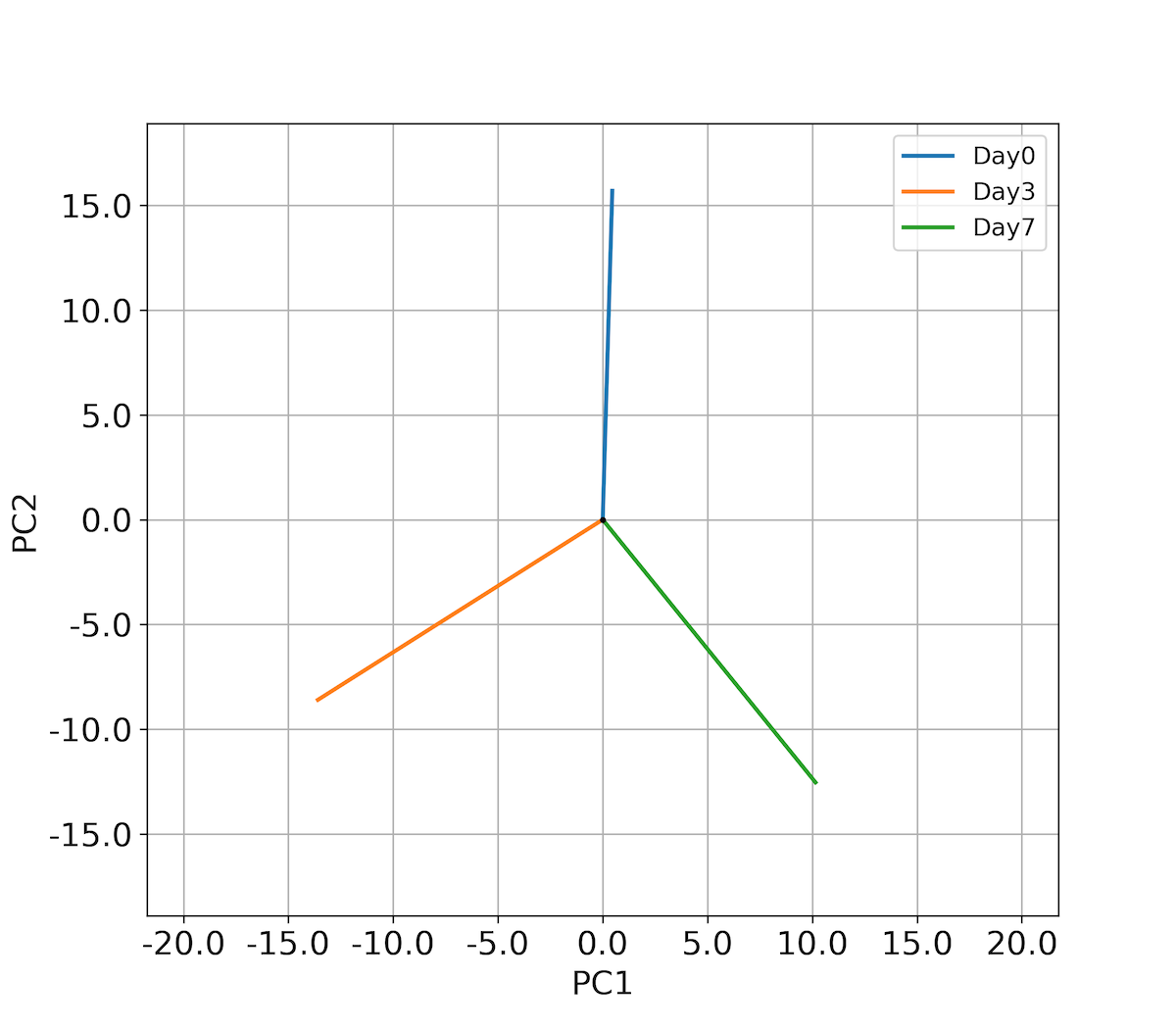}
        \caption{Latent time embedding trained on PBMC data.}
        \label{t_emb3}
    \end{center}
    \end{subfigure}    
    \caption{PCA representation of trained latent time embeddings.}
    \label{t_emb4}
    \end{center}
\end{figure}

\pagebreak{}

\end{appendices}

\end{document}